\documentclass[longauth]{aa}

\usepackage{graphicx}
\usepackage{natbib}
\usepackage{multirow}
\usepackage{placeins}

\usepackage{txfonts}
\usepackage{hyperref}
\hypersetup{
    colorlinks=true,
    linkcolor=blue,
    filecolor=blue,      
    citecolor=blue,
    urlcolor=blue
}

\makeatletter
\renewcommand*\aa@pageof{, page \thepage{} of \pageref*{LastPage}}
\makeatother

\usepackage[switch, modulo]{lineno}

\usepackage{orcidlink}
\usepackage{euclid}
\let\orcid\orcidlink

\newcommand{\eq}{Eq{.}~}

\newcommand{\fg}{Fig{.}~}
\newcommand{\fgs}{Figs{.}~}
\newcommand{\sct}{Sect{.}~}
\newcommand{\scts}{Sects{.}~}

\newcommand\T{\rule{0pt}{2.6ex}}       
\newcommand\B{\rule[-1.2ex]{0pt}{0pt}} 

\begin{document}

\title{Euclid Quick Data Release (Q1)}
\subtitle{Exploring galaxy morphology across cosmic time through S\'ersic fits}   

\author{Euclid Collaboration: L.~Quilley\orcid{0009-0008-8375-8605}\thanks{\email{louis.quilley@univ-lyon1.fr}}\inst{\ref{aff1}}
\and I.~Damjanov\orcid{0000-0003-4797-5246}\inst{\ref{aff2}}
\and V.~de~Lapparent\orcid{0009-0007-1622-1974}\inst{\ref{aff3}}
\and A.~Paulino-Afonso\orcid{0000-0002-0943-0694}\inst{\ref{aff4},\ref{aff5}}
\and H.~Dom\'inguez~S\'anchez\orcid{0000-0002-9013-1316}\inst{\ref{aff6}}
\and A.~Ferr\'e-Mateu\orcid{0000-0002-6411-220X}\inst{\ref{aff7},\ref{aff8}}
\and M.~Huertas-Company\orcid{0000-0002-1416-8483}\inst{\ref{aff7},\ref{aff9},\ref{aff10},\ref{aff11}}
\and M.~K\"ummel\orcid{0000-0003-2791-2117}\inst{\ref{aff12}}
\and D.~Delley\orcid{0000-0002-4958-7469}\inst{\ref{aff13}}
\and C.~Spiniello\orcid{0000-0002-3909-6359}\inst{\ref{aff14}}
\and M.~Baes\orcid{0000-0002-3930-2757}\inst{\ref{aff15}}
\and L.~Wang\orcid{0000-0002-6736-9158}\inst{\ref{aff16},\ref{aff17}}
\and U.~Kuchner\orcid{0000-0002-0035-5202}\inst{\ref{aff18}}
\and F.~Tarsitano\orcid{0000-0002-5919-0238}\inst{\ref{aff19}}
\and R.~Ragusa\inst{\ref{aff20}}
\and M.~Siudek\orcid{0000-0002-2949-2155}\inst{\ref{aff9},\ref{aff21}}
\and C.~Tortora\orcid{0000-0001-7958-6531}\inst{\ref{aff20}}
\and N.~Aghanim\orcid{0000-0002-6688-8992}\inst{\ref{aff22}}
\and B.~Altieri\orcid{0000-0003-3936-0284}\inst{\ref{aff23}}
\and A.~Amara\inst{\ref{aff24}}
\and S.~Andreon\orcid{0000-0002-2041-8784}\inst{\ref{aff25}}
\and N.~Auricchio\orcid{0000-0003-4444-8651}\inst{\ref{aff26}}
\and H.~Aussel\orcid{0000-0002-1371-5705}\inst{\ref{aff27}}
\and C.~Baccigalupi\orcid{0000-0002-8211-1630}\inst{\ref{aff28},\ref{aff29},\ref{aff30},\ref{aff31}}
\and M.~Baldi\orcid{0000-0003-4145-1943}\inst{\ref{aff32},\ref{aff26},\ref{aff33}}
\and A.~Balestra\orcid{0000-0002-6967-261X}\inst{\ref{aff34}}
\and S.~Bardelli\orcid{0000-0002-8900-0298}\inst{\ref{aff26}}
\and P.~Battaglia\orcid{0000-0002-7337-5909}\inst{\ref{aff26}}
\and R.~Bender\orcid{0000-0001-7179-0626}\inst{\ref{aff13},\ref{aff12}}
\and A.~Biviano\orcid{0000-0002-0857-0732}\inst{\ref{aff29},\ref{aff28}}
\and A.~Bonchi\orcid{0000-0002-2667-5482}\inst{\ref{aff35}}
\and D.~Bonino\orcid{0000-0002-3336-9977}\inst{\ref{aff36}}
\and E.~Branchini\orcid{0000-0002-0808-6908}\inst{\ref{aff37},\ref{aff38},\ref{aff25}}
\and M.~Brescia\orcid{0000-0001-9506-5680}\inst{\ref{aff39},\ref{aff20}}
\and J.~Brinchmann\orcid{0000-0003-4359-8797}\inst{\ref{aff5},\ref{aff40}}
\and S.~Camera\orcid{0000-0003-3399-3574}\inst{\ref{aff41},\ref{aff42},\ref{aff36}}
\and G.~Ca\~nas-Herrera\orcid{0000-0003-2796-2149}\inst{\ref{aff43},\ref{aff44},\ref{aff45}}
\and V.~Capobianco\orcid{0000-0002-3309-7692}\inst{\ref{aff36}}
\and C.~Carbone\orcid{0000-0003-0125-3563}\inst{\ref{aff46}}
\and J.~Carretero\orcid{0000-0002-3130-0204}\inst{\ref{aff47},\ref{aff48}}
\and S.~Casas\orcid{0000-0002-4751-5138}\inst{\ref{aff49}}
\and F.~J.~Castander\orcid{0000-0001-7316-4573}\inst{\ref{aff21},\ref{aff50}}
\and M.~Castellano\orcid{0000-0001-9875-8263}\inst{\ref{aff51}}
\and G.~Castignani\orcid{0000-0001-6831-0687}\inst{\ref{aff26}}
\and S.~Cavuoti\orcid{0000-0002-3787-4196}\inst{\ref{aff20},\ref{aff52}}
\and K.~C.~Chambers\orcid{0000-0001-6965-7789}\inst{\ref{aff53}}
\and A.~Cimatti\inst{\ref{aff54}}
\and C.~Colodro-Conde\inst{\ref{aff7}}
\and G.~Congedo\orcid{0000-0003-2508-0046}\inst{\ref{aff55}}
\and C.~J.~Conselice\orcid{0000-0003-1949-7638}\inst{\ref{aff56}}
\and L.~Conversi\orcid{0000-0002-6710-8476}\inst{\ref{aff57},\ref{aff23}}
\and Y.~Copin\orcid{0000-0002-5317-7518}\inst{\ref{aff58}}
\and A.~Costille\inst{\ref{aff59}}
\and F.~Courbin\orcid{0000-0003-0758-6510}\inst{\ref{aff60},\ref{aff61}}
\and H.~M.~Courtois\orcid{0000-0003-0509-1776}\inst{\ref{aff62}}
\and M.~Cropper\orcid{0000-0003-4571-9468}\inst{\ref{aff63}}
\and A.~Da~Silva\orcid{0000-0002-6385-1609}\inst{\ref{aff64},\ref{aff65}}
\and H.~Degaudenzi\orcid{0000-0002-5887-6799}\inst{\ref{aff19}}
\and G.~De~Lucia\orcid{0000-0002-6220-9104}\inst{\ref{aff29}}
\and A.~M.~Di~Giorgio\orcid{0000-0002-4767-2360}\inst{\ref{aff66}}
\and C.~Dolding\orcid{0009-0003-7199-6108}\inst{\ref{aff63}}
\and H.~Dole\orcid{0000-0002-9767-3839}\inst{\ref{aff22}}
\and C.~A.~J.~Duncan\orcid{0009-0003-3573-0791}\inst{\ref{aff56}}
\and X.~Dupac\inst{\ref{aff23}}
\and S.~Dusini\orcid{0000-0002-1128-0664}\inst{\ref{aff67}}
\and S.~Escoffier\orcid{0000-0002-2847-7498}\inst{\ref{aff68}}
\and M.~Fabricius\orcid{0000-0002-7025-6058}\inst{\ref{aff13},\ref{aff12}}
\and M.~Farina\orcid{0000-0002-3089-7846}\inst{\ref{aff66}}
\and R.~Farinelli\inst{\ref{aff26}}
\and F.~Faustini\orcid{0000-0001-6274-5145}\inst{\ref{aff51},\ref{aff35}}
\and S.~Ferriol\inst{\ref{aff58}}
\and P.~Fosalba\orcid{0000-0002-1510-5214}\inst{\ref{aff50},\ref{aff21}}
\and S.~Fotopoulou\orcid{0000-0002-9686-254X}\inst{\ref{aff69}}
\and M.~Frailis\orcid{0000-0002-7400-2135}\inst{\ref{aff29}}
\and E.~Franceschi\orcid{0000-0002-0585-6591}\inst{\ref{aff26}}
\and S.~Galeotta\orcid{0000-0002-3748-5115}\inst{\ref{aff29}}
\and K.~George\orcid{0000-0002-1734-8455}\inst{\ref{aff12}}
\and B.~Gillis\orcid{0000-0002-4478-1270}\inst{\ref{aff55}}
\and C.~Giocoli\orcid{0000-0002-9590-7961}\inst{\ref{aff26},\ref{aff33}}
\and J.~Gracia-Carpio\inst{\ref{aff13}}
\and B.~R.~Granett\orcid{0000-0003-2694-9284}\inst{\ref{aff25}}
\and A.~Grazian\orcid{0000-0002-5688-0663}\inst{\ref{aff34}}
\and F.~Grupp\inst{\ref{aff13},\ref{aff12}}
\and S.~V.~H.~Haugan\orcid{0000-0001-9648-7260}\inst{\ref{aff70}}
\and H.~Hoekstra\orcid{0000-0002-0641-3231}\inst{\ref{aff45}}
\and W.~Holmes\inst{\ref{aff71}}
\and I.~M.~Hook\orcid{0000-0002-2960-978X}\inst{\ref{aff72}}
\and F.~Hormuth\inst{\ref{aff73}}
\and A.~Hornstrup\orcid{0000-0002-3363-0936}\inst{\ref{aff74},\ref{aff75}}
\and P.~Hudelot\inst{\ref{aff3}}
\and K.~Jahnke\orcid{0000-0003-3804-2137}\inst{\ref{aff76}}
\and M.~Jhabvala\inst{\ref{aff77}}
\and B.~Joachimi\orcid{0000-0001-7494-1303}\inst{\ref{aff78}}
\and E.~Keih\"anen\orcid{0000-0003-1804-7715}\inst{\ref{aff79}}
\and S.~Kermiche\orcid{0000-0002-0302-5735}\inst{\ref{aff68}}
\and A.~Kiessling\orcid{0000-0002-2590-1273}\inst{\ref{aff71}}
\and B.~Kubik\orcid{0009-0006-5823-4880}\inst{\ref{aff58}}
\and M.~Kunz\orcid{0000-0002-3052-7394}\inst{\ref{aff80}}
\and H.~Kurki-Suonio\orcid{0000-0002-4618-3063}\inst{\ref{aff81},\ref{aff82}}
\and Q.~Le~Boulc'h\inst{\ref{aff83}}
\and A.~M.~C.~Le~Brun\orcid{0000-0002-0936-4594}\inst{\ref{aff84}}
\and D.~Le~Mignant\orcid{0000-0002-5339-5515}\inst{\ref{aff59}}
\and P.~Liebing\inst{\ref{aff63}}
\and S.~Ligori\orcid{0000-0003-4172-4606}\inst{\ref{aff36}}
\and P.~B.~Lilje\orcid{0000-0003-4324-7794}\inst{\ref{aff70}}
\and V.~Lindholm\orcid{0000-0003-2317-5471}\inst{\ref{aff81},\ref{aff82}}
\and I.~Lloro\orcid{0000-0001-5966-1434}\inst{\ref{aff85}}
\and G.~Mainetti\orcid{0000-0003-2384-2377}\inst{\ref{aff83}}
\and D.~Maino\inst{\ref{aff86},\ref{aff46},\ref{aff87}}
\and E.~Maiorano\orcid{0000-0003-2593-4355}\inst{\ref{aff26}}
\and O.~Mansutti\orcid{0000-0001-5758-4658}\inst{\ref{aff29}}
\and S.~Marcin\inst{\ref{aff88}}
\and O.~Marggraf\orcid{0000-0001-7242-3852}\inst{\ref{aff89}}
\and M.~Martinelli\orcid{0000-0002-6943-7732}\inst{\ref{aff51},\ref{aff90}}
\and N.~Martinet\orcid{0000-0003-2786-7790}\inst{\ref{aff59}}
\and F.~Marulli\orcid{0000-0002-8850-0303}\inst{\ref{aff91},\ref{aff26},\ref{aff33}}
\and R.~Massey\orcid{0000-0002-6085-3780}\inst{\ref{aff92}}
\and S.~Maurogordato\inst{\ref{aff93}}
\and H.~J.~McCracken\orcid{0000-0002-9489-7765}\inst{\ref{aff3}}
\and E.~Medinaceli\orcid{0000-0002-4040-7783}\inst{\ref{aff26}}
\and S.~Mei\orcid{0000-0002-2849-559X}\inst{\ref{aff94},\ref{aff95}}
\and M.~Melchior\inst{\ref{aff96}}
\and Y.~Mellier\inst{\ref{aff97},\ref{aff3}}
\and M.~Meneghetti\orcid{0000-0003-1225-7084}\inst{\ref{aff26},\ref{aff33}}
\and E.~Merlin\orcid{0000-0001-6870-8900}\inst{\ref{aff51}}
\and G.~Meylan\inst{\ref{aff98}}
\and A.~Mora\orcid{0000-0002-1922-8529}\inst{\ref{aff99}}
\and M.~Moresco\orcid{0000-0002-7616-7136}\inst{\ref{aff91},\ref{aff26}}
\and L.~Moscardini\orcid{0000-0002-3473-6716}\inst{\ref{aff91},\ref{aff26},\ref{aff33}}
\and R.~Nakajima\orcid{0009-0009-1213-7040}\inst{\ref{aff89}}
\and C.~Neissner\orcid{0000-0001-8524-4968}\inst{\ref{aff100},\ref{aff48}}
\and S.-M.~Niemi\inst{\ref{aff43}}
\and J.~W.~Nightingale\orcid{0000-0002-8987-7401}\inst{\ref{aff101}}
\and C.~Padilla\orcid{0000-0001-7951-0166}\inst{\ref{aff100}}
\and S.~Paltani\orcid{0000-0002-8108-9179}\inst{\ref{aff19}}
\and F.~Pasian\orcid{0000-0002-4869-3227}\inst{\ref{aff29}}
\and K.~Pedersen\inst{\ref{aff102}}
\and W.~J.~Percival\orcid{0000-0002-0644-5727}\inst{\ref{aff103},\ref{aff104},\ref{aff105}}
\and V.~Pettorino\inst{\ref{aff43}}
\and S.~Pires\orcid{0000-0002-0249-2104}\inst{\ref{aff27}}
\and G.~Polenta\orcid{0000-0003-4067-9196}\inst{\ref{aff35}}
\and M.~Poncet\inst{\ref{aff106}}
\and L.~A.~Popa\inst{\ref{aff107}}
\and L.~Pozzetti\orcid{0000-0001-7085-0412}\inst{\ref{aff26}}
\and F.~Raison\orcid{0000-0002-7819-6918}\inst{\ref{aff13}}
\and R.~Rebolo\inst{\ref{aff7},\ref{aff108},\ref{aff8}}
\and A.~Renzi\orcid{0000-0001-9856-1970}\inst{\ref{aff109},\ref{aff67}}
\and J.~Rhodes\orcid{0000-0002-4485-8549}\inst{\ref{aff71}}
\and G.~Riccio\inst{\ref{aff20}}
\and E.~Romelli\orcid{0000-0003-3069-9222}\inst{\ref{aff29}}
\and M.~Roncarelli\orcid{0000-0001-9587-7822}\inst{\ref{aff26}}
\and R.~Saglia\orcid{0000-0003-0378-7032}\inst{\ref{aff12},\ref{aff13}}
\and Z.~Sakr\orcid{0000-0002-4823-3757}\inst{\ref{aff110},\ref{aff111},\ref{aff112}}
\and A.~G.~S\'anchez\orcid{0000-0003-1198-831X}\inst{\ref{aff13}}
\and D.~Sapone\orcid{0000-0001-7089-4503}\inst{\ref{aff113}}
\and B.~Sartoris\orcid{0000-0003-1337-5269}\inst{\ref{aff12},\ref{aff29}}
\and J.~A.~Schewtschenko\orcid{0000-0002-4913-6393}\inst{\ref{aff55}}
\and P.~Schneider\orcid{0000-0001-8561-2679}\inst{\ref{aff89}}
\and M.~Scodeggio\inst{\ref{aff46}}
\and A.~Secroun\orcid{0000-0003-0505-3710}\inst{\ref{aff68}}
\and G.~Seidel\orcid{0000-0003-2907-353X}\inst{\ref{aff76}}
\and M.~Seiffert\orcid{0000-0002-7536-9393}\inst{\ref{aff71}}
\and S.~Serrano\orcid{0000-0002-0211-2861}\inst{\ref{aff50},\ref{aff114},\ref{aff21}}
\and P.~Simon\inst{\ref{aff89}}
\and C.~Sirignano\orcid{0000-0002-0995-7146}\inst{\ref{aff109},\ref{aff67}}
\and G.~Sirri\orcid{0000-0003-2626-2853}\inst{\ref{aff33}}
\and L.~Stanco\orcid{0000-0002-9706-5104}\inst{\ref{aff67}}
\and J.~Steinwagner\orcid{0000-0001-7443-1047}\inst{\ref{aff13}}
\and P.~Tallada-Cresp\'{i}\orcid{0000-0002-1336-8328}\inst{\ref{aff47},\ref{aff48}}
\and A.~N.~Taylor\inst{\ref{aff55}}
\and H.~I.~Teplitz\orcid{0000-0002-7064-5424}\inst{\ref{aff115}}
\and I.~Tereno\inst{\ref{aff64},\ref{aff116}}
\and N.~Tessore\orcid{0000-0002-9696-7931}\inst{\ref{aff78}}
\and S.~Toft\orcid{0000-0003-3631-7176}\inst{\ref{aff117},\ref{aff118}}
\and R.~Toledo-Moreo\orcid{0000-0002-2997-4859}\inst{\ref{aff119}}
\and F.~Torradeflot\orcid{0000-0003-1160-1517}\inst{\ref{aff48},\ref{aff47}}
\and I.~Tutusaus\orcid{0000-0002-3199-0399}\inst{\ref{aff111}}
\and L.~Valenziano\orcid{0000-0002-1170-0104}\inst{\ref{aff26},\ref{aff120}}
\and J.~Valiviita\orcid{0000-0001-6225-3693}\inst{\ref{aff81},\ref{aff82}}
\and T.~Vassallo\orcid{0000-0001-6512-6358}\inst{\ref{aff12},\ref{aff29}}
\and G.~Verdoes~Kleijn\orcid{0000-0001-5803-2580}\inst{\ref{aff17}}
\and A.~Veropalumbo\orcid{0000-0003-2387-1194}\inst{\ref{aff25},\ref{aff38},\ref{aff37}}
\and Y.~Wang\orcid{0000-0002-4749-2984}\inst{\ref{aff115}}
\and J.~Weller\orcid{0000-0002-8282-2010}\inst{\ref{aff12},\ref{aff13}}
\and A.~Zacchei\orcid{0000-0003-0396-1192}\inst{\ref{aff29},\ref{aff28}}
\and G.~Zamorani\orcid{0000-0002-2318-301X}\inst{\ref{aff26}}
\and F.~M.~Zerbi\inst{\ref{aff25}}
\and I.~A.~Zinchenko\orcid{0000-0002-2944-2449}\inst{\ref{aff12}}
\and E.~Zucca\orcid{0000-0002-5845-8132}\inst{\ref{aff26}}
\and V.~Allevato\orcid{0000-0001-7232-5152}\inst{\ref{aff20}}
\and M.~Ballardini\orcid{0000-0003-4481-3559}\inst{\ref{aff121},\ref{aff122},\ref{aff26}}
\and M.~Bolzonella\orcid{0000-0003-3278-4607}\inst{\ref{aff26}}
\and E.~Bozzo\orcid{0000-0002-8201-1525}\inst{\ref{aff19}}
\and C.~Burigana\orcid{0000-0002-3005-5796}\inst{\ref{aff123},\ref{aff120}}
\and R.~Cabanac\orcid{0000-0001-6679-2600}\inst{\ref{aff111}}
\and A.~Cappi\inst{\ref{aff26},\ref{aff93}}
\and D.~Di~Ferdinando\inst{\ref{aff33}}
\and J.~A.~Escartin~Vigo\inst{\ref{aff13}}
\and L.~Gabarra\orcid{0000-0002-8486-8856}\inst{\ref{aff14}}
\and W.~G.~Hartley\inst{\ref{aff19}}
\and J.~Mart\'{i}n-Fleitas\orcid{0000-0002-8594-569X}\inst{\ref{aff99}}
\and S.~Matthew\orcid{0000-0001-8448-1697}\inst{\ref{aff55}}
\and N.~Mauri\orcid{0000-0001-8196-1548}\inst{\ref{aff54},\ref{aff33}}
\and R.~B.~Metcalf\orcid{0000-0003-3167-2574}\inst{\ref{aff91},\ref{aff26}}
\and A.~Pezzotta\orcid{0000-0003-0726-2268}\inst{\ref{aff124},\ref{aff13}}
\and M.~P\"ontinen\orcid{0000-0001-5442-2530}\inst{\ref{aff81}}
\and C.~Porciani\orcid{0000-0002-7797-2508}\inst{\ref{aff89}}
\and I.~Risso\orcid{0000-0003-2525-7761}\inst{\ref{aff125}}
\and V.~Scottez\inst{\ref{aff97},\ref{aff126}}
\and M.~Sereno\orcid{0000-0003-0302-0325}\inst{\ref{aff26},\ref{aff33}}
\and M.~Tenti\orcid{0000-0002-4254-5901}\inst{\ref{aff33}}
\and M.~Viel\orcid{0000-0002-2642-5707}\inst{\ref{aff28},\ref{aff29},\ref{aff31},\ref{aff30},\ref{aff127}}
\and M.~Wiesmann\orcid{0009-0000-8199-5860}\inst{\ref{aff70}}
\and Y.~Akrami\orcid{0000-0002-2407-7956}\inst{\ref{aff128},\ref{aff129}}
\and I.~T.~Andika\orcid{0000-0001-6102-9526}\inst{\ref{aff130},\ref{aff131}}
\and S.~Anselmi\orcid{0000-0002-3579-9583}\inst{\ref{aff67},\ref{aff109},\ref{aff132}}
\and M.~Archidiacono\orcid{0000-0003-4952-9012}\inst{\ref{aff86},\ref{aff87}}
\and F.~Atrio-Barandela\orcid{0000-0002-2130-2513}\inst{\ref{aff133}}
\and C.~Benoist\inst{\ref{aff93}}
\and K.~Benson\inst{\ref{aff63}}
\and D.~Bertacca\orcid{0000-0002-2490-7139}\inst{\ref{aff109},\ref{aff34},\ref{aff67}}
\and M.~Bethermin\orcid{0000-0002-3915-2015}\inst{\ref{aff134}}
\and L.~Bisigello\orcid{0000-0003-0492-4924}\inst{\ref{aff34}}
\and A.~Blanchard\orcid{0000-0001-8555-9003}\inst{\ref{aff111}}
\and L.~Blot\orcid{0000-0002-9622-7167}\inst{\ref{aff135},\ref{aff84}}
\and S.~Borgani\orcid{0000-0001-6151-6439}\inst{\ref{aff136},\ref{aff28},\ref{aff29},\ref{aff30},\ref{aff127}}
\and M.~L.~Brown\orcid{0000-0002-0370-8077}\inst{\ref{aff56}}
\and S.~Bruton\orcid{0000-0002-6503-5218}\inst{\ref{aff137}}
\and F.~Buitrago\orcid{0000-0002-2861-9812}\inst{\ref{aff138},\ref{aff116}}
\and A.~Calabro\orcid{0000-0003-2536-1614}\inst{\ref{aff51}}
\and B.~Camacho~Quevedo\orcid{0000-0002-8789-4232}\inst{\ref{aff50},\ref{aff21}}
\and F.~Caro\inst{\ref{aff51}}
\and C.~S.~Carvalho\inst{\ref{aff116}}
\and T.~Castro\orcid{0000-0002-6292-3228}\inst{\ref{aff29},\ref{aff30},\ref{aff28},\ref{aff127}}
\and Y.~Charles\inst{\ref{aff59}}
\and F.~Cogato\orcid{0000-0003-4632-6113}\inst{\ref{aff91},\ref{aff26}}
\and T.~Contini\orcid{0000-0003-0275-938X}\inst{\ref{aff111}}
\and A.~R.~Cooray\orcid{0000-0002-3892-0190}\inst{\ref{aff139}}
\and O.~Cucciati\orcid{0000-0002-9336-7551}\inst{\ref{aff26}}
\and S.~Davini\orcid{0000-0003-3269-1718}\inst{\ref{aff38}}
\and F.~De~Paolis\orcid{0000-0001-6460-7563}\inst{\ref{aff140},\ref{aff141},\ref{aff142}}
\and G.~Desprez\orcid{0000-0001-8325-1742}\inst{\ref{aff17}}
\and A.~D\'iaz-S\'anchez\orcid{0000-0003-0748-4768}\inst{\ref{aff143}}
\and J.~J.~Diaz\inst{\ref{aff7}}
\and S.~Di~Domizio\orcid{0000-0003-2863-5895}\inst{\ref{aff37},\ref{aff38}}
\and J.~M.~Diego\orcid{0000-0001-9065-3926}\inst{\ref{aff6}}
\and P.-A.~Duc\orcid{0000-0003-3343-6284}\inst{\ref{aff134}}
\and A.~Enia\orcid{0000-0002-0200-2857}\inst{\ref{aff32},\ref{aff26}}
\and Y.~Fang\inst{\ref{aff12}}
\and A.~G.~Ferrari\orcid{0009-0005-5266-4110}\inst{\ref{aff33}}
\and P.~G.~Ferreira\orcid{0000-0002-3021-2851}\inst{\ref{aff14}}
\and A.~Finoguenov\orcid{0000-0002-4606-5403}\inst{\ref{aff81}}
\and A.~Franco\orcid{0000-0002-4761-366X}\inst{\ref{aff141},\ref{aff140},\ref{aff142}}
\and K.~Ganga\orcid{0000-0001-8159-8208}\inst{\ref{aff94}}
\and J.~Garc\'ia-Bellido\orcid{0000-0002-9370-8360}\inst{\ref{aff128}}
\and T.~Gasparetto\orcid{0000-0002-7913-4866}\inst{\ref{aff29}}
\and V.~Gautard\inst{\ref{aff144}}
\and R.~Gavazzi\orcid{0000-0002-5540-6935}\inst{\ref{aff59},\ref{aff3}}
\and E.~Gaztanaga\orcid{0000-0001-9632-0815}\inst{\ref{aff21},\ref{aff50},\ref{aff145}}
\and F.~Giacomini\orcid{0000-0002-3129-2814}\inst{\ref{aff33}}
\and G.~Gozaliasl\orcid{0000-0002-0236-919X}\inst{\ref{aff146},\ref{aff81}}
\and M.~Guidi\orcid{0000-0001-9408-1101}\inst{\ref{aff32},\ref{aff26}}
\and C.~M.~Gutierrez\orcid{0000-0001-7854-783X}\inst{\ref{aff147}}
\and A.~Hall\orcid{0000-0002-3139-8651}\inst{\ref{aff55}}
\and C.~Hern\'andez-Monteagudo\orcid{0000-0001-5471-9166}\inst{\ref{aff8},\ref{aff7}}
\and H.~Hildebrandt\orcid{0000-0002-9814-3338}\inst{\ref{aff148}}
\and J.~Hjorth\orcid{0000-0002-4571-2306}\inst{\ref{aff102}}
\and O.~Ilbert\orcid{0000-0002-7303-4397}\inst{\ref{aff59}}
\and J.~J.~E.~Kajava\orcid{0000-0002-3010-8333}\inst{\ref{aff149},\ref{aff150}}
\and Y.~Kang\orcid{0009-0000-8588-7250}\inst{\ref{aff19}}
\and V.~Kansal\orcid{0000-0002-4008-6078}\inst{\ref{aff151},\ref{aff152}}
\and D.~Karagiannis\orcid{0000-0002-4927-0816}\inst{\ref{aff121},\ref{aff153}}
\and K.~Kiiveri\inst{\ref{aff79}}
\and C.~C.~Kirkpatrick\inst{\ref{aff79}}
\and S.~Kruk\orcid{0000-0001-8010-8879}\inst{\ref{aff23}}
\and L.~Legrand\orcid{0000-0003-0610-5252}\inst{\ref{aff154},\ref{aff155}}
\and M.~Lembo\orcid{0000-0002-5271-5070}\inst{\ref{aff121},\ref{aff122}}
\and F.~Lepori\orcid{0009-0000-5061-7138}\inst{\ref{aff156}}
\and G.~Leroy\orcid{0009-0004-2523-4425}\inst{\ref{aff157},\ref{aff92}}
\and G.~F.~Lesci\orcid{0000-0002-4607-2830}\inst{\ref{aff91},\ref{aff26}}
\and J.~Lesgourgues\orcid{0000-0001-7627-353X}\inst{\ref{aff49}}
\and L.~Leuzzi\orcid{0009-0006-4479-7017}\inst{\ref{aff91},\ref{aff26}}
\and T.~I.~Liaudat\orcid{0000-0002-9104-314X}\inst{\ref{aff158}}
\and A.~Loureiro\orcid{0000-0002-4371-0876}\inst{\ref{aff159},\ref{aff160}}
\and J.~Macias-Perez\orcid{0000-0002-5385-2763}\inst{\ref{aff161}}
\and G.~Maggio\orcid{0000-0003-4020-4836}\inst{\ref{aff29}}
\and M.~Magliocchetti\orcid{0000-0001-9158-4838}\inst{\ref{aff66}}
\and E.~A.~Magnier\orcid{0000-0002-7965-2815}\inst{\ref{aff53}}
\and C.~Mancini\orcid{0000-0002-4297-0561}\inst{\ref{aff46}}
\and F.~Mannucci\orcid{0000-0002-4803-2381}\inst{\ref{aff162}}
\and R.~Maoli\orcid{0000-0002-6065-3025}\inst{\ref{aff163},\ref{aff51}}
\and C.~J.~A.~P.~Martins\orcid{0000-0002-4886-9261}\inst{\ref{aff4},\ref{aff5}}
\and L.~Maurin\orcid{0000-0002-8406-0857}\inst{\ref{aff22}}
\and M.~Miluzio\inst{\ref{aff23},\ref{aff164}}
\and P.~Monaco\orcid{0000-0003-2083-7564}\inst{\ref{aff136},\ref{aff29},\ref{aff30},\ref{aff28}}
\and C.~Moretti\orcid{0000-0003-3314-8936}\inst{\ref{aff31},\ref{aff127},\ref{aff29},\ref{aff28},\ref{aff30}}
\and G.~Morgante\inst{\ref{aff26}}
\and K.~Naidoo\orcid{0000-0002-9182-1802}\inst{\ref{aff145}}
\and A.~Navarro-Alsina\orcid{0000-0002-3173-2592}\inst{\ref{aff89}}
\and S.~Nesseris\orcid{0000-0002-0567-0324}\inst{\ref{aff128}}
\and F.~Passalacqua\orcid{0000-0002-8606-4093}\inst{\ref{aff109},\ref{aff67}}
\and K.~Paterson\orcid{0000-0001-8340-3486}\inst{\ref{aff76}}
\and L.~Patrizii\inst{\ref{aff33}}
\and A.~Pisani\orcid{0000-0002-6146-4437}\inst{\ref{aff68},\ref{aff165}}
\and D.~Potter\orcid{0000-0002-0757-5195}\inst{\ref{aff156}}
\and S.~Quai\orcid{0000-0002-0449-8163}\inst{\ref{aff91},\ref{aff26}}
\and M.~Radovich\orcid{0000-0002-3585-866X}\inst{\ref{aff34}}
\and P.-F.~Rocci\inst{\ref{aff22}}
\and S.~Sacquegna\orcid{0000-0002-8433-6630}\inst{\ref{aff140},\ref{aff141},\ref{aff142}}
\and M.~Sahl\'en\orcid{0000-0003-0973-4804}\inst{\ref{aff166}}
\and D.~B.~Sanders\orcid{0000-0002-1233-9998}\inst{\ref{aff53}}
\and E.~Sarpa\orcid{0000-0002-1256-655X}\inst{\ref{aff31},\ref{aff127},\ref{aff30}}
\and C.~Scarlata\orcid{0000-0002-9136-8876}\inst{\ref{aff167}}
\and J.~Schaye\orcid{0000-0002-0668-5560}\inst{\ref{aff45}}
\and A.~Schneider\orcid{0000-0001-7055-8104}\inst{\ref{aff156}}
\and D.~Sciotti\orcid{0009-0008-4519-2620}\inst{\ref{aff51},\ref{aff90}}
\and E.~Sellentin\inst{\ref{aff168},\ref{aff45}}
\and F.~Shankar\orcid{0000-0001-8973-5051}\inst{\ref{aff169}}
\and L.~C.~Smith\orcid{0000-0002-3259-2771}\inst{\ref{aff170}}
\and S.~A.~Stanford\orcid{0000-0003-0122-0841}\inst{\ref{aff171}}
\and K.~Tanidis\orcid{0000-0001-9843-5130}\inst{\ref{aff14}}
\and G.~Testera\inst{\ref{aff38}}
\and R.~Teyssier\orcid{0000-0001-7689-0933}\inst{\ref{aff165}}
\and S.~Tosi\orcid{0000-0002-7275-9193}\inst{\ref{aff37},\ref{aff38},\ref{aff25}}
\and A.~Troja\orcid{0000-0003-0239-4595}\inst{\ref{aff109},\ref{aff67}}
\and M.~Tucci\inst{\ref{aff19}}
\and C.~Valieri\inst{\ref{aff33}}
\and A.~Venhola\orcid{0000-0001-6071-4564}\inst{\ref{aff172}}
\and D.~Vergani\orcid{0000-0003-0898-2216}\inst{\ref{aff26}}
\and G.~Verza\orcid{0000-0002-1886-8348}\inst{\ref{aff173}}
\and P.~Vielzeuf\orcid{0000-0003-2035-9339}\inst{\ref{aff68}}
\and N.~A.~Walton\orcid{0000-0003-3983-8778}\inst{\ref{aff170}}
\and J.~G.~Sorce\orcid{0000-0002-2307-2432}\inst{\ref{aff174},\ref{aff22}}
\and E.~Soubrie\orcid{0000-0001-9295-1863}\inst{\ref{aff22}}
\and D.~Scott\orcid{0000-0002-6878-9840}\inst{\ref{aff175}}}

\institute{Centre de Recherche Astrophysique de Lyon, UMR5574, CNRS, Universit\'e Claude Bernard Lyon 1, ENS de Lyon, 69230, Saint-Genis-Laval, France\label{aff1}
\and
Department of Astronomy \& Physics and Institute for Computational Astrophysics, Saint Mary's University, 923 Robie Street, Halifax, Nova Scotia, B3H 3C3, Canada\label{aff2}
\and
Institut d'Astrophysique de Paris, UMR 7095, CNRS, and Sorbonne Universit\'e, 98 bis boulevard Arago, 75014 Paris, France\label{aff3}
\and
Centro de Astrof\'{\i}sica da Universidade do Porto, Rua das Estrelas, 4150-762 Porto, Portugal\label{aff4}
\and
Instituto de Astrof\'isica e Ci\^encias do Espa\c{c}o, Universidade do Porto, CAUP, Rua das Estrelas, PT4150-762 Porto, Portugal\label{aff5}
\and
Instituto de F\'isica de Cantabria, Edificio Juan Jord\'a, Avenida de los Castros, 39005 Santander, Spain\label{aff6}
\and
Instituto de Astrof\'{\i}sica de Canarias, V\'{\i}a L\'actea, 38205 La Laguna, Tenerife, Spain\label{aff7}
\and
Universidad de La Laguna, Departamento de Astrof\'{\i}sica, 38206 La Laguna, Tenerife, Spain\label{aff8}
\and
Instituto de Astrof\'isica de Canarias (IAC); Departamento de Astrof\'isica, Universidad de La Laguna (ULL), 38200, La Laguna, Tenerife, Spain\label{aff9}
\and
Universit\'e PSL, Observatoire de Paris, Sorbonne Universit\'e, CNRS, LERMA, 75014, Paris, France\label{aff10}
\and
Universit\'e Paris-Cit\'e, 5 Rue Thomas Mann, 75013, Paris, France\label{aff11}
\and
Universit\"ats-Sternwarte M\"unchen, Fakult\"at f\"ur Physik, Ludwig-Maximilians-Universit\"at M\"unchen, Scheinerstrasse 1, 81679 M\"unchen, Germany\label{aff12}
\and
Max Planck Institute for Extraterrestrial Physics, Giessenbachstr. 1, 85748 Garching, Germany\label{aff13}
\and
Department of Physics, Oxford University, Keble Road, Oxford OX1 3RH, UK\label{aff14}
\and
Sterrenkundig Observatorium, Universiteit Gent, Krijgslaan 281 S9, 9000 Gent, Belgium\label{aff15}
\and
SRON Netherlands Institute for Space Research, Landleven 12, 9747 AD, Groningen, The Netherlands\label{aff16}
\and
Kapteyn Astronomical Institute, University of Groningen, PO Box 800, 9700 AV Groningen, The Netherlands\label{aff17}
\and
School of Physics and Astronomy, University of Nottingham, University Park, Nottingham NG7 2RD, UK\label{aff18}
\and
Department of Astronomy, University of Geneva, ch. d'Ecogia 16, 1290 Versoix, Switzerland\label{aff19}
\and
INAF-Osservatorio Astronomico di Capodimonte, Via Moiariello 16, 80131 Napoli, Italy\label{aff20}
\and
Institute of Space Sciences (ICE, CSIC), Campus UAB, Carrer de Can Magrans, s/n, 08193 Barcelona, Spain\label{aff21}
\and
Universit\'e Paris-Saclay, CNRS, Institut d'astrophysique spatiale, 91405, Orsay, France\label{aff22}
\and
ESAC/ESA, Camino Bajo del Castillo, s/n., Urb. Villafranca del Castillo, 28692 Villanueva de la Ca\~nada, Madrid, Spain\label{aff23}
\and
School of Mathematics and Physics, University of Surrey, Guildford, Surrey, GU2 7XH, UK\label{aff24}
\and
INAF-Osservatorio Astronomico di Brera, Via Brera 28, 20122 Milano, Italy\label{aff25}
\and
INAF-Osservatorio di Astrofisica e Scienza dello Spazio di Bologna, Via Piero Gobetti 93/3, 40129 Bologna, Italy\label{aff26}
\and
Universit\'e Paris-Saclay, Universit\'e Paris Cit\'e, CEA, CNRS, AIM, 91191, Gif-sur-Yvette, France\label{aff27}
\and
IFPU, Institute for Fundamental Physics of the Universe, via Beirut 2, 34151 Trieste, Italy\label{aff28}
\and
INAF-Osservatorio Astronomico di Trieste, Via G. B. Tiepolo 11, 34143 Trieste, Italy\label{aff29}
\and
INFN, Sezione di Trieste, Via Valerio 2, 34127 Trieste TS, Italy\label{aff30}
\and
SISSA, International School for Advanced Studies, Via Bonomea 265, 34136 Trieste TS, Italy\label{aff31}
\and
Dipartimento di Fisica e Astronomia, Universit\`a di Bologna, Via Gobetti 93/2, 40129 Bologna, Italy\label{aff32}
\and
INFN-Sezione di Bologna, Viale Berti Pichat 6/2, 40127 Bologna, Italy\label{aff33}
\and
INAF-Osservatorio Astronomico di Padova, Via dell'Osservatorio 5, 35122 Padova, Italy\label{aff34}
\and
Space Science Data Center, Italian Space Agency, via del Politecnico snc, 00133 Roma, Italy\label{aff35}
\and
INAF-Osservatorio Astrofisico di Torino, Via Osservatorio 20, 10025 Pino Torinese (TO), Italy\label{aff36}
\and
Dipartimento di Fisica, Universit\`a di Genova, Via Dodecaneso 33, 16146, Genova, Italy\label{aff37}
\and
INFN-Sezione di Genova, Via Dodecaneso 33, 16146, Genova, Italy\label{aff38}
\and
Department of Physics "E. Pancini", University Federico II, Via Cinthia 6, 80126, Napoli, Italy\label{aff39}
\and
Faculdade de Ci\^encias da Universidade do Porto, Rua do Campo de Alegre, 4150-007 Porto, Portugal\label{aff40}
\and
Dipartimento di Fisica, Universit\`a degli Studi di Torino, Via P. Giuria 1, 10125 Torino, Italy\label{aff41}
\and
INFN-Sezione di Torino, Via P. Giuria 1, 10125 Torino, Italy\label{aff42}
\and
European Space Agency/ESTEC, Keplerlaan 1, 2201 AZ Noordwijk, The Netherlands\label{aff43}
\and
Institute Lorentz, Leiden University, Niels Bohrweg 2, 2333 CA Leiden, The Netherlands\label{aff44}
\and
Leiden Observatory, Leiden University, Einsteinweg 55, 2333 CC Leiden, The Netherlands\label{aff45}
\and
INAF-IASF Milano, Via Alfonso Corti 12, 20133 Milano, Italy\label{aff46}
\and
Centro de Investigaciones Energ\'eticas, Medioambientales y Tecnol\'ogicas (CIEMAT), Avenida Complutense 40, 28040 Madrid, Spain\label{aff47}
\and
Port d'Informaci\'{o} Cient\'{i}fica, Campus UAB, C. Albareda s/n, 08193 Bellaterra (Barcelona), Spain\label{aff48}
\and
Institute for Theoretical Particle Physics and Cosmology (TTK), RWTH Aachen University, 52056 Aachen, Germany\label{aff49}
\and
Institut d'Estudis Espacials de Catalunya (IEEC),  Edifici RDIT, Campus UPC, 08860 Castelldefels, Barcelona, Spain\label{aff50}
\and
INAF-Osservatorio Astronomico di Roma, Via Frascati 33, 00078 Monteporzio Catone, Italy\label{aff51}
\and
INFN section of Naples, Via Cinthia 6, 80126, Napoli, Italy\label{aff52}
\and
Institute for Astronomy, University of Hawaii, 2680 Woodlawn Drive, Honolulu, HI 96822, USA\label{aff53}
\and
Dipartimento di Fisica e Astronomia "Augusto Righi" - Alma Mater Studiorum Universit\`a di Bologna, Viale Berti Pichat 6/2, 40127 Bologna, Italy\label{aff54}
\and
Institute for Astronomy, University of Edinburgh, Royal Observatory, Blackford Hill, Edinburgh EH9 3HJ, UK\label{aff55}
\and
Jodrell Bank Centre for Astrophysics, Department of Physics and Astronomy, University of Manchester, Oxford Road, Manchester M13 9PL, UK\label{aff56}
\and
European Space Agency/ESRIN, Largo Galileo Galilei 1, 00044 Frascati, Roma, Italy\label{aff57}
\and
Universit\'e Claude Bernard Lyon 1, CNRS/IN2P3, IP2I Lyon, UMR 5822, Villeurbanne, F-69100, France\label{aff58}
\and
Aix-Marseille Universit\'e, CNRS, CNES, LAM, Marseille, France\label{aff59}
\and
Institut de Ci\`{e}ncies del Cosmos (ICCUB), Universitat de Barcelona (IEEC-UB), Mart\'{i} i Franqu\`{e}s 1, 08028 Barcelona, Spain\label{aff60}
\and
Instituci\'o Catalana de Recerca i Estudis Avan\c{c}ats (ICREA), Passeig de Llu\'{\i}s Companys 23, 08010 Barcelona, Spain\label{aff61}
\and
UCB Lyon 1, CNRS/IN2P3, IUF, IP2I Lyon, 4 rue Enrico Fermi, 69622 Villeurbanne, France\label{aff62}
\and
Mullard Space Science Laboratory, University College London, Holmbury St Mary, Dorking, Surrey RH5 6NT, UK\label{aff63}
\and
Departamento de F\'isica, Faculdade de Ci\^encias, Universidade de Lisboa, Edif\'icio C8, Campo Grande, PT1749-016 Lisboa, Portugal\label{aff64}
\and
Instituto de Astrof\'isica e Ci\^encias do Espa\c{c}o, Faculdade de Ci\^encias, Universidade de Lisboa, Campo Grande, 1749-016 Lisboa, Portugal\label{aff65}
\and
INAF-Istituto di Astrofisica e Planetologia Spaziali, via del Fosso del Cavaliere, 100, 00100 Roma, Italy\label{aff66}
\and
INFN-Padova, Via Marzolo 8, 35131 Padova, Italy\label{aff67}
\and
Aix-Marseille Universit\'e, CNRS/IN2P3, CPPM, Marseille, France\label{aff68}
\and
School of Physics, HH Wills Physics Laboratory, University of Bristol, Tyndall Avenue, Bristol, BS8 1TL, UK\label{aff69}
\and
Institute of Theoretical Astrophysics, University of Oslo, P.O. Box 1029 Blindern, 0315 Oslo, Norway\label{aff70}
\and
Jet Propulsion Laboratory, California Institute of Technology, 4800 Oak Grove Drive, Pasadena, CA, 91109, USA\label{aff71}
\and
Department of Physics, Lancaster University, Lancaster, LA1 4YB, UK\label{aff72}
\and
Felix Hormuth Engineering, Goethestr. 17, 69181 Leimen, Germany\label{aff73}
\and
Technical University of Denmark, Elektrovej 327, 2800 Kgs. Lyngby, Denmark\label{aff74}
\and
Cosmic Dawn Center (DAWN), Denmark\label{aff75}
\and
Max-Planck-Institut f\"ur Astronomie, K\"onigstuhl 17, 69117 Heidelberg, Germany\label{aff76}
\and
NASA Goddard Space Flight Center, Greenbelt, MD 20771, USA\label{aff77}
\and
Department of Physics and Astronomy, University College London, Gower Street, London WC1E 6BT, UK\label{aff78}
\and
Department of Physics and Helsinki Institute of Physics, Gustaf H\"allstr\"omin katu 2, 00014 University of Helsinki, Finland\label{aff79}
\and
Universit\'e de Gen\`eve, D\'epartement de Physique Th\'eorique and Centre for Astroparticle Physics, 24 quai Ernest-Ansermet, CH-1211 Gen\`eve 4, Switzerland\label{aff80}
\and
Department of Physics, P.O. Box 64, 00014 University of Helsinki, Finland\label{aff81}
\and
Helsinki Institute of Physics, Gustaf H{\"a}llstr{\"o}min katu 2, University of Helsinki, Helsinki, Finland\label{aff82}
\and
Centre de Calcul de l'IN2P3/CNRS, 21 avenue Pierre de Coubertin 69627 Villeurbanne Cedex, France\label{aff83}
\and
Laboratoire d'etude de l'Univers et des phenomenes eXtremes, Observatoire de Paris, Universit\'e PSL, Sorbonne Universit\'e, CNRS, 92190 Meudon, France\label{aff84}
\and
SKA Observatory, Jodrell Bank, Lower Withington, Macclesfield, Cheshire SK11 9FT, UK\label{aff85}
\and
Dipartimento di Fisica "Aldo Pontremoli", Universit\`a degli Studi di Milano, Via Celoria 16, 20133 Milano, Italy\label{aff86}
\and
INFN-Sezione di Milano, Via Celoria 16, 20133 Milano, Italy\label{aff87}
\and
University of Applied Sciences and Arts of Northwestern Switzerland, School of Computer Science, 5210 Windisch, Switzerland\label{aff88}
\and
Universit\"at Bonn, Argelander-Institut f\"ur Astronomie, Auf dem H\"ugel 71, 53121 Bonn, Germany\label{aff89}
\and
INFN-Sezione di Roma, Piazzale Aldo Moro, 2 - c/o Dipartimento di Fisica, Edificio G. Marconi, 00185 Roma, Italy\label{aff90}
\and
Dipartimento di Fisica e Astronomia "Augusto Righi" - Alma Mater Studiorum Universit\`a di Bologna, via Piero Gobetti 93/2, 40129 Bologna, Italy\label{aff91}
\and
Department of Physics, Institute for Computational Cosmology, Durham University, South Road, Durham, DH1 3LE, UK\label{aff92}
\and
Universit\'e C\^{o}te d'Azur, Observatoire de la C\^{o}te d'Azur, CNRS, Laboratoire Lagrange, Bd de l'Observatoire, CS 34229, 06304 Nice cedex 4, France\label{aff93}
\and
Universit\'e Paris Cit\'e, CNRS, Astroparticule et Cosmologie, 75013 Paris, France\label{aff94}
\and
CNRS-UCB International Research Laboratory, Centre Pierre Binetruy, IRL2007, CPB-IN2P3, Berkeley, USA\label{aff95}
\and
University of Applied Sciences and Arts of Northwestern Switzerland, School of Engineering, 5210 Windisch, Switzerland\label{aff96}
\and
Institut d'Astrophysique de Paris, 98bis Boulevard Arago, 75014, Paris, France\label{aff97}
\and
Institute of Physics, Laboratory of Astrophysics, Ecole Polytechnique F\'ed\'erale de Lausanne (EPFL), Observatoire de Sauverny, 1290 Versoix, Switzerland\label{aff98}
\and
Aurora Technology for European Space Agency (ESA), Camino bajo del Castillo, s/n, Urbanizacion Villafranca del Castillo, Villanueva de la Ca\~nada, 28692 Madrid, Spain\label{aff99}
\and
Institut de F\'{i}sica d'Altes Energies (IFAE), The Barcelona Institute of Science and Technology, Campus UAB, 08193 Bellaterra (Barcelona), Spain\label{aff100}
\and
School of Mathematics, Statistics and Physics, Newcastle University, Herschel Building, Newcastle-upon-Tyne, NE1 7RU, UK\label{aff101}
\and
DARK, Niels Bohr Institute, University of Copenhagen, Jagtvej 155, 2200 Copenhagen, Denmark\label{aff102}
\and
Waterloo Centre for Astrophysics, University of Waterloo, Waterloo, Ontario N2L 3G1, Canada\label{aff103}
\and
Department of Physics and Astronomy, University of Waterloo, Waterloo, Ontario N2L 3G1, Canada\label{aff104}
\and
Perimeter Institute for Theoretical Physics, Waterloo, Ontario N2L 2Y5, Canada\label{aff105}
\and
Centre National d'Etudes Spatiales -- Centre spatial de Toulouse, 18 avenue Edouard Belin, 31401 Toulouse Cedex 9, France\label{aff106}
\and
Institute of Space Science, Str. Atomistilor, nr. 409 M\u{a}gurele, Ilfov, 077125, Romania\label{aff107}
\and
Consejo Superior de Investigaciones Cientificas, Calle Serrano 117, 28006 Madrid, Spain\label{aff108}
\and
Dipartimento di Fisica e Astronomia "G. Galilei", Universit\`a di Padova, Via Marzolo 8, 35131 Padova, Italy\label{aff109}
\and
Institut f\"ur Theoretische Physik, University of Heidelberg, Philosophenweg 16, 69120 Heidelberg, Germany\label{aff110}
\and
Institut de Recherche en Astrophysique et Plan\'etologie (IRAP), Universit\'e de Toulouse, CNRS, UPS, CNES, 14 Av. Edouard Belin, 31400 Toulouse, France\label{aff111}
\and
Universit\'e St Joseph; Faculty of Sciences, Beirut, Lebanon\label{aff112}
\and
Departamento de F\'isica, FCFM, Universidad de Chile, Blanco Encalada 2008, Santiago, Chile\label{aff113}
\and
Satlantis, University Science Park, Sede Bld 48940, Leioa-Bilbao, Spain\label{aff114}
\and
Infrared Processing and Analysis Center, California Institute of Technology, Pasadena, CA 91125, USA\label{aff115}
\and
Instituto de Astrof\'isica e Ci\^encias do Espa\c{c}o, Faculdade de Ci\^encias, Universidade de Lisboa, Tapada da Ajuda, 1349-018 Lisboa, Portugal\label{aff116}
\and
Cosmic Dawn Center (DAWN)\label{aff117}
\and
Niels Bohr Institute, University of Copenhagen, Jagtvej 128, 2200 Copenhagen, Denmark\label{aff118}
\and
Universidad Polit\'ecnica de Cartagena, Departamento de Electr\'onica y Tecnolog\'ia de Computadoras,  Plaza del Hospital 1, 30202 Cartagena, Spain\label{aff119}
\and
INFN-Bologna, Via Irnerio 46, 40126 Bologna, Italy\label{aff120}
\and
Dipartimento di Fisica e Scienze della Terra, Universit\`a degli Studi di Ferrara, Via Giuseppe Saragat 1, 44122 Ferrara, Italy\label{aff121}
\and
Istituto Nazionale di Fisica Nucleare, Sezione di Ferrara, Via Giuseppe Saragat 1, 44122 Ferrara, Italy\label{aff122}
\and
INAF, Istituto di Radioastronomia, Via Piero Gobetti 101, 40129 Bologna, Italy\label{aff123}
\and
INAF - Osservatorio Astronomico di Brera, via Emilio Bianchi 46, 23807 Merate, Italy\label{aff124}
\and
INAF-Osservatorio Astronomico di Brera, Via Brera 28, 20122 Milano, Italy, and INFN-Sezione di Genova, Via Dodecaneso 33, 16146, Genova, Italy\label{aff125}
\and
ICL, Junia, Universit\'e Catholique de Lille, LITL, 59000 Lille, France\label{aff126}
\and
ICSC - Centro Nazionale di Ricerca in High Performance Computing, Big Data e Quantum Computing, Via Magnanelli 2, Bologna, Italy\label{aff127}
\and
Instituto de F\'isica Te\'orica UAM-CSIC, Campus de Cantoblanco, 28049 Madrid, Spain\label{aff128}
\and
CERCA/ISO, Department of Physics, Case Western Reserve University, 10900 Euclid Avenue, Cleveland, OH 44106, USA\label{aff129}
\and
Technical University of Munich, TUM School of Natural Sciences, Physics Department, James-Franck-Str.~1, 85748 Garching, Germany\label{aff130}
\and
Max-Planck-Institut f\"ur Astrophysik, Karl-Schwarzschild-Str.~1, 85748 Garching, Germany\label{aff131}
\and
Laboratoire Univers et Th\'eorie, Observatoire de Paris, Universit\'e PSL, Universit\'e Paris Cit\'e, CNRS, 92190 Meudon, France\label{aff132}
\and
Departamento de F{\'\i}sica Fundamental. Universidad de Salamanca. Plaza de la Merced s/n. 37008 Salamanca, Spain\label{aff133}
\and
Universit\'e de Strasbourg, CNRS, Observatoire astronomique de Strasbourg, UMR 7550, 67000 Strasbourg, France\label{aff134}
\and
Center for Data-Driven Discovery, Kavli IPMU (WPI), UTIAS, The University of Tokyo, Kashiwa, Chiba 277-8583, Japan\label{aff135}
\and
Dipartimento di Fisica - Sezione di Astronomia, Universit\`a di Trieste, Via Tiepolo 11, 34131 Trieste, Italy\label{aff136}
\and
California Institute of Technology, 1200 E California Blvd, Pasadena, CA 91125, USA\label{aff137}
\and
Departamento de F\'{i}sica Te\'{o}rica, At\'{o}mica y \'{O}ptica, Universidad de Valladolid, 47011 Valladolid, Spain\label{aff138}
\and
Department of Physics \& Astronomy, University of California Irvine, Irvine CA 92697, USA\label{aff139}
\and
Department of Mathematics and Physics E. De Giorgi, University of Salento, Via per Arnesano, CP-I93, 73100, Lecce, Italy\label{aff140}
\and
INFN, Sezione di Lecce, Via per Arnesano, CP-193, 73100, Lecce, Italy\label{aff141}
\and
INAF-Sezione di Lecce, c/o Dipartimento Matematica e Fisica, Via per Arnesano, 73100, Lecce, Italy\label{aff142}
\and
Departamento F\'isica Aplicada, Universidad Polit\'ecnica de Cartagena, Campus Muralla del Mar, 30202 Cartagena, Murcia, Spain\label{aff143}
\and
CEA Saclay, DFR/IRFU, Service d'Astrophysique, Bat. 709, 91191 Gif-sur-Yvette, France\label{aff144}
\and
Institute of Cosmology and Gravitation, University of Portsmouth, Portsmouth PO1 3FX, UK\label{aff145}
\and
Department of Computer Science, Aalto University, PO Box 15400, Espoo, FI-00 076, Finland\label{aff146}
\and
Instituto de Astrof\'\i sica de Canarias, c/ Via Lactea s/n, La Laguna 38200, Spain. Departamento de Astrof\'\i sica de la Universidad de La Laguna, Avda. Francisco Sanchez, La Laguna, 38200, Spain\label{aff147}
\and
Ruhr University Bochum, Faculty of Physics and Astronomy, Astronomical Institute (AIRUB), German Centre for Cosmological Lensing (GCCL), 44780 Bochum, Germany\label{aff148}
\and
Department of Physics and Astronomy, Vesilinnantie 5, 20014 University of Turku, Finland\label{aff149}
\and
Serco for European Space Agency (ESA), Camino bajo del Castillo, s/n, Urbanizacion Villafranca del Castillo, Villanueva de la Ca\~nada, 28692 Madrid, Spain\label{aff150}
\and
ARC Centre of Excellence for Dark Matter Particle Physics, Melbourne, Australia\label{aff151}
\and
Centre for Astrophysics \& Supercomputing, Swinburne University of Technology,  Hawthorn, Victoria 3122, Australia\label{aff152}
\and
Department of Physics and Astronomy, University of the Western Cape, Bellville, Cape Town, 7535, South Africa\label{aff153}
\and
DAMTP, Centre for Mathematical Sciences, Wilberforce Road, Cambridge CB3 0WA, UK\label{aff154}
\and
Kavli Institute for Cosmology Cambridge, Madingley Road, Cambridge, CB3 0HA, UK\label{aff155}
\and
Department of Astrophysics, University of Zurich, Winterthurerstrasse 190, 8057 Zurich, Switzerland\label{aff156}
\and
Department of Physics, Centre for Extragalactic Astronomy, Durham University, South Road, Durham, DH1 3LE, UK\label{aff157}
\and
IRFU, CEA, Universit\'e Paris-Saclay 91191 Gif-sur-Yvette Cedex, France\label{aff158}
\and
Oskar Klein Centre for Cosmoparticle Physics, Department of Physics, Stockholm University, Stockholm, SE-106 91, Sweden\label{aff159}
\and
Astrophysics Group, Blackett Laboratory, Imperial College London, London SW7 2AZ, UK\label{aff160}
\and
Univ. Grenoble Alpes, CNRS, Grenoble INP, LPSC-IN2P3, 53, Avenue des Martyrs, 38000, Grenoble, France\label{aff161}
\and
INAF-Osservatorio Astrofisico di Arcetri, Largo E. Fermi 5, 50125, Firenze, Italy\label{aff162}
\and
Dipartimento di Fisica, Sapienza Universit\`a di Roma, Piazzale Aldo Moro 2, 00185 Roma, Italy\label{aff163}
\and
HE Space for European Space Agency (ESA), Camino bajo del Castillo, s/n, Urbanizacion Villafranca del Castillo, Villanueva de la Ca\~nada, 28692 Madrid, Spain\label{aff164}
\and
Department of Astrophysical Sciences, Peyton Hall, Princeton University, Princeton, NJ 08544, USA\label{aff165}
\and
Theoretical astrophysics, Department of Physics and Astronomy, Uppsala University, Box 515, 751 20 Uppsala, Sweden\label{aff166}
\and
Minnesota Institute for Astrophysics, University of Minnesota, 116 Church St SE, Minneapolis, MN 55455, USA\label{aff167}
\and
Mathematical Institute, University of Leiden, Einsteinweg 55, 2333 CA Leiden, The Netherlands\label{aff168}
\and
School of Physics \& Astronomy, University of Southampton, Highfield Campus, Southampton SO17 1BJ, UK\label{aff169}
\and
Institute of Astronomy, University of Cambridge, Madingley Road, Cambridge CB3 0HA, UK\label{aff170}
\and
Department of Physics and Astronomy, University of California, Davis, CA 95616, USA\label{aff171}
\and
Space physics and astronomy research unit, University of Oulu, Pentti Kaiteran katu 1, FI-90014 Oulu, Finland\label{aff172}
\and
Center for Computational Astrophysics, Flatiron Institute, 162 5th Avenue, 10010, New York, NY, USA\label{aff173}
\and
Univ. Lille, CNRS, Centrale Lille, UMR 9189 CRIStAL, 59000 Lille, France\label{aff174}
\and
Department of Physics and Astronomy, University of British Columbia, Vancouver, BC V6T 1Z1, Canada\label{aff175}}

\titlerunning{Exploring galaxy morphology across cosmic time through S\'ersic fits}
\authorrunning{Euclid Collaboration: L.~Quilley et al.}

\abstract 
{We present the results of the single-component S\'ersic profile fitting for the magnitude-limited sample of \IE$<23$ galaxies within the 63.1 deg$^2$ area of the Euclid Quick Data Release (Q1). The associated morphological catalogue includes two sets of structural parameters fitted using \texttt{SourceXtractor++}: one for VIS \IE images and one for a combination of three NISP images in \YE, \JE, and \HE bands. We compared the resulting S\'ersic parameters to other morphological measurements provided in the Q1 data release and to the equivalent parameters based on higher-resolution \HST imaging. These comparisons confirmed the consistency and the reliability of the fits to Q1 data. Our analysis of colour gradients shows that NISP profiles systematically have smaller effective radii ($R_{\rm e}$) and larger S\'ersic indices ($n$) than in VIS. In addition, we highlight trends in NISP-to-VIS parameter ratios with both magnitude and $n_{\rm VIS}$. From the 2D bimodality of the $(u-r)$ colour-$\log(n)$ plane, we defined a $(u-r)_{\rm lim}(n)$ that separates early- and late-type galaxies (ETGs and LTGs). We used the two sub-populations to examine the variations of $n$ across well-known scaling relations at $z<1$. The ETGs display a steeper size--stellar mass relation than the LTGs, indicating a difference in the main drivers of their mass assembly. Similarly, LTGs and ETGs occupy different parts of the stellar mass--star-formation rate plane, with ETGs at higher masses than LTGs and further below the main sequence of star-forming galaxies. This clear separation highlights the link known between the shutdown of star formation and morphological transformations in the \Euclid imaging data set. In conclusion, our analysis demonstrates both the robustness of the S\'ersic fits available in the Q1 morphological catalogue and the wealth of information they provide for studies of galaxy evolution with \Euclid.}

   
\maketitle

\section{\label{sc:Intro} Introduction}

The study of galaxy morphology is a topic as old as galaxy science itself, which started with the Hubble sequence \citep{Hubble-1926-extragalactic-nebulae}, where galaxies were classified according to their shapes and features. This sequence challenges any theory of galaxy formation and evolution to explain how such diversity arose across the history of the Universe.

Morphology has remained a topical issue for all galaxy scientists due to its many connections to other aspects of galaxy formation. The galaxy-morphology density relation \citep{Dressler-1980-morphology-density-relation} first quantified that different morphological types of galaxies are located in different environments with respect to large-scale structures. However, the colour bimodality of galaxy populations is related to the morphology, since red and blue galaxies are predominantly early- and late-type galaxies \citep{Strateva-2001-color-bimodality, Baldry-2004-color-bimodality, Allen-2006-MGC-BD-decomp}. The correlation between the morphology and star-forming state of galaxies was further investigated to understand the causal relation linking morphological transformations from late- to early-type galaxies and the shutdown of star formation, known as the quenching of galaxies \citep{Schawinski-2014-GV-2-pathways, Bremer-2018-GAMA-survey-morph-transf-GV, Quilley-2022-bimodality, de-Sa-Freitas-2022-quenching-bursting-colours-morphology, Dimauro-2022-bulge-growth}.
Notably, in observational studies the growth of the bulges of galaxies has been shown to correlate with the quenching of galaxies \citep{Lang-2014-bulge-growth-quenching-CANDELS, Bluck-2014-bulge-mass, Bremer-2018-GAMA-survey-morph-transf-GV, Bluck-2022-quenching-bulge-disk-ML, Dimauro-2022-bulge-growth, Quilley-2022-bimodality}. In that regard, \cite{Martig-2009-morphological-quenching} proposed a morphological quenching process in which the bulge of a galaxy, when it becomes massive enough, can stabilise the disc against cloud fragmentation and prevent further star formation. But other phenomena could also explain these trends. For example, black hole growth \citep{Bluck-2022-quenching-bulge-disk-ML, Brownson-2022-quenching-kinematics-bulge-AGN} is correlated with bulge growth and eventually leads to quenching by active galactic nucleus (AGN) feedback \citep{Silk-Rees-1998-AGN-mechanical-feedback, Croton-2006-AGN-thermal-feedback, Chen-2020-quenching-halo-AGN}.

There are several ways to characterise galaxy morphologies. The earliest analyses relied on visual classification \citep{Hubble-1926-extragalactic-nebulae, Nair-Abraham-2010-catalog-morph-class, Baillard-2011-EFIGI}; however, it is a time-consuming human-based task. Automatic and quantitative methods were thus later developed. Notably, parametric surface-brightness fitting based on the S\'ersic function \citep{Sersic-1963-sersic-model} has become a widely used method to characterise galaxy morphology and obtain estimates of their sizes and shapes \citep[among many others]{Trujillo-2004-size-luminosity-mass-z-3, Kelvin-2012-GAMA-sersic-fits, Morishita-2014-size-morpho-GOODS-HST, Mowla-2019-size-mass-CANDELS, Kartatelpe-2023-CEERS-gal-morpho-JWST, Lee-2024-morpho-JWST-fields}. For spiral and lenticular galaxies, the use of two distinct profiles, usually a S\'ersic profile to model the bulge and an exponential profile (S\'ersic profile with a fixed S\'ersic index $n$ of 1) for the disc, has led to improvements in fitting the full surface brightness profiles of galaxies, and most importantly, the approach provides more physically meaningful parameters \citep{Meert-2015-SDSS-BD-catalog, Lange-2016-bulge-disk-decomposition-GAMA, Margalef-Bentabol-2016-bulge-disk-decomposition-CANDELS-z-3, Dimauro-2018-bulge-disk-decomposition-CANDELS, Fischer-2019-SDSS-IV-MANGA-morpho-catalog, Casura-2022-bulge-disk-decomposition-GAMA, Hashemizadeh-2022-DEVILS-bulge-emergence-since-z-1, Quilley-2022-bimodality}.

The \Euclid space mission \citep{EuclidSkyOverview} is envisioned to observe an effective sky area of roughly 14\,000\,${\rm deg^2}$ in order to map the evolution of the large-scale structure and will therefore observe a number of resolved galaxies several orders of magnitude larger than other previous missions, such as the \HST and the \textit{James Webb} Space Telescope. Its coverage will range from the red optical region, using the Visible Imager (VIS; with a single broadband filter \IE, \citealt{EuclidSkyVIS, Q1-TP002}), to the near infrared (NIR), using the Near-Infrared Spectrometer and Photometer (NISP). The latter enables both imaging in the NIR through three filters (\YE, \JE, and \HE) and low-resolution NIR spectroscopy (\citealt{EuclidSkyNISP,Schirmer-EP18, Q1-TP003}). 
Therefore, the large sky area of the \Euclid mission combined with the exquisite spatial resolution of its instruments -- pixel sizes of \ang{;;0.1} and \ang{;;0.3} and a point spread function (PSF) full width at half maximum (FWHM) of \ang{;;0.16} and \mbox{\ang{;;0.48}} for the optical and NIR, respectively -- and the unparalleled surface brightness sensitivities expected to be reached -- approximately 29.8\,mag\,arcsec$^{-2}$ for the Euclid Wide Survey (EWS; \citealt{Scaramella-EP1}) and 31.8\,mag\,arcsec$^{-2}$ for the Euclid Deep Survey (EDS) -- make it uniquely suited for groundbreaking discoveries. 

Preliminary work has already been conducted to evaluate how beneficial \Euclid will be to morphological analyses combining statistics, precision, and redshift coverage. Using deep generative models, \cite{Bretonniere-EP13} estimated that the EWS and EDS would be able to resolve the inner structure of galaxies down to a surface brightness of 22.5\,mag\,arcsec$^{-2}$ and 24.9\, mag\,arcsec$^{-2}$, respectively. Furthermore, the Euclid Morphology Challenge (EMC) compared and estimated the performance of five luminosity-fitting codes in retrieving the photometry \citep{Merlin-EP25} and structural parameters \citep{Bretonniere-EP26}. It concluded that single-S\'ersic fits and bulge and disc decomposition were reliable up to an \IE of 23 and 21, respectively. The \texttt{SourceXtractor++} software \citep{Bertin-2020-SourceXtractor-plus-plus, Kummel-2022-SE-use} showed a high level of performance and was therefore chosen to perform the fits included in the \Euclid data processing function.

Moreover, a first morphological analysis based on \Euclid data using luminosity profile-fitting was performed in \citet[Q25a hereafter]{Quilley25}, where galaxies in the Early Release Observations (ERO) of the Perseus cluster \citep{EROPerseusOverview} were modelled. The analysis focused not on the cluster itself but on the thousands of background galaxies present. The study of Q25a represents the first demonstration of how the quality of \Euclid images can enable morphological analyses, and it notably measured a bulge-disc colour dichotomy as well as colour gradients within discs for galaxies up to $z<0.6$ (Q25a).  

In this work, we first present the catalogue obtained by performing S\'ersic fits on all the galaxies detected in the first Euclid Quick Data Release (Q1), and we then leverage them to reassert the importance of morphological transformations in galaxy evolution. These new Q1 data are described in \sct\ref{sc:Data} together with the galaxy sample that is used throughout this analysis. In \sct\ref{sc:Methodo}, we detail the choices made in the configuration of the S\'ersic fits. In \sct\ref{sc:sersic-params}, we present the distributions of the best-fit S\'ersic parameters. In \sct\ref{sc:compare-other-methods}, we compare the S\'ersic parameters to other morphological indicators available in the Q1 data or from previous HST studies to discuss their consistencies. In \sct\ref{sc:results-color-variation}, we investigate the variations of the S\'ersic parameters with observing band by comparing the values derived from the VIS image to those obtained from the three NISP images, which allowed us to probe galaxy colour gradients. In \sct\ref{sc:results-evol-plots}, we emphasise the role of morphology in galaxy evolution, first by establishing the bimodality of galaxy populations in terms of colour, mass, and morphology and then by examining the variation of the S\'ersic index across key scaling relations such as the size--stellar mass and the star-formation rate (SFR)--stellar mass relations.

\section{\label{sc:Data} Data}

This work focuses on data from \citet[Q1]{Q1cite}, which covered over 63.1\,deg$^{2}$ of the sky during three sessions in 2024. They targeted the different Euclid Deep Fields (EDFs): EDF-North, \,22.9\,deg$^{2}$, EDF-Fornax, \,12.1\,deg$^{2}$), and EDF-South, \,28.1\,deg$^{2}$. The Q1 data are representative of standard EWS observations conducted as a single Reference Observation Sequence. For additional details on the Q1 data and its processing, we refer the reader to \cite{Q1-TP001}.

The analysis presented in this work focuses on the Q1 tiles from the EDFs. Each tile has a field of view of \mbox{0.57\,deg$^{2}$} (\mbox{\ang{0.75}$\,\times$\,\ang{0.75}}). Observations followed a dithered sequence: simultaneous imaging in the \IE filter and NIR slit-less grism spectra, followed by NIR imaging through the \YE, \HE, and \JE filters. The telescope was dithered between observations, and the sequence was repeated four times. Exposure times were 1864\,s for the \IE filter and 348.8\,s for each NIR filter, and photometric calibration uncertainty was constrained to less than 10$\%$, with zero points of $I_{\rm AB}$=23.9 for all bands. AB magnitudes are referred to simply as magnitudes throughout this work.

To construct the final sample of galaxies from the Q1 data set, we used photometric data extracted from the photometric catalogues\footnote{\url{http://st-dm.pages.euclid-sgs.uk/data-product-doc/dm10/index.html}} produced by the OU-MER\footnote{OU-MER aims to produce object catalogues from the merging of all the multi-wavelength data, both from \Euclid and ground-based complementary observations.} pipeline \citep{Q1-TP004}. In addition to \Euclid data, it includes external data from ground-based surveys that provide optical photometry in bands narrower than \IE that are useful for photometric redshift computation and spectral energy distribution (SED)-fitting \citep{Q1-TP005}. More specifically, $griz$ photometry from DES \citep{Abbott-2021-DES-DR2} is used in EDF-S and EDF-F, whereas for the EDF-N, $ugriz$ photometry is provided by CFIS ($u$ and $r$, \citealt{Ibata-2017-CFIS-u-band}), Pan-STARRS ($i$, \citealt{Magnier-2020-Pan-STARRS}) and HSC ($g$ and $z$, \citealt{Aihara-2018-HSC-overview}).
A series of selection criteria were then applied to ensure the robustness and reliability of the analysis, as outlined below:
 \begin{itemize}
    \item[$\ast$] {\tt VIS$\_$DET}\,=\,1 to ensure that galaxies are detected in the VIS filter;
    \item[$\ast$] {\tt spurious$\_$flags}\,=\,0 to prevent spurious sources from contaminating our sample;
    \item[$\ast$] {\tt point$\_$like$\_$prob}\,$\leq$\,0.1 (which ranges from 0 to 1) to discard point-like sources;
    \item[$\ast$] \IE$\leq$\,24.5 for the largest sample available, but we mostly restrain it to \IE$<23$ for reliability purposes and when not specified otherwise.
 \end{itemize}

By applying these selection criteria, we obtained a sample consisting of 1\,312\,068 galaxies up to \IE$<23$ that are used in \sct\ref{sc:sersic-params}. In this section, we further identify a series of criteria based on S\'ersic parameters that allowed us to discard unsuccessful fits (those with extreme parameters). The criteria for keeping a fit are the following:
\begin{itemize}
    \item[$\ast$] $0.05 < q < 1.00$ with $q$ the S\'ersic axis ratio;
    \item[$\ast$] $0.302 < n < 5.45$ with $n$ the S\'ersic index. 
\end{itemize}
They are applied throughout the rest of the paper, starting in \sct\ref{sc:compare-other-methods}. This cleaning reduced the sample to 1\,209\,661 galaxies.
Similarly, we further identify in \sct\ref{sc:sersic-vs-iso} other spurious fits that can be discarded by keeping only fits satisfying
\begin{itemize}
    \item[$\ast$] $ 0.1\,a < R_\mathrm{e} < 20\,a$, with $a$ as the isophotal semi-major axis and $R_\mathrm{e}$ as the effective radius of the S\'ersic profile. 
\end{itemize}
This led to a sample of 1\,207\,997 galaxies with supposedly clean S\'ersic fits at $\IE < 23$.

In \scts\ref{sc:results-color-variation} and \ref{sc:results-evol-plots}, where galaxies are investigated as a function of redshift, we use the median of the redshift posterior distribution derived through the spectral energy distribution fitting (see \citealt{Q1-TP005}). Therefore, we further added the following selection criteria to ensure the reliability of the photometric redshift as well as of the physical parameters we derived:
\begin{itemize}
    \item[$\ast$] The offset between the median redshift derived by the SED fitting to determine the physical properties of galaxies {\tt PHZ$\_$PP$\_$MEDIAN$\_$Z} and the photometric redshift {\tt PHZ$\_$MEDIAN} derived for cosmology is less than 0.2.
    \item[$\ast$] The offset between the median {\tt PHZ$\_$PP$\_$MEDIAN$\_$Z} and the mode {\tt PHZ$\_$PP$\_$MODE$\_$Z} of the redshift posterior distribution is less than 0.2.
    \item[$\ast$] {\tt phys$\_$param$\_$flags}\,=\,0. 
\end{itemize}
With these additional selection criteria, we obtained a final sample of 774\,837 galaxies with redshifts up to 1.3. The above redshift criterion could bias our final sample by preferentially excluding galaxies with degenerate photo-$z$ solutions, such as Balmer-break galaxies, so we reproduced the analysis of \sct\ref{sc:results-evol-plots} without applying it and found similar results.

The current sample also contains additional physical parameters of galaxies, most importantly the SFR and stellar masses ($M_\ast$), which are used throughout \sct\ref{sc:results-evol-plots}. \cite{Q1-TP005} includes a detailed description of the modelling used for deriving these parameters; here we provide a brief overview. Galaxy physical parameters are based on photometric measurements in \Euclid NISP and ground-based $griz$ bands, a grid of model SEDs generated by \texttt{Bagpipes}, and the \texttt{NNPZ} algorithm.  For each observed galaxy, the algorithm identifies 30 closest model SEDs in the multidimensional space of photometric fluxes, with the likelihood defined by the distance between the model and the observations. The catalogue contains for each galaxy property both the maximum likelihood model values (mode) and the weighted median value for the marginalised posterior derived from the 30 neighbours (median). We use in the current analysis median values for SFR and stellar masses, in agreement with the choice made for photo-$z$, and tested that similar results were obtained with mode values.

\begin{figure*}[ht]
\centering
\includegraphics[width=\columnwidth]{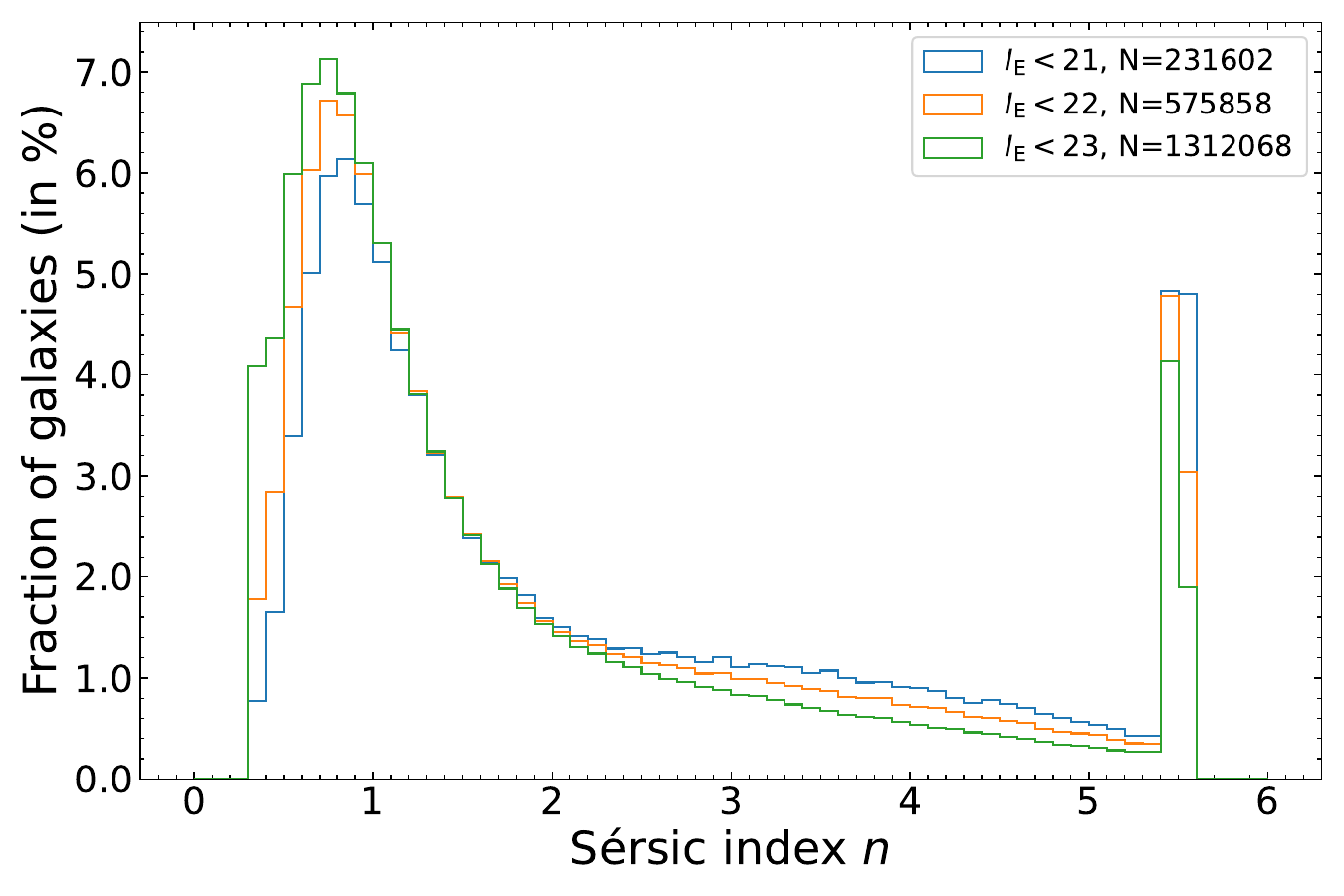}
\includegraphics[width=\columnwidth]{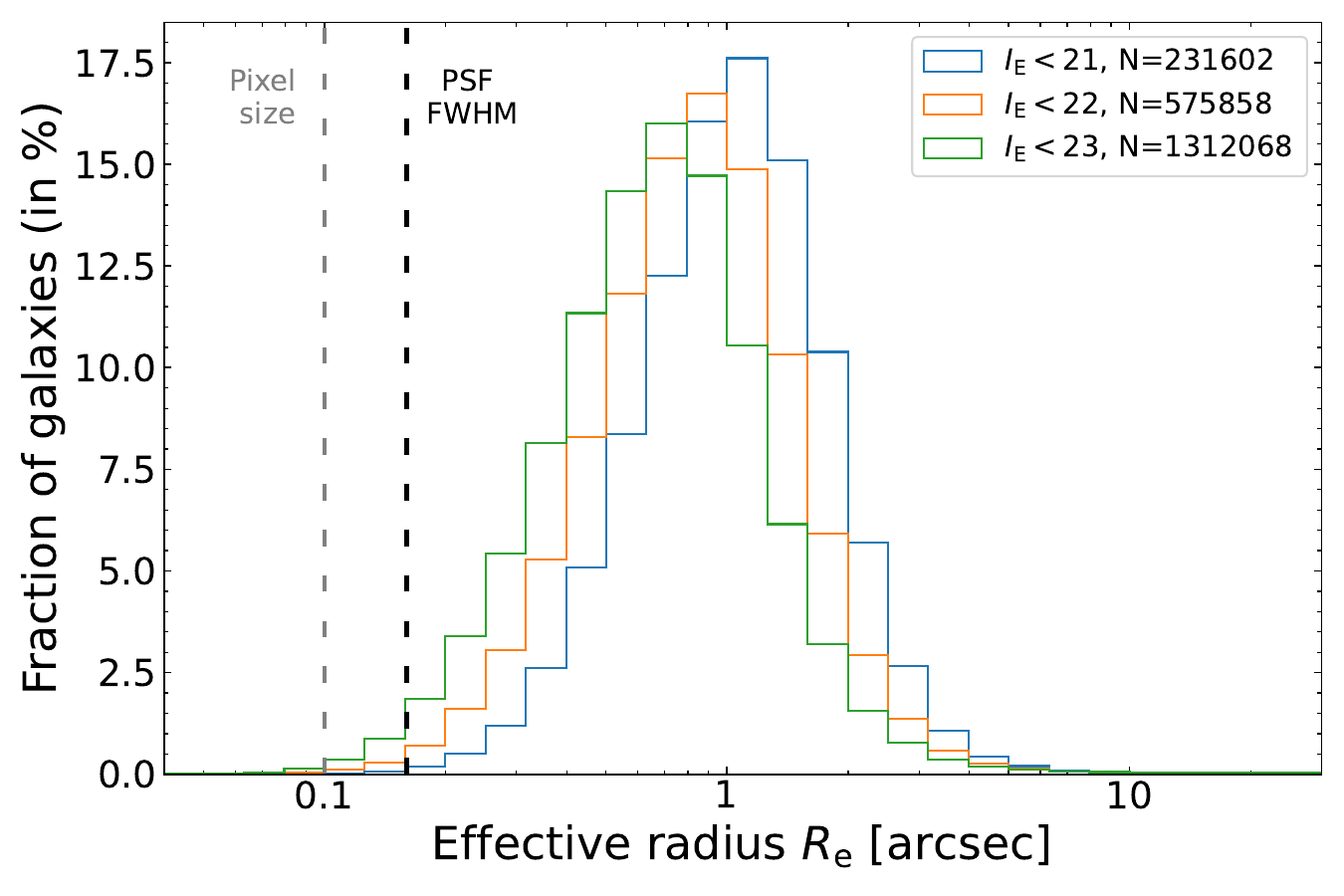}
\includegraphics[width=\columnwidth]{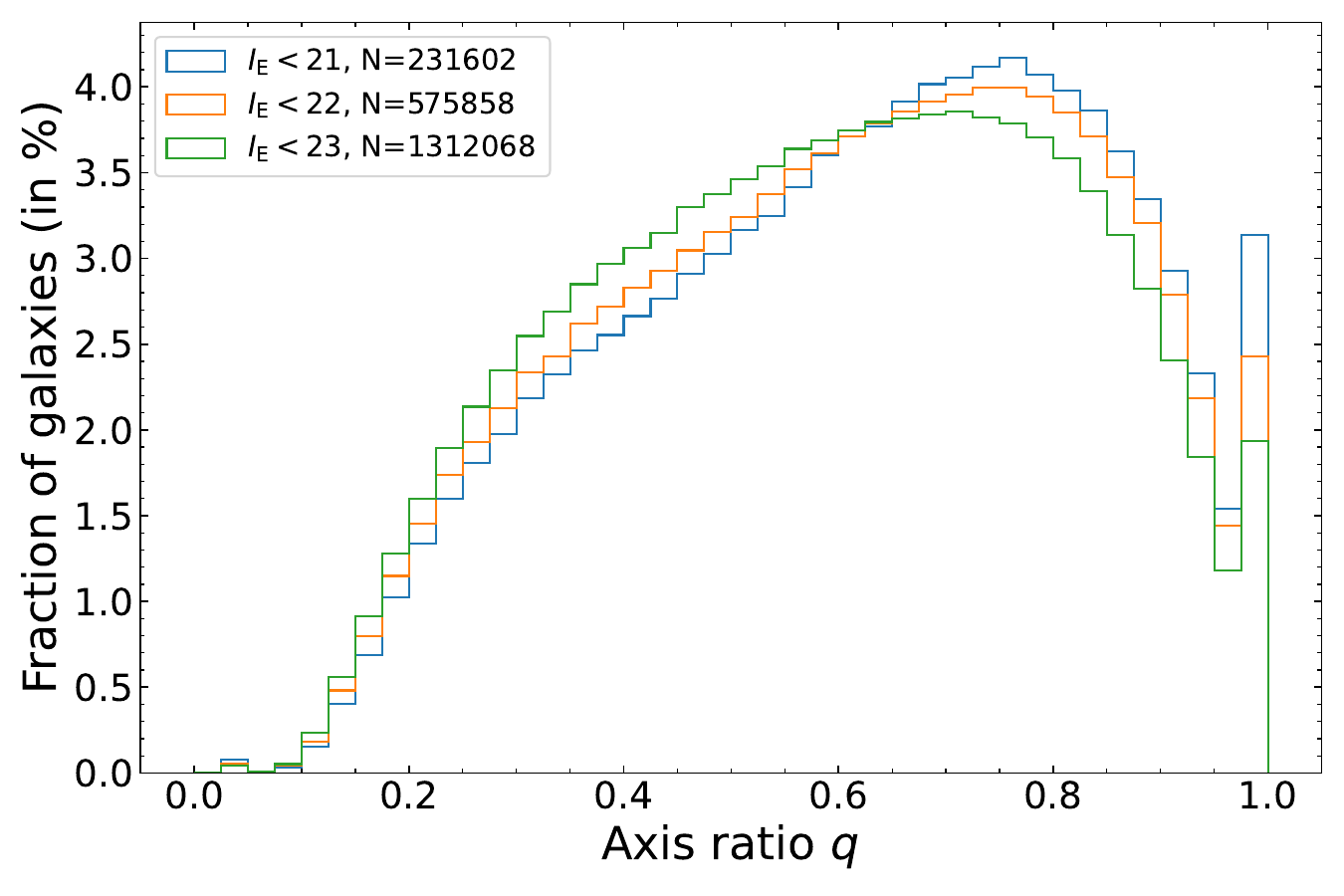}
\includegraphics[width=\columnwidth]{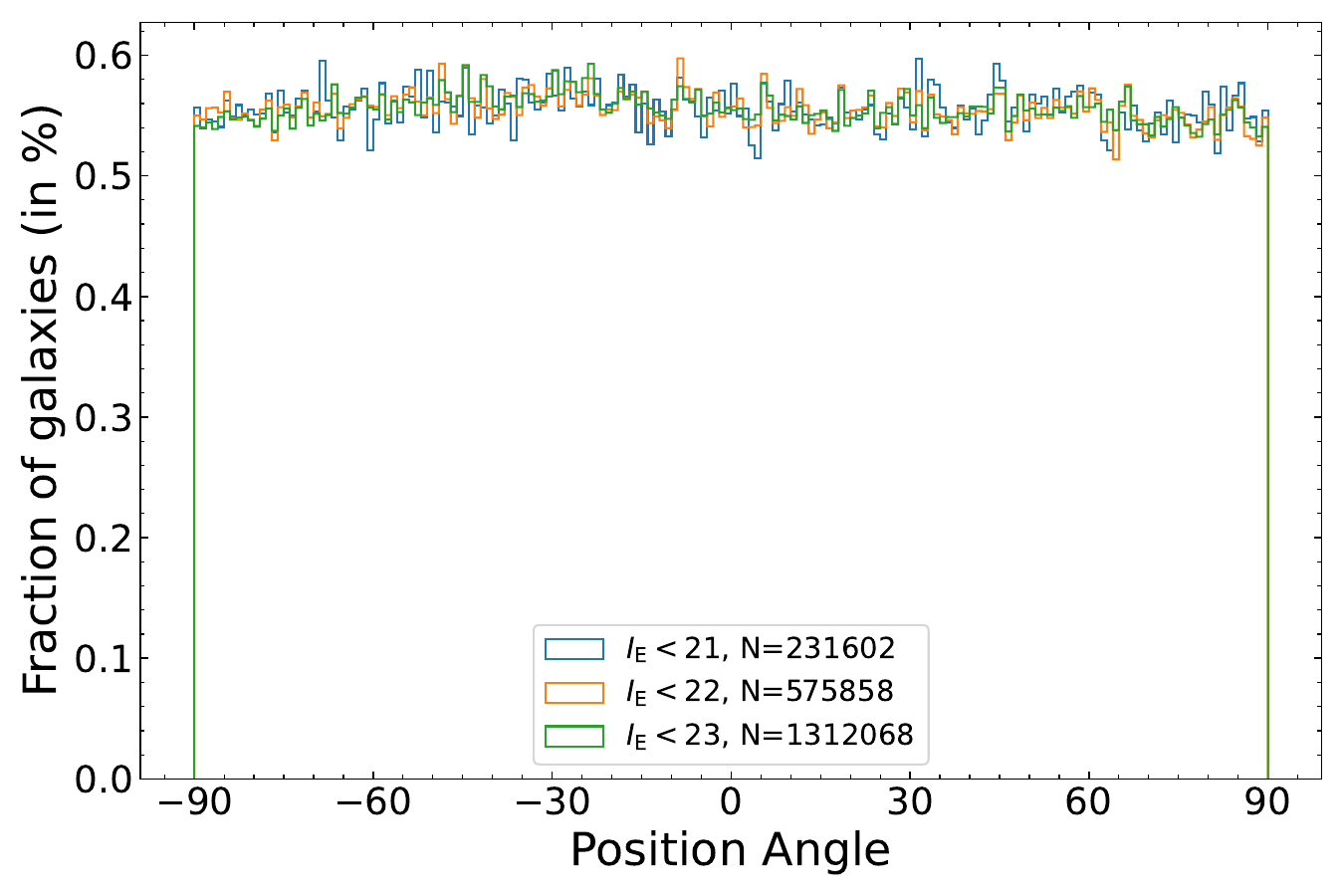}
\caption{Fractional distributions of parameters of the fitted S\'ersic profiles. From left to right and from top to bottom: S\'ersic index ($n$), effective radius ($R_{\rm e}$), axis ratio ($q$), and position angle. The blue, orange, and green curves correspond to samples down to a limiting \IE of 21, 22, and 23, respectively. The number of galaxies in each sample is indicated in each insert. The $n$ values peak around 0.8, and then sharply decrease until $n\sim2$, there is a tail of higher values, and an `artificial' peak at the limit of the parameter space at $n=5.5$ (see text). The distributions of $R_{\rm e}$ peak around \ang{;;1.0}, with lower values for fainter magnitudes. The pixel sizes and PSF FWHM are indicated as vertical dotted lines. 
}
\label{fig:distribs-sersic-params-percent}
\end{figure*}

\section{\label{sc:Methodo} Methodology}

The software \texttt{SourceXtractor++}, whose level of performance was assessed in the EMC, is used to fit elliptically symmetric 2D S\'ersic profiles \citep{Sersic-1963-sersic-model} to all galaxies detected in \Euclid images. We recall here the one-dimensional S\'ersic function, which describes the variation of the light intensity, $I$, as a function of the angular radius, $r$:
\begin{equation}
    I(r) = I_\mathrm{e} \exp\left\{ -b_n\left[\left( \frac{r}{R_\mathrm{e}}\right) ^{1/n} -1\right]\right\}\,,
    \label{eq-sersic}
\end{equation}
with $R_\mathrm{e}$ as the major-axis of the elliptical profile that encloses half of the total light (or effective radius), $I_\mathrm{e} = I(R_\mathrm{e})$ as the light intensity at $R_\mathrm{e}$, with the S\'ersic index $n$ characterising the steepness of the profile, and with $b_n$ as a normalisation parameter depending solely on $n$. Because we fit 2D elliptical profiles, the parameters fully describing the profile also include a position angle and an axis ratio, $q$.

The S\'ersic fitting procedure uses two sets of structural parameters ($R_\mathrm{e}$, $n$, and $q$), one for the VIS image alone, and another one common for the three NISP images. The fits were performed with a common position angle for the two models. This model fitting on the \Euclid images provides the aforementioned parameters as well as model photometry in the \IE, \YE, \JE, and \HE bands. Then, for the external data images, the structural parameters of the S\'ersic profiles obtained from the VIS image are used and only magnitudes are allowed to vary to obtain model photometry from these optical bands. This configuration was decided in order to fully benefit from the enhanced spatial resolution of VIS while obtaining colour information, through the comparison of VIS and near-IR parameters (see \sct\ref{sc:results-color-variation}).

The exact ranges and initial values for all parameters are given below:
\begin{itemize}
    \item We initialised $R_\mathrm{e}$ at the isophotal semi-major axis, $a$ (see \sct\ref{sc:sersic-vs-iso} for details on all isophotal parameters), and it exponentially spans the [0.01\,$a$, 30\,$a$] interval.
    \item We initialised $n$ at 1, and it linearly spans the [0.3, 5.5] range.
    \item We initialised $q$ at the isophotal axis ratio, and it exponentially spans the [0.03, 1] interval.
    \item We initialised $q$ at the isophotal position angle without any bounds, and it was reprojected onto the $[-\ang{90},\,\ang{90}]$ range.
    \item The $x$ and $y$ coordinates of the centre of the profile were fixed using the built-in function \texttt{o.centroid} from \texttt{SourceXtractor++}.
\end{itemize}

The fitting was performed on all pixels in the square region with the side equal to the maximum between the vertical and horizontal extent of the segmentation area, multiplied by a scaling factor of 1.4, to include lower level isophotes as well as some sky background, and with 4 pixels added for the padding of small sources. All models are convolved with the PSF before comparison with the data, and the PSF is position dependent, but does not take into account any other source property (see \sct 5 of \citealt{Q1-TP004} for more details). Specifically, despite the wide wavelength coverage of the \IE band, its PSF is achromatic and thus does not vary with the galaxy SED within the band. This can lead to biases, depending on whether or not galaxy observed emission in the \IE band is skewed towards a specific side. Since the PSF FWHM increases  with $\lambda$ (linearly on first order, in the diffraction-limited regime), using an average VIS PSF underestimates (overestimates) the PSF impact in redder (bluer) galaxies. Redder galaxies would then have their $R_e$ overestimated and $n$ underestimated. Future data releases of \Euclid may use an improved chromatic PSF to solve this issue. In \cite{Liu-2023-chromatic-PSF}, the PSF colour bias was quantified in the $griz$ optical bands, hence matching the \IE wavelengths, and it resulted in mean size variation of the order of 0.1\%, so we can consider them negligible in the current study. Moreover, there is also a colour gradient bias to consider, as galaxies display internal colour gradients. This higher-order effect cannot be addressed by a chromatic PSF alone, as a single-galaxy SED is not sufficient to capture the differential colour effects across different parts of a galaxy, particularly the difference between a red bulge and a blue disk. As of now, we believe that these effects are secondary compared to the absence of first-order chromatic dependence of the PSF. Similarly to what we argued above, for a bulge containing close to half the total light, this would lead to an overestimation of $R_e$, with a decreasing bias for decreasing bulge-to-total ratio. The high $n$ values driven by the peaky bulge profile would also be underestimated. The impact of these colour and colour gradient biases on our results are discussed in \sct\ref{sc:results-evol-plots}.

For reproducibility purposes, we provide the configuration files used within the \Euclid pipeline as well as an example image, with appropriate documentation.\footnote{\url{https://cloud.physik.lmu.de/index.php/s/3K4KemBsw5y9yqd}} This allows anyone to run {\tt SourceXtractor++} single-S\'ersic fits on \Euclid images in the configuration described throughout this section.

\section{\label{sc:sersic-params}Distributions of the structural parameters}

We present in \fg\ref{fig:distribs-sersic-params-percent}, the fractional distributions of $n$, $R_{\rm e}$, $q$, and position angles of the S\'ersic profiles fitted on the VIS images out to \IE of 21, 22, and 23 (the total number of objects are indicated in the inserts). The distributions of $n$ peak around 0.75--0.80 and decrease up to the maximum value of 5.5, at which a second peak is observed. The latter corresponds to fits reaching the limit of the parameter space (see \sct\ref{sc:Methodo}). Therefore, fits with $n>5.45$ should be flagged and removed from any S\'ersic-based analysis. $R_{\rm e}$ show distributions peaking around \ang{;;1.0}, with lower peaks, and overall histograms shifting towards smaller $R_{\rm e}$ for fainter magnitudes. 94.1\% of the sources with \IE$<23$ have an effective diameter larger than 3 times the PSF FWHM. The axis ratios show broad distributions with peaks around 0.71, 0.75, and 0.76 at the three magnitude limits, framed by a sharp peak at $q=1.00$ and a small bump around $q=0.03$ corresponding to the bounds of the parameter space. Visual inspection of the fits with $q < 0.05$ shows that these sources correspond to stars not correctly masked, diffraction spikes, or cosmic rays, so they should be flagged and removed from the analysis. However, the situation for $q=1.00$ is different because those objects are galaxies that visually appear rather round and sometimes with slight asymmetries. However, the best-fit solution returned is not necessarily the absolute best-fit possible and instead corresponds to a local minimum in the parameter space due to the limit at $q=1.00$. All the outliers identified in this section are removed in the rest of the analysis. 

\begin{figure}[ht]
\centering
\includegraphics[width=\columnwidth]{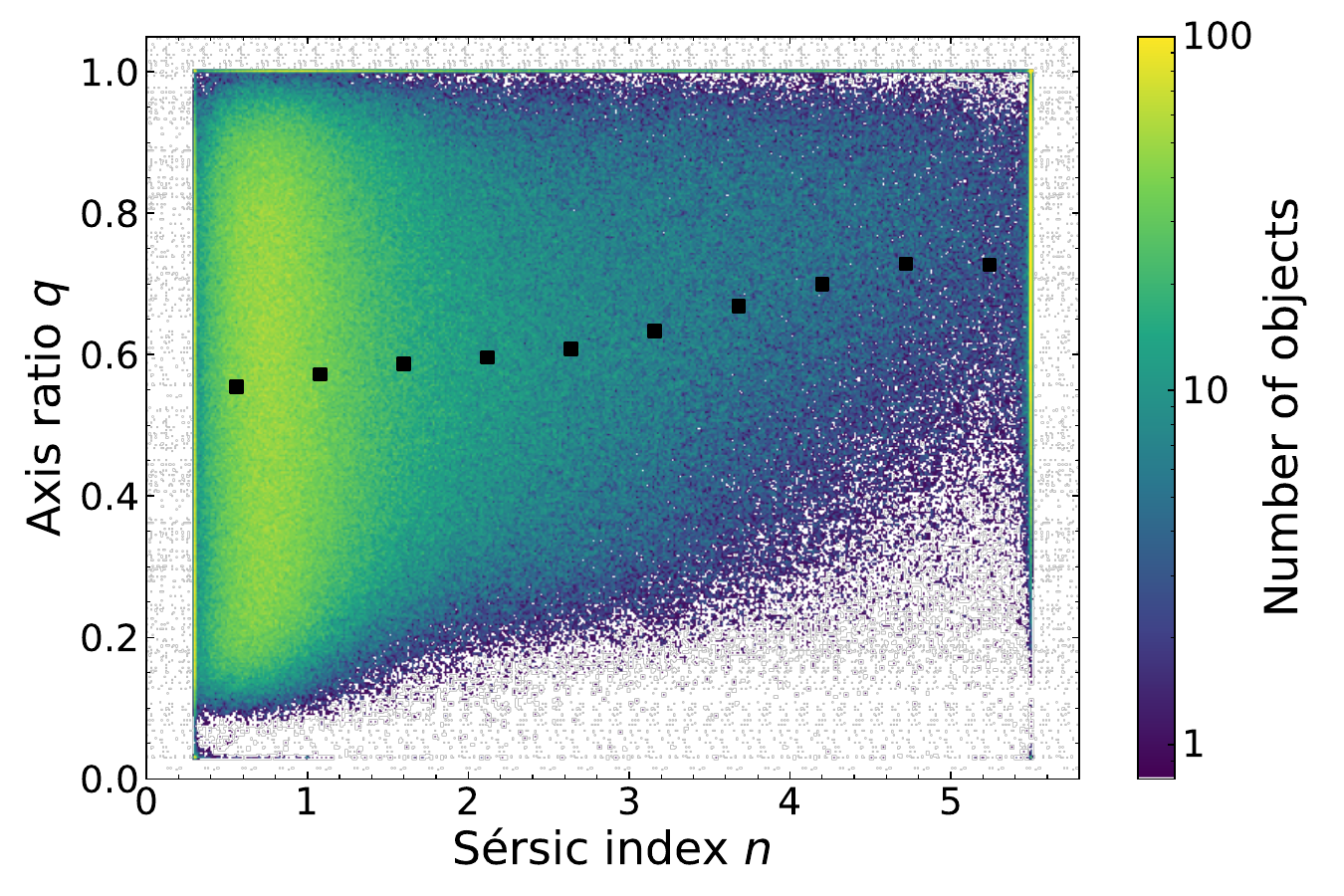}
\caption{Axis ratio $q$ versus S\'ersic index $n$ of the fitted profiles. Black squares indicate the running median $q$ in bins of $n$. Pure discs ($n\sim1$) can be seen at all inclinations, from edge-on to face-on, leading to a wide range of $q\in[0.1,\, 1.0]$, with a lower density of object for the most extreme cases at its edges. For higher values of $n$, either the bulge `bulging out' from the disc or more spheroidal galaxies prevent very low $q$.}
\label{fig:axis-ratio-vs-n}
\end{figure}

The $q$ measured in projection onto the sky depend on the tri-dimensional shape of the galaxies, characterised by their axes $a$, $b$, and $c$, as well as the orientation at which we observe them (see e.g. \citealt{Van-der-Wel-2014-geometry-SF-galaxies, Pandya-2024-3D-geometry-high-z-galaxies-CEERS}). Disc galaxies can be observed at all inclinations, from face-on to edge-on. They can thus display $q$ going from the physical one of the disc, $b/a$, close to 0.8--0.9 (since discs are almost but not perfectly circular; \citealt{Padilla-2008-shape-galaxies-SDSS, Rodriguez-2013-shape-galaxies-SDSS-Zoo}), to the ratio between their disc thickness and radius (taking values between $c/a$ and $c/b$; \citealt{Favaro-2025-intrinsic-flattening-disk}). Figure\,\ref{fig:axis-ratio-vs-n} shows the distribution of $q$ versus $n$ and confirms that for pure disc morphologies ($n \lesssim 1$), all $q$ within [0.1,\,1] are found. For higher $n$ the lower bound of the envelope of points increases to higher $q$. This is first due to the emergence of bulges in spiral galaxies ($n\sim$2--3), which are also included in the S\'ersic fit and bias the profile towards larger $q$ than those for the discs, as shown by Q25a. This trend continues for $n\gtrsim 4$, corresponding to early-type morphologies, especially elliptical galaxies, which are predominantly oblate spheroids, and thus their values of $c/a$ are higher than for discs \citep{Padilla-2008-shape-galaxies-SDSS,Rodriguez-2013-shape-galaxies-SDSS-Zoo}. 

\begin{figure*}[ht]
\centering
\includegraphics[width=2\columnwidth]{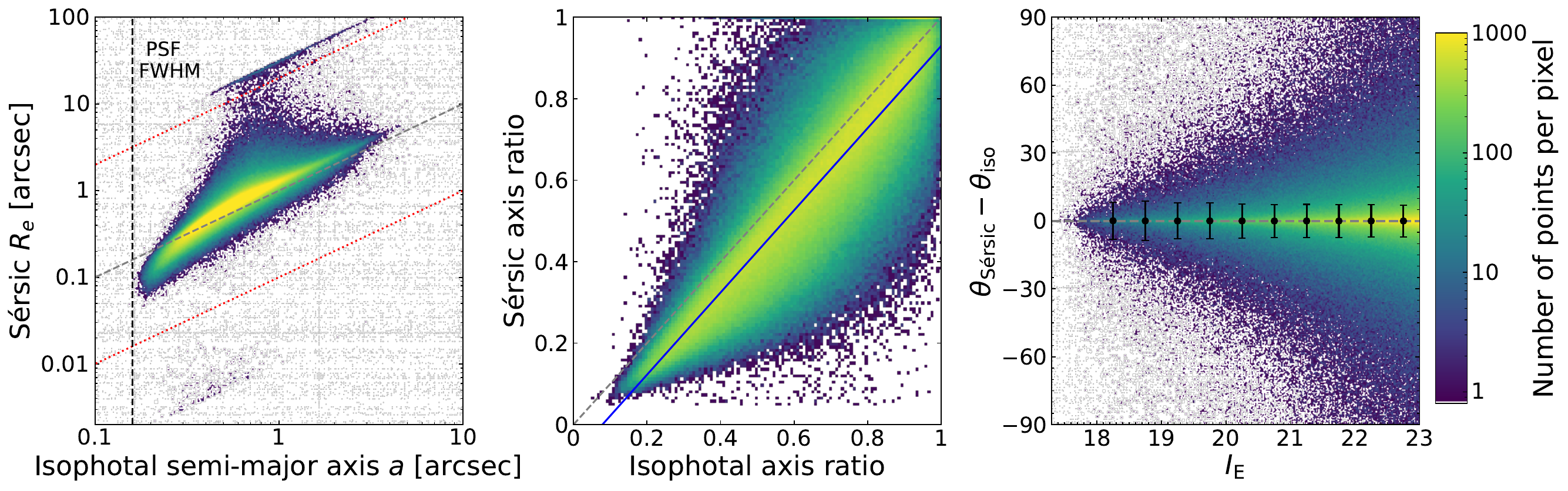}
\caption{Comparison between structural parameters of the S\'ersic profiles and isophotal measures for \IE$\le23$, with a colour-map indicating the density of points. \textit{Left:} S\'ersic effective radius $R_{\rm e}$ versus isophotal semi-major axis, $a$, with the identity line in dashed grey and red dotted lines identifying outliers (see text). \textit{Middle:} S\'ersic $q$ versus isophotal $q$. The identity line is shown with a dashed grey line, whereas the blue solid line is a fit to the data. \textit{Right}: Difference between the S\'ersic and isophotal position angles. The dashed grey line indicates a null difference, which is observed at all magnitudes, as displayed by the median difference in magnitude bins as black points and associated dispersion as error bars. Overall, there is consistency between the S\'ersic and isophotal structural parameters.
}
\label{fig:vs_iso_params}
\end{figure*}

\section{\label{sc:compare-other-methods}Comparison with other morphological measurements}

The \Euclid pipeline also includes other measurements of galaxy morphology (see \citealt{Q1-TP004} and \citealt{Q1-SP047}). In this section we compare the results of the S\'ersic fits to these other morphological measurements, as well as S\'ersic fits from previous analyses, in order to highlight their consistencies or discrepancies.

\subsection{\label{sc:sersic-vs-iso} S\'ersic versus isophotal parameters}

As part of the detection of the sources, the isophotal magnitude above the detection isophote is computed, as well as the semi-major and semi-minor axes, $q$, and the position angle of the semi-major axis, all being calculated on the spatial distribution of pixels detected above the detection threshold in the filtered background-subtracted image\footnote{The mathematical expressions for these parameters is available at \url{https://sextractor.readthedocs.io/en/latest/Position.html}} \citep{Q1-TP004}. Figure\,\ref{fig:vs_iso_params} shows the comparison between the fitted S\'ersic parameters and the corresponding isophotal measurements.

In the left panel, one can see that for most galaxies with \IE$\le23$, the S\'ersic $R_\mathrm{e}$ is correlated with the isophotal semi-major axis $a$. We note that we obtained similar results as figure 1 of \cite{Vulcani-2014-GAMA-color-gradients} comparing Kron radius to S\'ersic $R_\mathrm{e}$. On the on hand, the relation is curved for the smallest sources, due to the PSF impact on isophotal measures: $a$ remains above the PSF FWHM whereas the PSF-convolved S\'ersic fits can reach smaller sizes. On the other hand, $R_\mathrm{e}$ is systematically above $a$ for larger sources as the isophotal print (or similarly Kron apertures) misses flux in the outer regions of galaxies. Moreover, outliers can be easily identified in this plot, forming the lower and upper diagonal lines. Visual inspection of the samples defined as $R_\mathrm{e}<0.1\,a$, $R_\mathrm{e}>20\,a$ allowed us to confirm that these sources are poorly masked stars, cosmic rays, or poorly deblended objects in the halo of a foreground source. In all cases, these points need to be flagged and are removed from the sample for the rest of the analysis.

The middle panel of \fg\ref{fig:vs_iso_params} displays the relation between the S\'ersic profile axis ratio, $q_{\rm Sersic}$, and the isophotal axis ratio, $q_{\rm iso}$, showing that these two measures correlate with a Pearson coefficient of $r=0.88$ but that the S\'ersic profiles are on average slightly more elongated than the isophotal prints. Indeed, a linear fit yields
\begin{equation}
    q_{\rm Sersic} = (1.0060 \pm 0.0005) q_{\rm iso} - 0.0768 \pm 0.0003
\end{equation}
and hence an offset of 0.077. The $R^2$ score for this fit is 0.79, and the RMS dispersion around it is 0.10. Moreover, we computed the median difference between the S\'ersic and isophotal $q$ in bins of ellipticities of width $0.05$, obtaining values always between $-0.02$ and $-0.05$, with an associated dispersion around the median ranging from $0.02$ to $0.09$.

Finally, the right panel of \fg\ref{fig:vs_iso_params} shows the difference between the S\'ersic and isophotal position angles ($\theta_{\rm Sersic}$ and $\theta_{\rm iso}$), as a function of \IE. Median values displayed in black indicate that for any magnitude interval between 18 and 23, there is no bias, with median differences remaining below \ang{0.02}, except for the brightest magnitude interval [18,18.5] where it is \ang{0.55}. The apparently increasingly wider envelope of position angle differences at fainter magnitudes results from the lower signal-to-noise of the sources, and larger number of galaxies, causing more spread-out outliers, whereas the dispersion around the median remains stable and even slightly, and monotonically, decreases from \ang{9.2} to \ang{7.1} from the brightest to faintest magnitude intervals.

\subsection{\label{sc:sersic-vs-CAS} S\'ersic index versus concentration}

\begin{figure}[ht]
\centering
\includegraphics[width=\columnwidth]{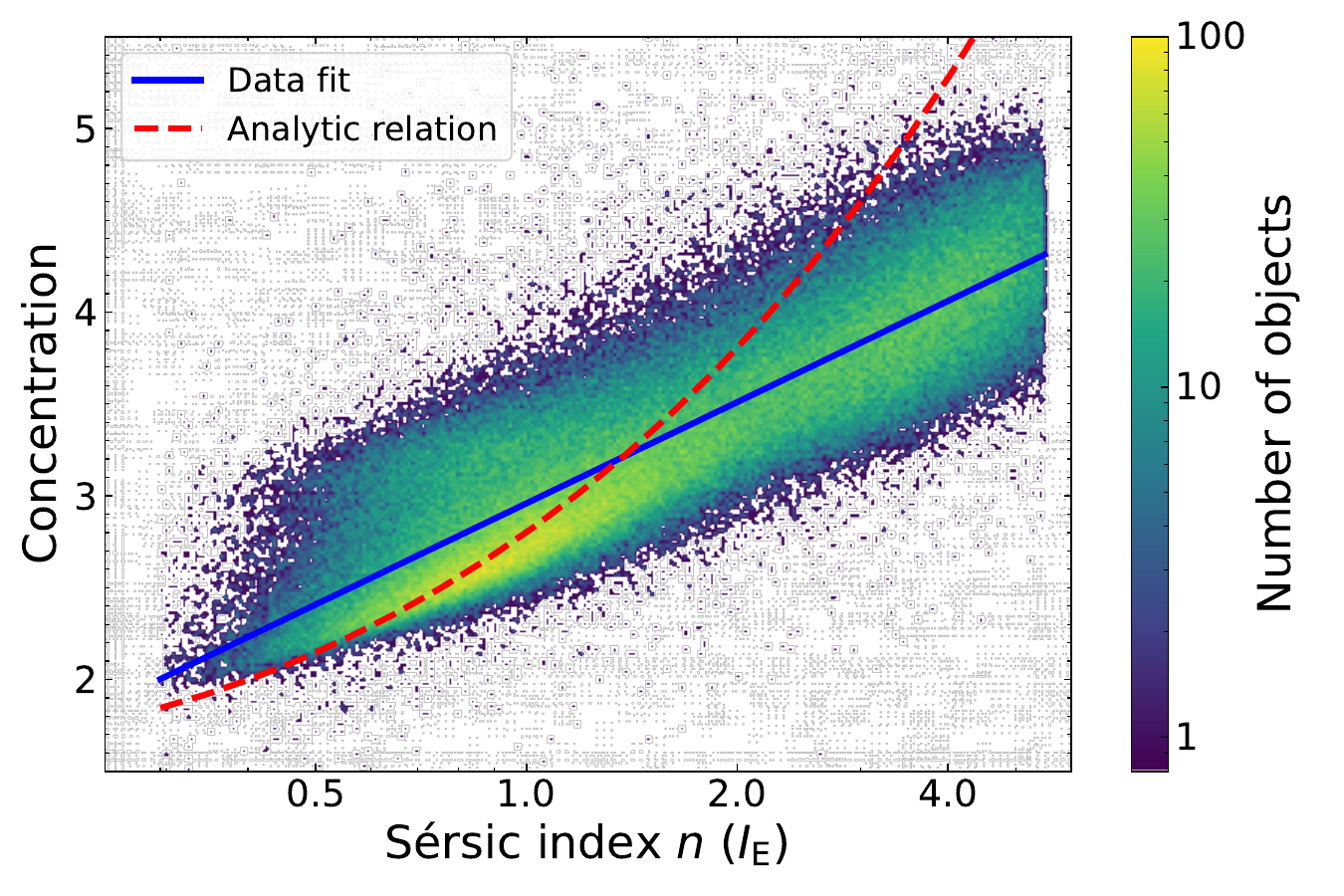}
\caption{Concentration versus $n$ for \Euclid galaxies with \IE$<~21$. The blue solid line indicates a linear fit of the data, whereas the red line is an analytical relation for pure S\'ersic profiles (see text).}
\label{fig:nsersic_vs_C}
\end{figure}

The concentration parameter, $C$, measures how concentrated the light distribution of a galaxy is in its inner regions, and is computed as $C=5\logten(r_{80}/r_{20})$, where $r_{80}$ and $r_{20}$ are the radii of circular apertures enclosing 80\% and 20\% of the total galaxy flux, respectively \citep{Conselice-2003-CAS-system, Q1-TP004}. Therefore, assuming that galaxies are well described by a S\'ersic profile, one would expect a correlation between these two parameters, i.e. high $n$ and concentration values for early-type galaxies, lower values for later morphologies (see \citealt{Andrae-2011-param-morphologies-potentials-pitfalls, Tarsitano-2018-DES-morpho-catalog, Wang-2024-PSF-effect-concentration-high-z}). Figure \ref{fig:nsersic_vs_C} shows that this expected correlation is obtained in the \Euclid morphological measurements for all galaxies with $\IE<21$, presenting a Pearson correlation coefficient of $r=0.82$ between $C$ and $\logten(n)$. Fitting a first degree polynomial for $C$ as a function of $\logten(n)$ yields
\begin{equation}
    C = 2.97 + 1.81 \logten(n)\,,
\end{equation}
with a $R^2$ score of 0.73, and a RMS dispersion of 0.32. This correlation therefore supports the reliability of both measures.
Moreover, for a pure S\'ersic profile, it is possible to compute analytically its concentration and it only depends on $n$ \citep{Graham-2005-sersic-considerations, Baes-Ciotti-2019-sersic-law-IV-energy-concentration}. This relation is shown as a dashed red line in \fg\ref{fig:nsersic_vs_C}. We note that for $n\sim 1$ it goes through the high density of points with the lowest $C$ values around 2.5. We suggest that this high density area corresponds to disc-like galaxies, whereas other objects at similar $n$ but higher $C$ might host a weak bulge or bar, which is not fitted by the S\'ersic profile but stills impact the concentration. For higher $n$ values, the analytical prediction leads to $C$ values well above what is measured because, while S\'ersic fits are convolved by the PSF, the computation of $C$ does not take it into account and is therefore biased (the PSF spreads the light and hence diminishes concentration). Combining the data fit and the analytical relation could allow one to derive $C$ values corrected for the PSF impact, as is done in \cite{Tarsitano-2018-DES-morpho-catalog}.

\subsection{\label{sc:sersic-vs-Zoobot} S\'ersic parameters versus \texttt{Zoobot}}

\begin{figure*}[ht]
\centering
\includegraphics[width=\columnwidth]{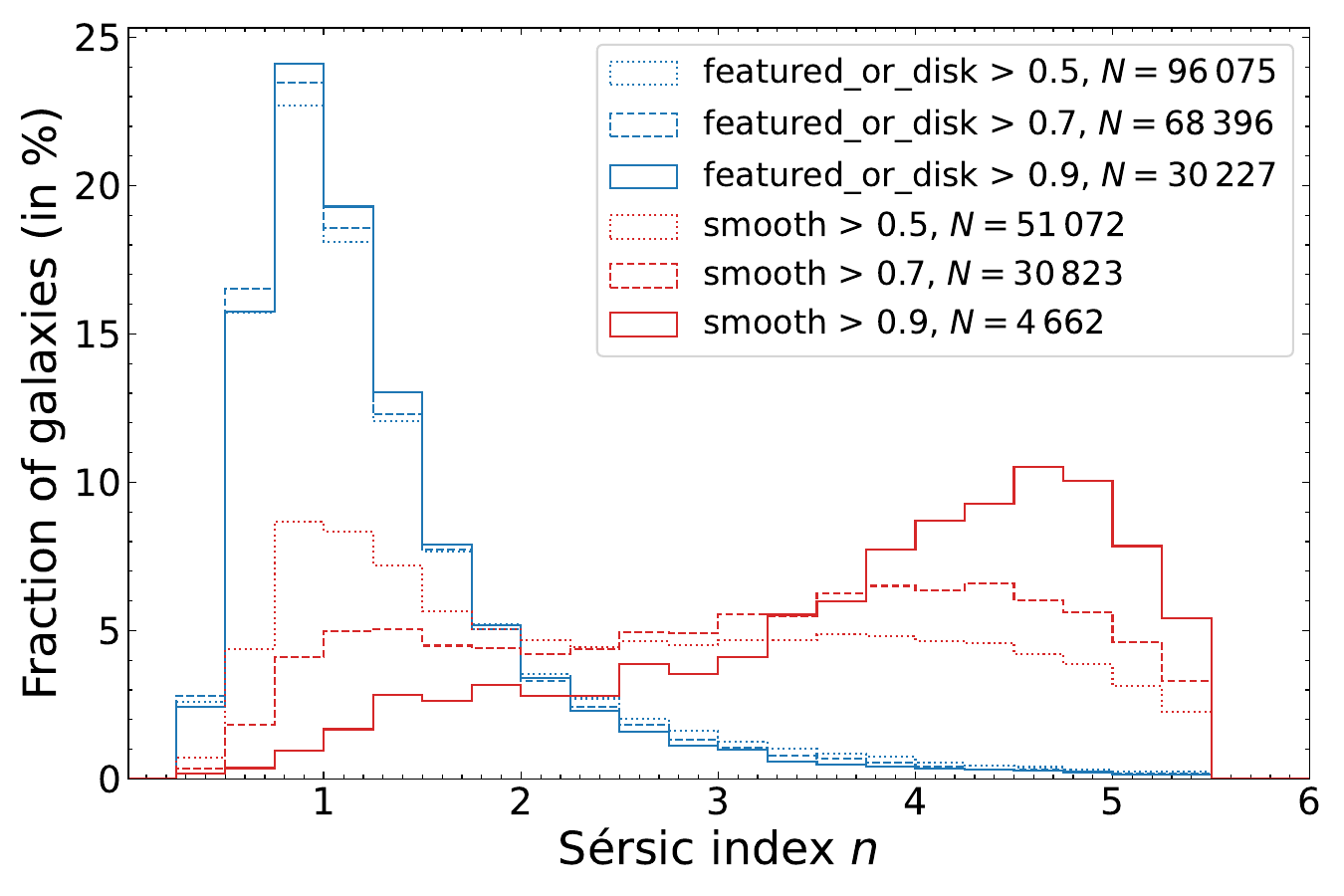}
\includegraphics[width=\columnwidth]{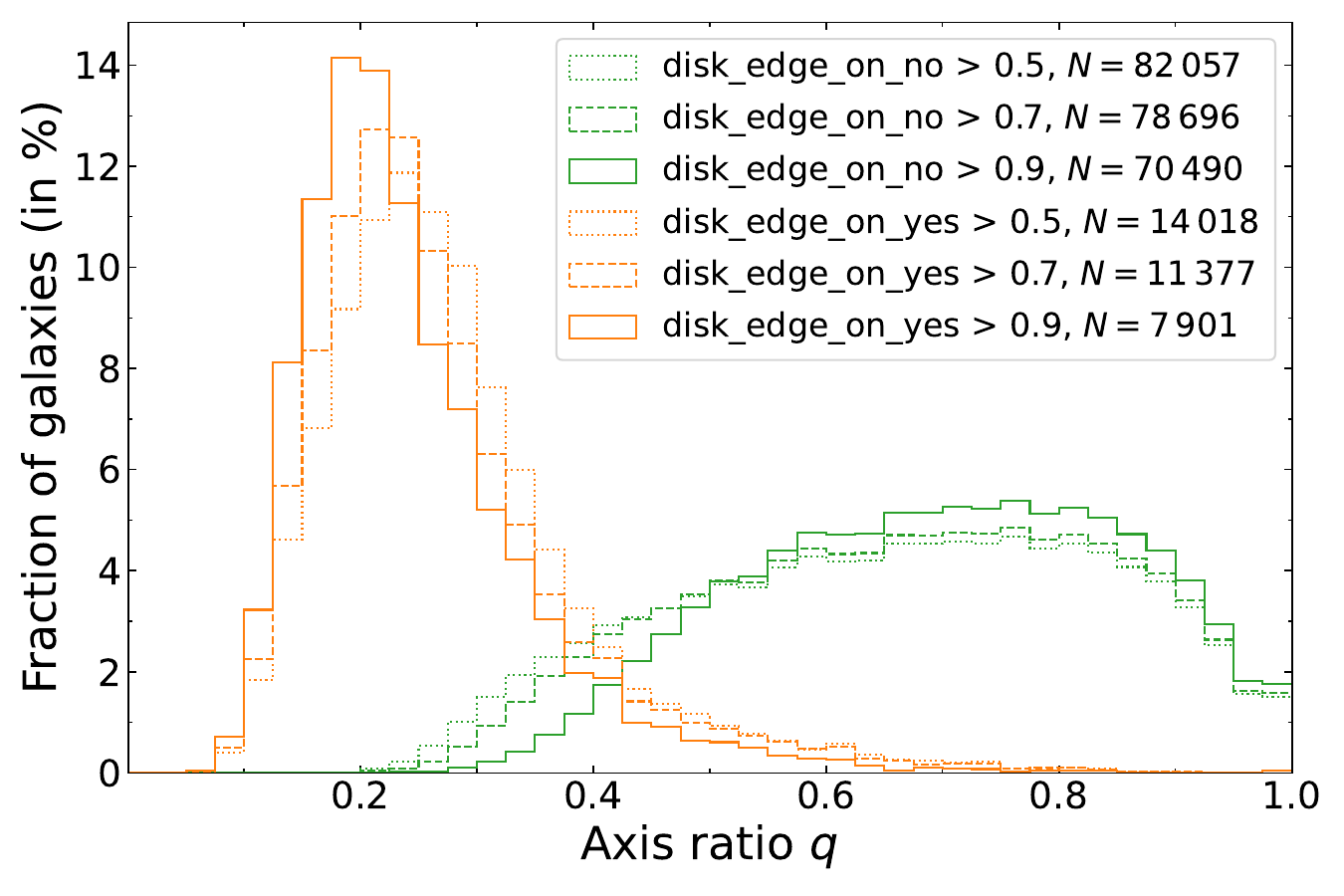}
\includegraphics[width=\columnwidth]{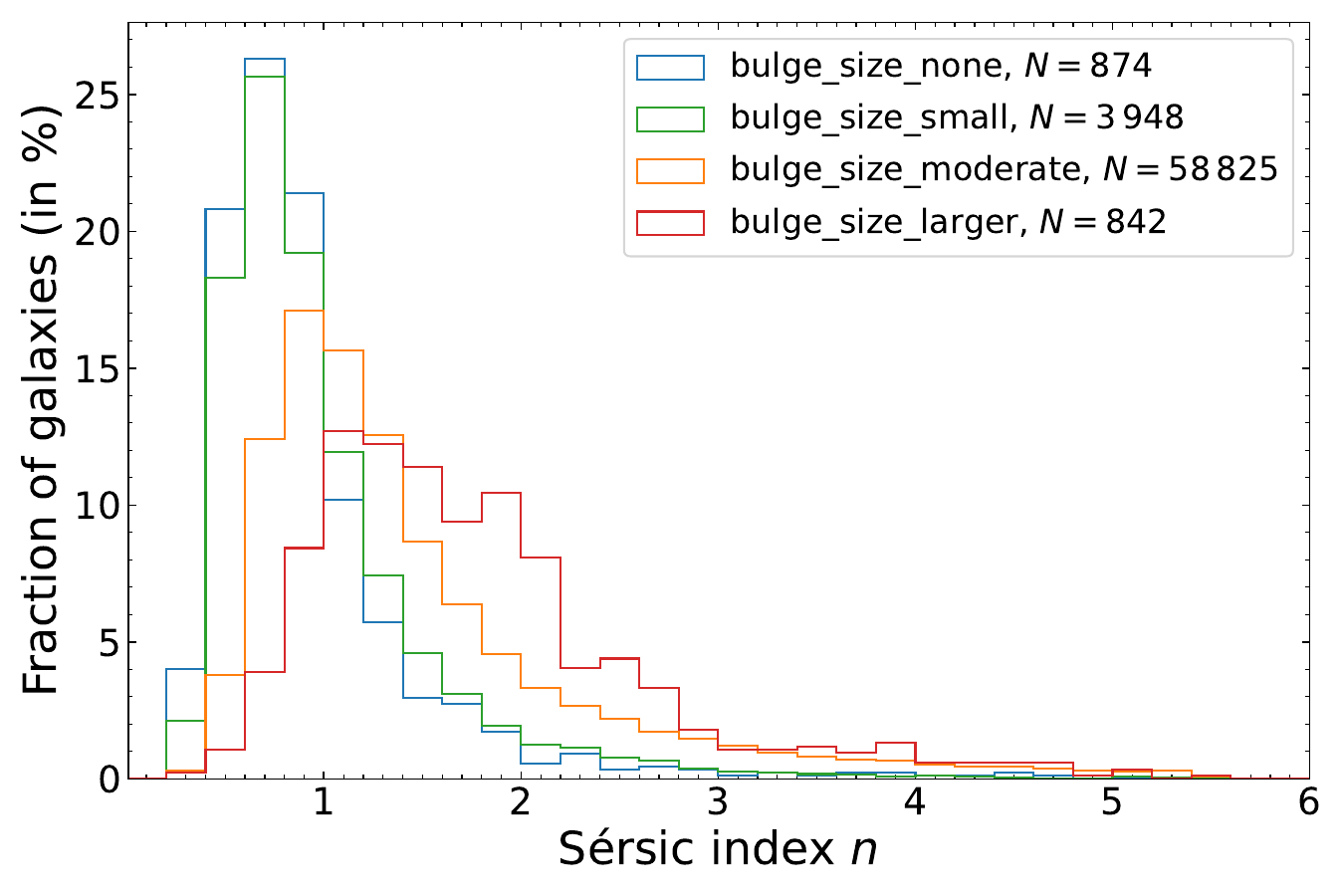}
\includegraphics[width=\columnwidth]{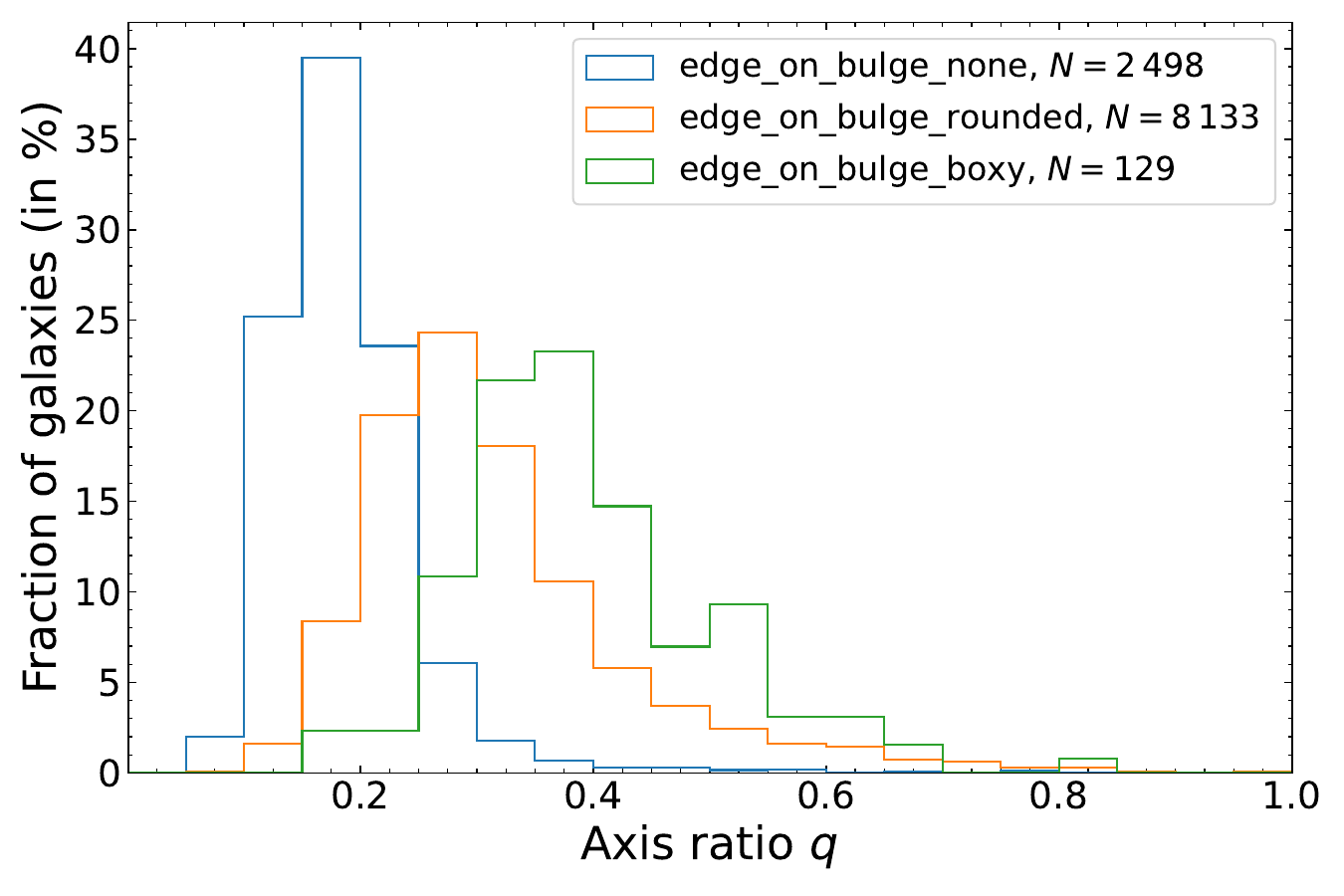}
\caption{Distribution of S\'ersic parameters measured on the VIS image for sub-samples of galaxies depending on their classification by the \texttt{Zoobot} labels. The different line styles correspond to the different shares of expected votes predicted by \texttt{Zoobot} to consider that a galaxy falls into a certain category. Sample sizes are systematically indicated in the legend, and a 50\% threshold of voters was adopted when not specified (bottom plots). \textit{Top left}: Distribution of $n$ for galaxies classified as either smooth (red lines) or displaying a disc or features (blue lines) shows a clear dichotomy with the latter sample being almost exclusively made of galaxies with $n\sim1$, whereas the former contains a lot of galaxies with high values of $n$ and increasingly so for higher selection threshold. \textit{Top right}: Distribution of $q$ for disc galaxies depending if they are edge-on or not. Again, a clear bimodality appears, with low $q$ for edge-on galaxies and higher values spread across a wider interval for non-edge-on (i.e. face-on to moderately inclined) galaxies. Using higher thresholds reinforces this dichotomy. \textit{Bottom left}: Distribution for disc or featured galaxies of $n$ for different categories of \texttt{Zoobot} bulge sizes. Histograms shift to higher values of $n$ for larger relative bulge sizes. \textit{Bottom right}: Distribution of $q$ for edge-on disc galaxies, depending on if they display no bulge, a boxy bulge, or a rounded bulge. The presence of a bulge, either rounded or boxy, leads to a higher $q$.}
\label{fig:sersic-vs-zoobot}
\end{figure*}

Next, we compared the results of S\'ersic fits to machine-learning based labels obtained using \texttt{Zoobot}, which is presented in \cite{Q1-SP047}. We followed the tree structure of the various questions asked to citizens and for which an expected fraction of voters was predicted by \texttt{Zoobot}. The first question asked was `Is the galaxy simply smooth and rounded, with no sign of a disk?' with two possible answers: `smooth' and `disc or features'. If volunteers answered the latter, they were then asked to answer if the disc is edge-on or not. If they answered `edge-on', they were then asked about the presence of a either a round or boxy bulge sticking out of the disc profile, whereas if they answered `not edge-on', they were asked about the bulge size relative to the whole galaxy. Other questions are beyond the scope of this section. The sample used in this section corresponds to galaxies for which such labels are available, with $\IE<20.5$ or a segmentation area larger than 1200\,pixels (see \citealt{Q1-SP047}).

The top-left panel of \fg\ref{fig:sersic-vs-zoobot} shows the distribution of $n$ for galaxies depending on whether they appear smooth (red curves), or whether they appear as a disc or display features (blue curves). We consider different shares of predicted votes (50\%, 70\%, and 90\%, as dotted, dashed, and solid lines respectively) to consider that a galaxy falls into a category. The resulting number of galaxies for all samples are indicated in the figure insert. In all cases, the $n$ distribution for featured or disc galaxies peaks around 1, sharply decreasing for higher values of $n$, with a median of 1.1, and associated dispersion of 0.5 (estimated as half the 16--84th percentile range). This is the expected behaviour as disc-like galaxies are known to be preferably fitted by an exponential profile, corresponding to $n=1$. Using a more stringent share of voters slightly decreases the tail of high $n$ values with 90\% quantiles of 2.5, 2.3, and 2.2 for the 50\%, 70\%, and 90\% share of votes. Therefore, to benefit from larger statistics, selecting for disc or featured galaxies in the subsequent plots of this section is done using a predicted share of voters above 50\%. Regarding galaxies voted as smooth, they should correspond to ellipticals, which are usually characterised by high $n$ (especially $n=4$, which is the de Vaucouleurs profile, \citealp{De-Vaucouleurs-1948-r14}). Again, we observe the expected trend as the distribution of $n$ includes higher $n$ galaxies than the disc or featured ones, or than the global distribution shown in the left panel of \fg\ref{fig:distribs-sersic-params-percent}. However, here the chosen threshold has a significant impact on the resulting distribution of $n$, with median $n$ increasing from 2.5 to 3.3 and 4.0 for more stringent thresholds.

The top-right panel of \fg\ref{fig:sersic-vs-zoobot} shows the distribution of $q_{\rm Sersic}$ for disc or featured galaxies depending if they were classified by \texttt{Zoobot} as being edge-on (orange) or not (green), still using the three thresholds of 50, 70, and 90\% of expected votes. We observe again a striking consistency between the S\'ersic parameters and \texttt{Zoobot} labels, since there is a clear dichotomy in the distributions of $q$ between the galaxies identified as edge-on or not, with median values of 0.26 and 0.67, and dispersions of 0.09 and 0.20, respectively (using the 50\% threshold). Moreover, the two histograms barely overlap for intermediate $q\sim 0.4$, with 90\% of not edge-on discs having $q>0.4$ whereas 88\% of edge-on ones have $q<0.4$. Requiring higher shares of votes to decide if a galaxy is edge-on or not slightly reinforces this separation between the two histograms, with, for instance, respective medians of 0.22 and 0.70 using a 90\% threshold, but at the expense of statistics. Furthermore, if edge-on discs are expected to have minimal values of $q$, the presence of a bulge `bulging out' from the disc results in a modelled $q$ higher than the actual disc $q$, as shown by Q25a. This would explain the tail of higher $q$ values. This hypothesis is confirmed in the bottom-right panel of \fg\ref{fig:sersic-vs-zoobot} using the subsequent \texttt{Zoobot} question regarding the presence of a bulge for galaxies selected as disc or featured, and as edge-on (always using a 50\% threshold). Indeed, the resulting $q$ distributions show that in the absence of a bulge, it peaks around a median value of 0.18, with a dispersion of 0.05, and displays a sharper decrease than for all edge-on galaxies, with only 4\% of galaxies having $q>0.3$. The presence of a rounded or boxy bulge shifts the distribution towards higher $q$, with respective medians of 0.29 and 0.37, as well as 46\% and 83\% of the sub-samples displaying $q>0.3$.

\begin{figure*}[ht]
\centering
\includegraphics[width=2\columnwidth]{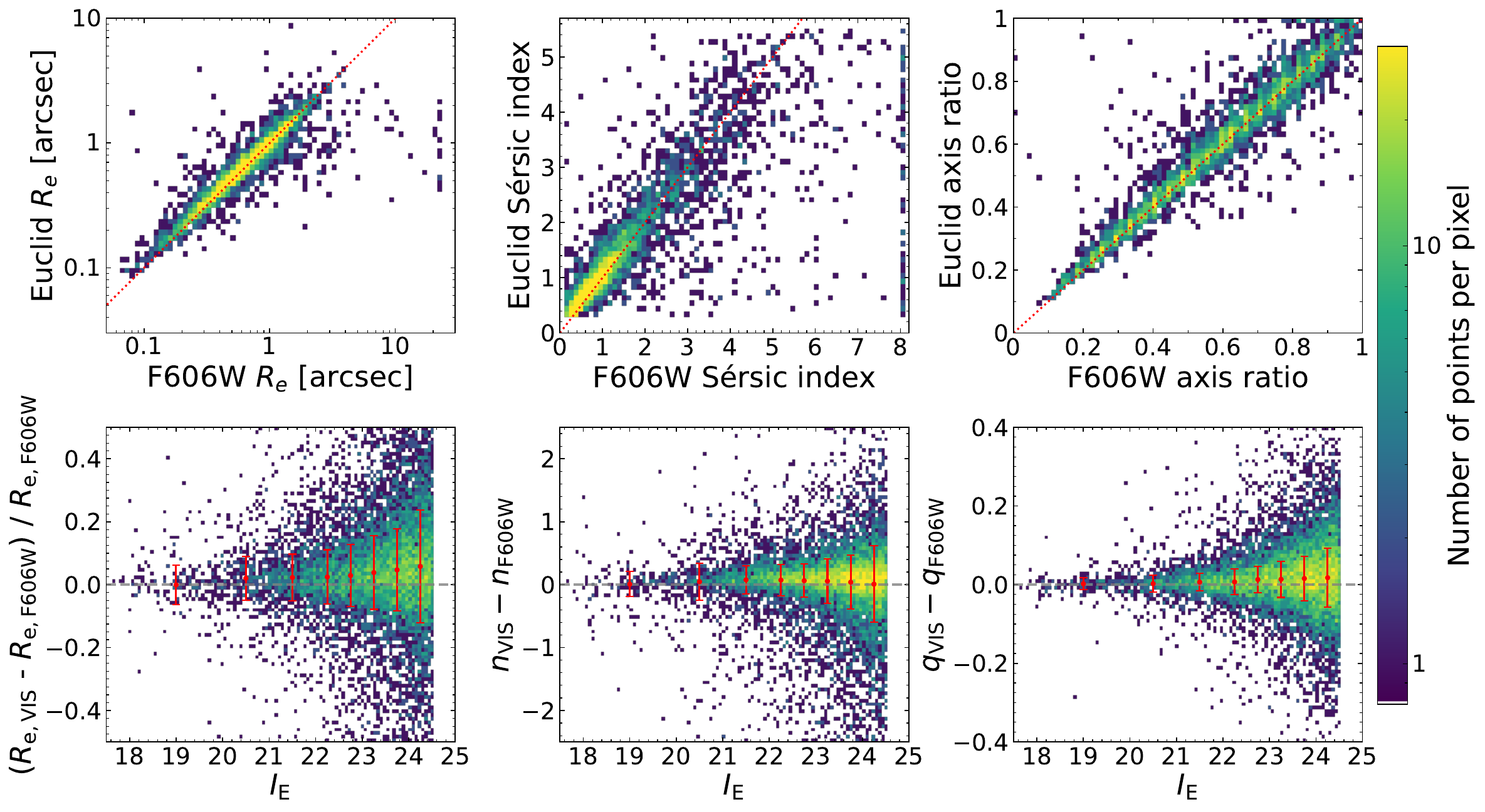}
\caption{Comparison between \Euclid and ACS-GC derived parameters. The top line shows the $R_\mathrm{e}$, $n$, and $q$ for galaxies with $\IE<23$. The bottom line includes galaxies with $\IE<24.5$ and shows the relative difference or difference between the same set of parameters as a function of the magnitude. The red points and error bars indicate median values and associated dispersions computed in bins of magnitude.}
\label{fig:compare_ACSGC}
\end{figure*}

Finally, in the bottom-left panel of \fg\ref{fig:sersic-vs-zoobot} we show the $n$ distributions of galaxies that are featured or discs not edge-on depending on their predicted bulge size, using for all labels a threshold of 50\% of predicted votes. The answers \texttt{bulge\_size\_large} and \texttt{bulge\_size\_dominant} have been grouped together as \texttt{bulge\_size\_larger} since they were rarer and because galaxies with very few expected votes in the three smaller bulge sizes often have split votes between these two larger categories, so none of them were above the 50\% threshold, but the expected votes repartition still indicated that \texttt{Zoobot} predicted a large bulge within these galaxies. Since the single S\'ersic profile fits both the light of the bulge and the disc, the more important the bulge flux is compared to the total galaxy flux, the higher the fitted $n$ should be. The current \texttt{Zoobot} question measures rather the ratio between the bulge and disc sizes, but this quantity is correlated to the bulge-to-total light ratio \citep{Quilley-2023-scaling-bulges-and-disks}. The distributions observed in the bottom-left panel of \fg\ref{fig:sersic-vs-zoobot} correspond to this prediction, with overall larger values of $n$ for larger bulge sizes predicted by \texttt{Zoobot}: the successive median values are 0.81, 0.85, 1.21, 1.60. The share of galaxies within the tail of high $n$ also increases, with 3.2\%, 3.6\%, 11.4\%, and 15.9\% of galaxies with $n > 2.5$, respectively. The lack of a significant difference between the \texttt{bulge\_size\_none} and \texttt{bulge\_size\_small}, especially compared to the gap seen between \texttt{bulge\_size\_small} and \texttt{bulge\_size\_moderate} can be explained by the fact that bulges have to reach some fraction of the total flux in order to impact and steepen the single S\'ersic fits, otherwise they are not modelled, as shown in Q25a.

Overall, the descriptions of galaxy morphology provided by parametric model-fitting and by machine-learning methods are consistent with each other, and the observed trends are those expected. This analysis also demonstrates the potential of using both information jointly.

\subsection{\label{sc:compare-ACSGC}Comparison with previous surveys}

In order to confirm the validity of S\'ersic parameters derived in the \Euclid pipeline, we compared them to previous measurements. We used the Advanced Camera for Surveys General Catalog \citep[ACS-GC hereafter]{Griffith-2012-ACS-catalog-sersic-fits}, containing the GEMS field, which overlaps with the EDF-F. In this case the S\'ersic fits were performed using HST images in the F606W band, with a pixel scale of \ang{;;0.03}, and using {\tt GALAPAGOS}. 

We performed a cross-match between our catalogue and ACS-GC. Two galaxies were considered to be the same object if the angular distance between them is smaller than \ang{;;1.0}. We found 11\,884 galaxies with $\IE<24.5$ and 3805 galaxies with $\IE<23$ in the 0.21\,deg$^2$ of overlapping area. Figure \ref{fig:compare_ACSGC} shows the comparison between the S\'ersic structural parameters derived from \Euclid and ACS-GC. The respectively derived $n$ and $q$ values are in agreement with each other, with median differences compatible with zero: They are below 0.1 for $n$ and below 0.01 for $q$. The $R_{\rm e}$ values display an offset, with VIS radii being larger than the ACS-GC radii, and increasingly so for fainter magnitudes: the median relative difference remains below 3\% for $\IE<23$ and increases to 6\% at the faintest magnitudes. This offset is likely to be a consequence of the different depths and resolutions of the images used in both analyses \citep{George-2024-two-rest-frame-wavelengths-galaxy-sizes-emerging-bulges}. For all three parameters, the scatter increases with magnitude.
The agreement found here, as well as the evidence detailed in the previous sub-sections, confirms the robustness of the S\'ersic fits implemented within the \Euclid pipeline. 

\section{\label{sc:results-color-variation}Variation of structural parameters with wavelength: Colour gradients}

\begin{figure*}[ht]
\centering
\includegraphics[width=2\columnwidth]{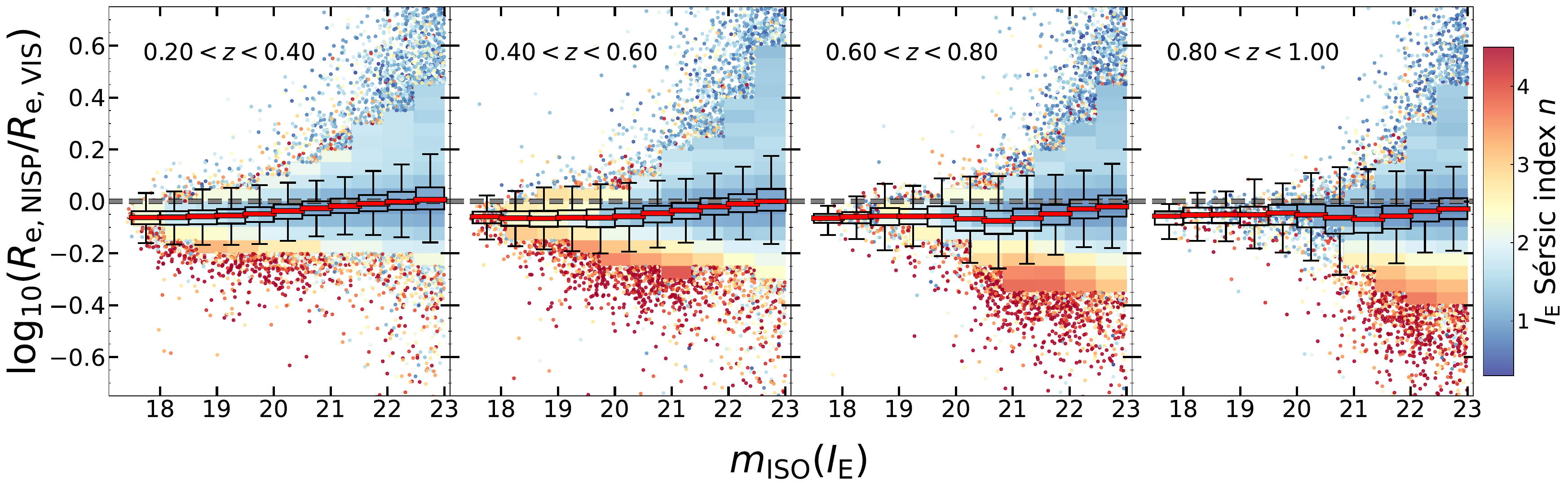}
\caption{Ratio between the effective radius based on the common S\'ersic model for three NISP images ($R_\mathrm{e,\, NISP}$) and that from the S\'ersic fit to the VIS image ($R_\mathrm{e,\, VIS}$) as a function of isophotal VIS magnitude. The four panels correspond to the redshift bins we used to trace evolutionary trends in \sct~\ref{sc:results-evol-plots}. The 2D histogram in each panel are colour-coded by the median $n_\mathrm{VIS}$ for galaxies in a given $[\Delta m_{\rm ISO} (\IE)=0.5$, $\Delta\logten(R_\mathrm{e,\, NISP}/R_\mathrm{e,\, VIS})=0.05]$ pixel. The 2D histogram includes pixels with $N>100$ galaxies. In more sparsely populated regions we instead display individual galaxies colour-coded by their $n_\mathrm{VIS}$. The box-and-whisker diagrams show median values (red solid lines), interquartile range (IQR, boxes) and the extent of the 1.5\,IQR range above the 75th and below the 25th percentile (whiskers) for $R_\mathrm{e,\, NISP}/R_\mathrm{e,\, VIS}$ in $\Delta m_{\rm ISO}=0.5$ bins of isophotal \IE. The grey dashed line corresponds to the constant ratio of $R_\mathrm{e,\, NISP}/R_\mathrm{e,\, VIS}=1$.}
\label{f7}
\end{figure*}

\begin{figure*}[ht]
\centering
\includegraphics[width=2\columnwidth]{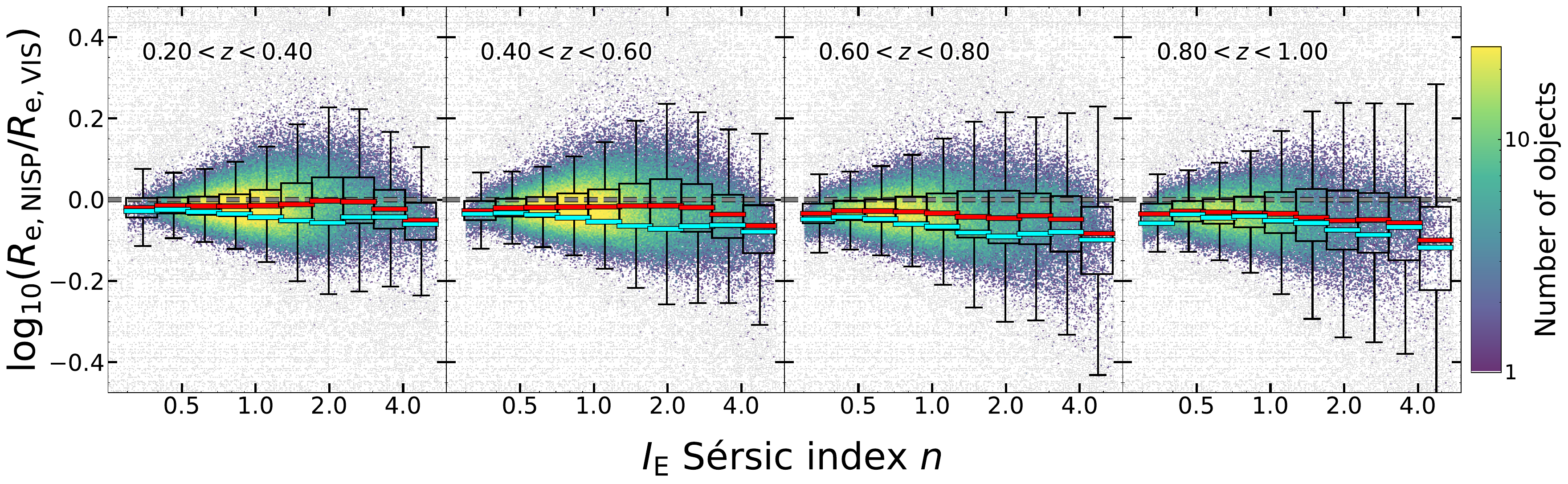}
\caption{NISP-to-VIS $R_\mathrm{e}$ ratio as a function of $n_\mathrm{VIS}$ for galaxies with isophotal VIS magnitude $m_{\rm ISO}(\IE)<23$. The four panels correspond to the same redshift bins as in \fg\ref{f7}. In each panel, the density map is colour-coded by the number of galaxies in a given pixel. The box-and-whisker diagrams show median values (red lines), interquartile range (IQR, boxes) and the extent of the 1.5\,IQR range above the 75th and below the 25th percentile (whiskers) for $R_\mathrm{e,\, NISP}/R_\mathrm{e,\, VIS}$ ratios in 10 bins of $n_\mathrm{VIS}$. For comparison, we show median $R_\mathrm{e,\, NISP}/R_\mathrm{e,\, VIS}$ ratios in the same $n_\mathrm{VIS}$ for galaxies with the isophotal VIS magnitude $m_{\rm ISO}(\IE)<21$ (cyan solid lines). The grey dashed line corresponds to the constant ratio of $R_\mathrm{e,\, NISP}/R_\mathrm{e,\, VIS}=1$.}
\label{f8}
\end{figure*}

\begin{figure*}[ht]
\centering
\includegraphics[width=2\columnwidth]{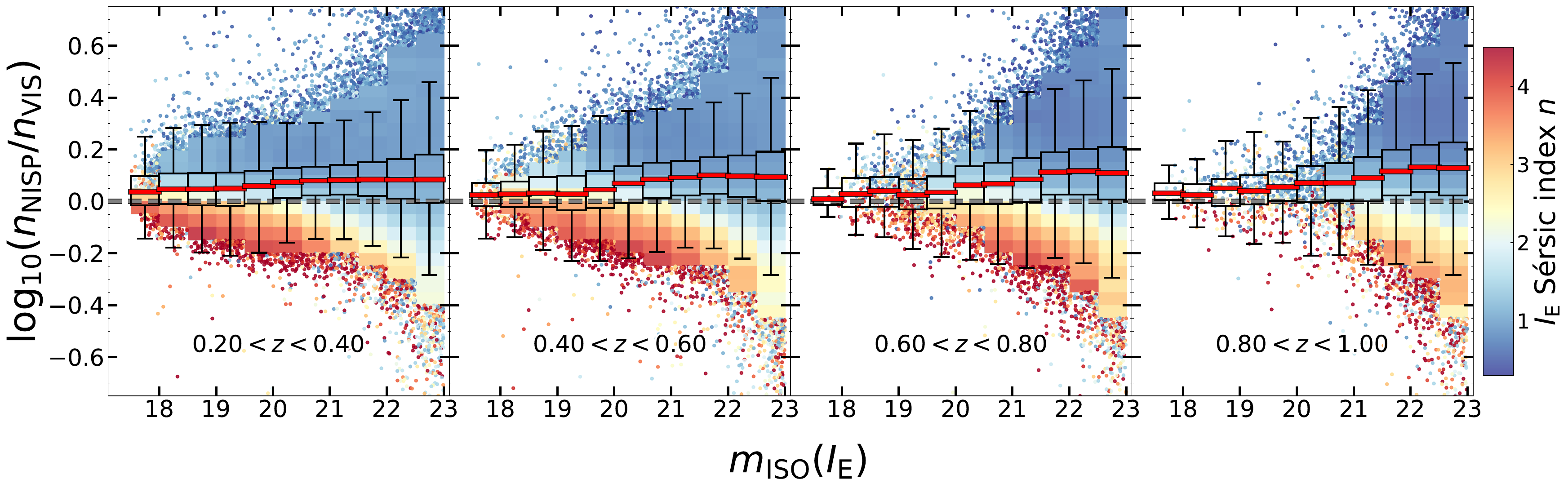}
\caption{Equivalent to \fg\ref{f7} but for the ratio of $n$ in NISP and VIS bands as a function of galaxy isophotal VIS magnitude.}
\label{f9}
\end{figure*}

\begin{figure*}[ht]
\centering
\includegraphics[width=2\columnwidth]{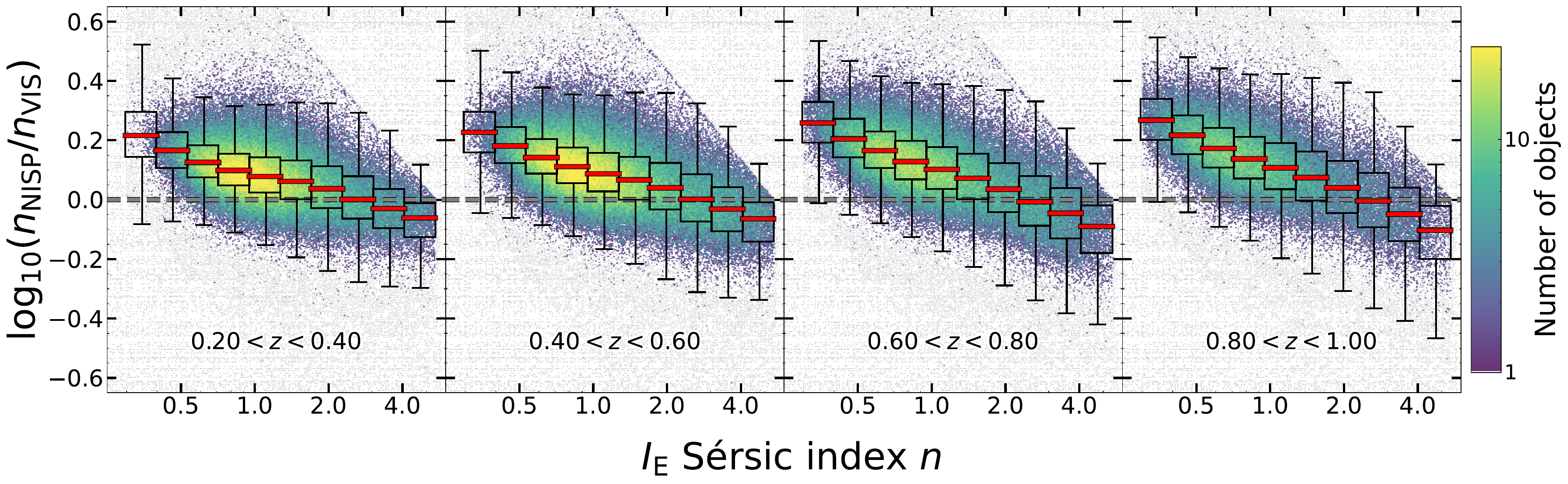}
\caption{Equivalent to \fg\ref{f8} but for the ratio of $n$ in NISP and VIS bands as a function of $n_\mathrm{VIS}$. We omit the binned median values of the index ratio for $m_{\rm ISO}(\IE)<21$ sample because they are nearly identical to the median values for the parent sample with $m_{\rm ISO}(\IE)<23$.}
\label{f10}
\end{figure*}

\begin{table*}
\centering
\caption[]{Redshift evolution of ratios of the fitted structural parameters.}
                \begin{tabular}{ccccccc}
                \hline \hline \noalign{\vskip 1.5pt}
                \multicolumn{3}{c}{} &\multicolumn{4}{c}{Redshift} \\
                \multicolumn{3}{c}{} & $0.2<z<0.4$ & $0.4<z<0.6$ & $0.6<z<0.8$ & $0.8<z<1$\\
                \hline
                        \multirow{4}{*}{$\frac{R_\mathrm{e,\, NISP}}{R_\mathrm{e,\, VIS}}$}&\multirow{2}{*}{$\IE<21$} &
            $n_\mathrm{VIS}<2.5$&$0.914^{+0.001}_{-0.001}$&$0.891^{+0.001}_{-0.001}$&$0.859^{+0.002}_{-0.002}$&$0.889^{+0.004}_{-0.005}$\T\B\\
            \cline{3-7}
           &&$n_\mathrm{VIS}>2.5$ &$0.894^{+0.001}_{-0.002}$&$0.856^{+0.003}_{-0.002}$&$0.821^{+0.005}_{-0.004}$&$0.820^{+0.020}_{-0.008}$\T\B\\  
            \cline{2-7}

            &\multirow{2}{*}{$\IE<23$}&$n_\mathrm{VIS}<2.5$&$0.968^{+0.001}_{-0.001}$&$0.959^{+0.001}_{-0.001}$&$0.940^{+0.001}_{-0.001}$&$0.926^{+0.001}_{-0.001}$\T\B\\
            \cline{3-7}

            &&$n_\mathrm{VIS}>2.5$ &$0.845^{+0.002}_{-0.002}$&$0.919^{+0.001}_{-0.002}$&$0.888^{+0.002}_{-0.002}$&$0.866^{+0.003}_{-0.003}$\T\B\\  
            \hline

            \multirow{4}{*}{$\frac{n_\mathrm{NISP}}{n_\mathrm{VIS}}$}&\multirow{2}{*}{$\IE<21$} &
             $n_\mathrm{VIS}<2.5$&$1.230^{+0.002}_{-0.002}$&$1.268^{+0.002}_{-0.002}$&$1.250^{+0.006}_{-0.004}$&$1.220^{+0.010}_{-0.010}$\T\B\\
             \cline{3-7}
            &&$n_\mathrm{VIS}>2.5$ &$0.939^{+0.003}_{-0.002}$&$0.947^{+0.002}_{-0.002}$&$0.963^{+0.005}_{-0.005}$&$0.950^{+0.010}_{-0.001}$\T\B\\  
             \cline{2-7}
             &\multirow{2}{*}{$\IE<23$}&$n_\mathrm{VIS}<2.5$&$1.231^{+0.001}_{-0.001}$&$1.275^{+0.001}_{-0.001}$&$1.345^{+0.001}_{-0.001}$&$1.401^{+0.002}_{-0.002}$\T\B\\

             \cline{3-7}
             &&$n_\mathrm{VIS}>2.5$ &$0.934^{+0.002}_{-0.002}$&$0.929^{+0.002}_{-0.002}$&$0.902^{+0.002}_{-0.003}$&$0.893^{+0.004}_{-0.003}$\T\B\\  
             \hline
                \end{tabular}
        
        \tablefoot{The median ratio of structural parameters $R_e$ and $n$ measured in NISP and VIS images for galaxies in two $m_{\rm ISO}(\IE)$-limited samples and four redshift bins of the $0.2<z<1$ interval. We separate both magnitude-limited samples into $n<2.5$ and $n>2.5$ subsets based on the single S\'ersic fits to the VIS images. The upper and lower limits correspond to the $1\, \sigma$ bootstrap confidence interval.}
        \label{table:ratios}
\end{table*}

The colours of galaxies vary spatially across their extent, with, in most cases, the inner region of the bulge showing redder colour than the disc \citep{Mollenhoff-2004-BD-spirals-UBVRI, Vika-2014-multiband-BD-decomposition-MegaMorph, Kennedy-2016-GAMA-color-gradients-vs-BT-color-BD, Casura-2022-bulge-disk-decomposition-GAMA}, but the variations go beyond that dichotomy and also arise within bulges or discs \citep[][Q25a]{Natali-1992-disk-gradient-NGC-628, Balcells-1994-bulge-color-gradients, de-Jong-1996-BD-decomposition-IV-stellar-dust-gradients, Pompei-Natali-1997-7-spiral-disk-gradients}. These colour gradients are the result of stellar population (age and metallicity) gradients within galaxies and differential dust attenuation. It is estimated that the relative contribution of stellar and dust gradients to the measured colour gradient is of the order of $80\%$ and $20\%$ respectively \citep{Baes-2024-Re-mass-light-wavelength}. Therefore, the observation and quantification of colour gradients provide constraints for the evolutionary scenarios of the galaxy stellar mass build-up. Moreover, studies of galaxy colour gradients provide a valuable insight into biases affecting the shear measurements for weak lensing analysis \citep{Semboloni-2013-shear-bias-galaxy-profile-color, Er-2018-color-gradient-bias-shear}.

In photometric studies, spatial variations are mostly investigated radially to test the inside-out or outside-in mass assembly models \citep[e.g.][]{Avila-Reese2018}. This can be done using a series of aperture magnitudes of increasing radius or, in the present case, modelling galaxy 2D profiles with the same analytic function in different bands, and inspecting how the parameters of the model vary between bands. For S\'ersic profiles, variations in R$_\mathrm{e}$ and $n$ with the observing band are indicative of colour gradients. 

In order to assess \Euclid's ability to detect colour gradients in large galaxy samples, we take advantage of the fact that the present data release provides two sets of S\'ersic parameters, one in \IE and one common to the three NISP bands. We note that the previous study of Q25a on the \Euclid ERO data has shown that the structural parameters of a common model for the \YE, \JE, and \HE bands lead to values close to an independent model for the \JE image, due to its central wavelength position between \YE and \HE bands and the smoothness of the variation of the structural parameters with wavelength in these neighbouring bands. 

Figures~\ref{f7}--\ref{f10} show trends in structural parameter ratios $R_\mathrm{e}$ and $n$ with \IE  and $n_{\rm VIS}$ across four $\Delta z=0.2$ redshift bins, which match the intervals explored in \sct\ref{sc:results-evol-plots}. Table~\ref{table:ratios} lists median values for different sample subsets in these bins. Smaller R$_\mathrm{e}$ and more concentrated profiles (higher $n$) in NISP compared to VIS are both indicative of redder-inside colours (and vice versa).

For the sample described in \sct\ref{sc:Data}, the median S\'ersic $R_{\rm e}$ in combined NISP bands is about $\sim10\%$ smaller than the R$_{\rm e,\,VIS}$ for all galaxies at $z<1$ up to \IE$\lesssim21$ (red solid lines in \fg\ref{f7}, see also Table~\ref{table:ratios}). The trend significantly weakens at fainter magnitudes (\IE$>21$) in the $0.2<z<0.4$ and $0.4<z<0.6$ intervals, whereas it increases at magnitudes brighter than $\IE\approx19.5$ in the two higher redshift intervals, then weakens starting from $\IE\approx21$.
A previous analysis of the \Euclid ERO data with two-component fits to galaxy light profiles showed that the decrease in $R_\mathrm{e}$ with the wavelength of observation is a combination of two effects: bulge-disc colour dichotomy and the colour gradients within disc component (Q25a), affecting the dominating population of spiral galaxies detected in the present sample. Thus, the ratio between effective radii in observed NISP and VIS  ($R_\mathrm{e,\, NISP}/R_\mathrm{e,\, VIS}$) in \fg\ref{f7} traces, on average, a more compact structure of longer-wavelength light profiles for all galaxies with \IE$\le21$, which can be reliably fitted with two-component (bulge+disc) S\'ersic models (see \citealp{Bretonniere-EP26}; Q25a). 

At fixed magnitude, the ratio distribution follows a trend with $n_\mathrm{VIS}$ (\fg~\ref{f7}). For \IE$\lesssim20$ and $z<0.8$, a broad range of median $R_\mathrm{e,\, NISP}/R_\mathrm{e,\, VIS}$ values ([0.4, 1.6]) of galaxies with $n\gtrsim2.5$ envelops the $R_\mathrm{e,\, NISP}/R_\mathrm{e,\, VIS}\lesssim1$ for disc ($n\lesssim1$) galaxies. Densely populated cells with low median values of $n_\mathrm{VIS}$ drive the median ratio of $\approx0.87$ at these magnitudes. At \IE$\sim21$, galaxies with $n\gtrsim2.5$ have $R_\mathrm{e,\, NISP}$ up to 60\% smaller than $R_\mathrm{e,\, VIS}$. In contrast, the $R_\mathrm{e,\, NISP}$ of disc galaxies ($n\lesssim1$) in the same \IE and redshift bins can be 2.5 times {\it larger} than $R_\mathrm{e,\, VIS}$.  As in brighter magnitude bins, $n\sim1$ (hence likely spiral) galaxies dominate the most populated cells at \IE$\sim21$ and drive the median ratio of radii in two (sets of) bands (Table~\ref{table:ratios}).

Combined magnitude bins show that the median ratio of galaxy $R_\mathrm{e}$ in the two observed wavelength regimes depends mildly on $n_{\rm VIS}$ (\fg\ref{f8}). The median offset between $R_\mathrm{e,\, NISP}/R_\mathrm{e,\, VIS}$ ranges from approximately $-5\%$ to $-10\%$ (of $R_\mathrm{e,\, VIS}$) for galaxies with \IE$<23$ and $n_\mathrm{VIS}$ from about $\sim1$ to 4, respectively (red solid lines in \fg\ref{f8}, Table~\ref{table:ratios}). 
Since the median ratio of the two radii at fixed $n$ is driven by the most numerous fainter galaxies (see the distribution of cells in \fg\ref{f7}), and because both simulated images and the analysis of ERO data indicate that the bulge-disc decomposition of \Euclid galaxies becomes challenging at $\IE\ge21$ (\citealp{Bretonniere-EP26}; Q25a), we also explore median values as a function of $n$ for the \IE$<21$ sample (cyan lines in \fg\ref{f8}, Table~\ref{table:ratios}). For this brighter subset, the median $R_\mathrm{e,\, NISP}/R_\mathrm{e,\, VIS}$ ratio is below the value for the parent sample in all $n$ bins across the redshift interval we probe, and the pattern of variations with $n_\mathrm{VIS}$ is similar between the four redshift intervals. The larger difference between two sets of ratios at from $n\sim1$ to $n\sim4$, and especially around $n\sim2$ highlights the average redder-inside colour gradient of galaxies that can be reliably fitted by the sum of bulge and disc profiles. Thus \Euclid Q1 data set reveals trends in galaxy structural properties that will inform future studies using more detailed parametric 2D models.    

The S\'ersic index of galaxies in our sample also changes with the observed wavelength range. Figure \ref{f9} illustrates the trend in the NISP-to-VIS $n$ ratio ($n_\mathrm{NISP}/n_\mathrm{VIS}$) with the galaxy isophotal \IE in the four $z<1$ redshift bins explored. The median values of this ratio (red solid lines) consistently exceed unity (grey dashed lines), with the profile concentration index measured in combined NISP bands being between $\sim$10\% and 30\% larger than the equivalent measurement in VIS band for \IE$<23$ galaxies (Table~\ref{table:ratios}). We note that the same trend has been observed for galaxies at $0.5<z<3$ in \cite{Martorano-2023-sersic-index-wavelength-dependence-JWST} using S\'ersic fits on JWST images.

At a fixed \IE, the distribution of individual $n_\mathrm{NISP}/n_\mathrm{VIS}$ ratios is a function of $n_\mathrm{VIS}$ (\fg\ref{f9}). The most centrally concentrated VIS light profiles ($n_\mathrm{VIS}\gtrsim 2.5$) exhibit NISP profiles with up to $\sim40\%$ and to $\sim60\%$ (for the \IE$<21$ subset and for the full \IE$<23$ sample) smaller $n$. Together with, on average, smaller R$_\mathrm{e,\,NISP}$ in galaxies with centrally concentrated VIS light profiles from \fg~\ref{f7}, this trend could illustrate the build-up of outer halos around central bulges in these systems. 
In contrast to galaxies with the most concentrated VIS light profile, the majority of disc galaxies ($n_{\rm VIS}<2$) have more centrally concentrated NISP profiles, with $n_{\rm NISP}$ up to $\times$\,2.5 or up to four times larger (for the \IE$<21$ subset and for the full \IE$<23$ sample, respectively) than $n_{\rm VIS}$. 
The median values of the $n_\mathrm{NISP}/n_\mathrm{VIS}$ in Table~\ref{table:ratios} quantify further the difference between $n_\mathrm{VIS}<2.5$ and $n_\mathrm{VIS}>2.5$ systems. This trend is expected if combined NISP images are more sensitive to the redder central (bulge-like, $n\sim4$) than VIS. However, we note that the highest ratio ($n_{\rm NISP}\gtrsim2.5\,n_{\rm VIS}$) are here attributed to galaxies with $n_{\rm VIS}<1$ (dark-blue cells in \fg~\ref{f9}).

We next combine all \IE$<23$ bins to explore the trend in $n$ ratio with $n_\mathrm{VIS}$ in \fg\ref{f10}. Galaxies in our sample follow the same median trend of decreasing $n_\mathrm{NISP}/n_\mathrm{VIS}$ ratio with increasing $n_\mathrm{VIS}$ in all four redshift bins explored. The majority of galaxies in our parent sample are disc-dominated ($n_\mathrm{VIS}<2$, \fg\ref{fig:distribs-sersic-params-percent}), showing more concentrated NISP profiles ($n_\mathrm{NISP}>n_\mathrm{VIS}$, \fg\ref{f9}). For disc-dominated galaxies with $n_\mathrm{VIS}\sim1$, $n_\mathrm{NISP}$ are roughly 25\% and 40\% larger than $n_\mathrm{VIS}$ in $z\sim0.3$ and $z\sim0.9$, respectively. For extreme $n_\mathrm{VIS}\lesssim0.5$ values, this difference increases to $\sim65\%$ at $0.2<z<0.6$ and $\sim85\%$ at $0.6<z<1$, respectively. 

The decreasing trend in the $n_\mathrm{NISP}/n_\mathrm{VIS}$ ratio with $n_\mathrm{VIS}$ is significant, exceeding the median trend with magnitude illustrated in \fg\ref{f9}. We note again, however, that the large differences at very low $n_\mathrm{VIS}$ values do not imply a drastic change in morphology from disc-like to bulge-like between VIS and NISP. Nevertheless, the $n_\mathrm{NISP}/n_\mathrm{VIS}$ ratio $>1$ for all galaxies with $n_\mathrm{VIS}<2.5$ (quantified also in Table~\ref{table:ratios}), suggest that a more concentrated inner structure (i.e. bulge) affects single S\'ersic light profiles of disc-dominated galaxies at longer wavelengths, regardless of the exact rest-frame wavelength probed with VIS and NISP. In contrast, the median ratios close to unity for galaxies with significant bulge-like structure ($n_\mathrm{VIS}\gtrsim2.5$) suggest more subtle effects for these early-type galaxies, in agreement with Q25a. 

Based on the \Euclid Q1 galaxy data set and their single-S\'ersic profile fits, a combination of the median trend of larger $n$ in NISP bands with respect to VIS ($n_\mathrm{NISP}/n_\mathrm{VIS}>1$, \fg\ref{f10}) for galaxies with prominent disc component ($n_\mathrm{VIS}<2$) and the median $R_\mathrm{e,\, NISP}/R_\mathrm{e,\, VIS}\lesssim1$ trend (\fg\ref{f8}) supports an inside-out mass assembly scenario \citep[e.g.][]{Munoz-Mateos2007,Pezzulli2015,Baker-2025-core-SF-disk}. Furthermore, the median $n$ ratio close to unity and the smaller $R_\mathrm{e,\, VIS}$ for bulge-dominated ($n_{\rm VIS}>2.5$) galaxies align with the predictions of the `two-phase' formation scenario with a rapid early phase ($z>2$) of in situ star-formation followed by an extended phase of (gas and/or star) accretion \citep[e.g.][]{Naab2009,Oser2010,Hilz2012,Rodriguez-Gomez2016}.
Statistical galaxy samples from future \Euclid data releases spanning a broad range in redshift, stellar mass, and star-formation activity and with structural parameters for both the bulge and disc component will be ideal platforms for probing multiple scenarios of galaxy mass assembly.

\begin{figure}[ht]
\centering
\includegraphics[width=\columnwidth]{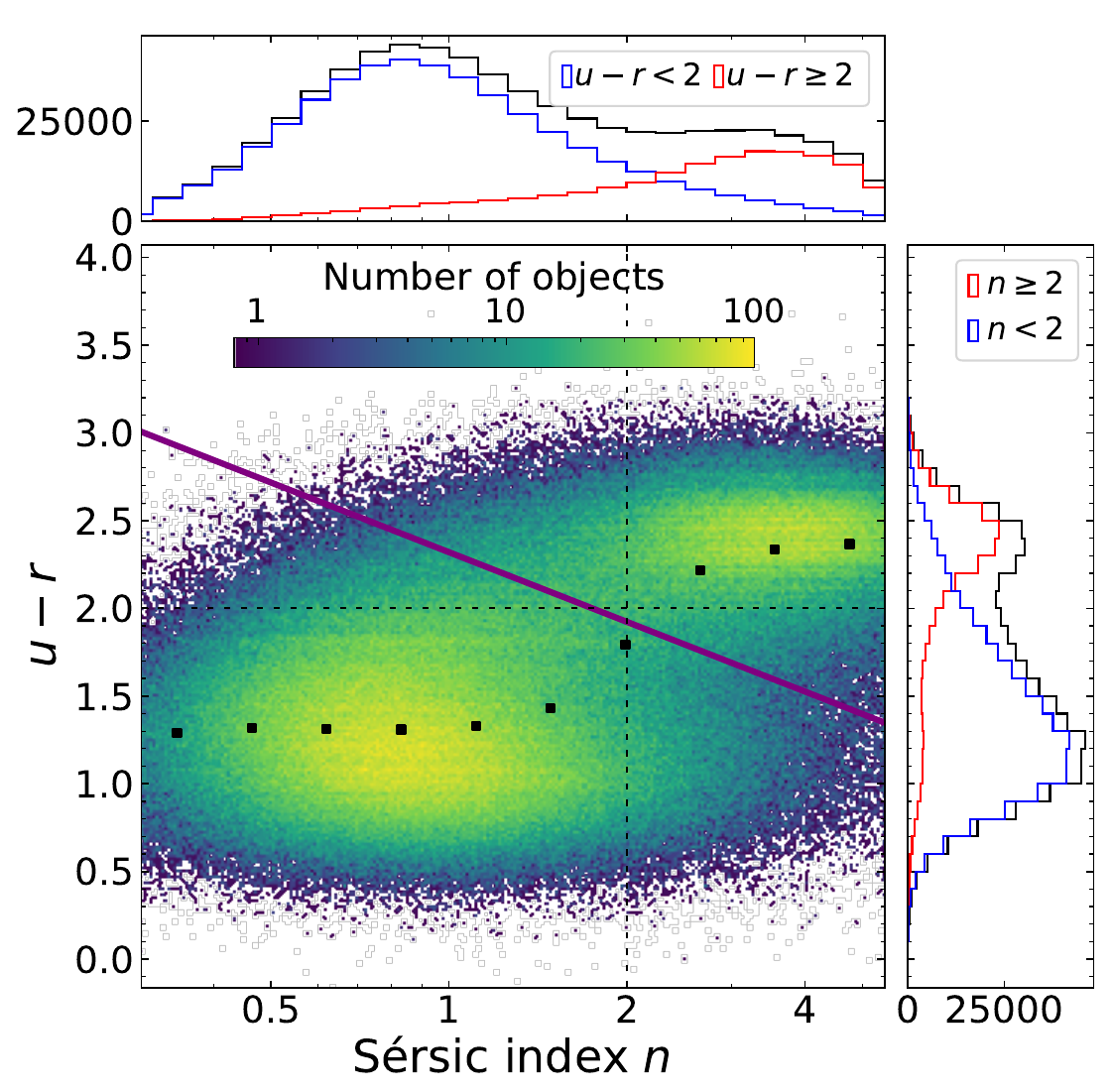}
\caption{Rest-frame $(u-r)$ colour versus $n$ for all galaxies with \IE$\leq23$. Black squares indicate the running median colour in bins of $n$. Histograms on the right and top plots show the 1D distributions for each quantity. A clear colour--morphology bimodality in the galaxy populations is displayed, with two density peaks corresponding respectively to blue spiral galaxies with $n\sim$\,1 and $(u-r)\sim$\,1.2, and red early-type galaxies of $n\sim$\,4 and $(u-r)\sim$\,2.4. We use this plane to define the early- and late-type galaxies sub-samples, separated by the solid purple line.}
\label{fig:colour-n-bimod}
\end{figure}

We also emphasise that for galaxies at different redshifts, variations in $R_\mathrm{e}$ and $n$ across bands have different implications, since galaxies at different redshifts are fitted with single S\'ersic profiles in different rest-frame bands. Therefore, the similar pattern of $R_\mathrm{e}$ and $n$ ratios between the observed VIS and NISP bands measured must be interpreted with caution since the wavelength and redshift effects are entangled.
For instance, at $z\approx1$, the \IE band probes mostly wavelengths below $4000\,\AA$ whereas the NISP central band \JE probes roughly the 6000--7500\,$\AA$ range. As bulge-to-total ratios that contribute to single-S\'ersic gradients increase more strongly from UV to optical than from optical to near-infrared, in the absence of redshift evolution, a steeper gradient is therefore expected in the $z\sim0.8$\,--1.0 sub-sample (hence ratios of $R_\mathrm{e}$ and $n$ deviating more from 1) than for the lowest redshift sub-sample. The fact that such an effect is unseen does not exclude redshift evolution in the galaxy internal colour distributions between $z\approx0$ to $1$, as it could be compensated by the effect of the shift in rest-frame bandwidth.
Interpreting the variations of the colour gradients requires precise redshift determination and careful analysis of stellar populations and dust effects, as well as the sample selection effects and the different PSF and angular resolution in VIS compared to NISP. A detailed study is left for future work.  

\section{\label{sc:results-evol-plots}Variations of scaling-laws with morphology for redshifts $z<1$}

In this section, we investigate how morphology correlates with the stellar mass and star-forming state of galaxies by examining the position of different galaxy morphologies across well-established scaling relations, namely the size--mass relation and the SFR--mass relation, which highlights the main sequence of star-forming galaxies. Redshifts, absolute magnitudes, stellar masses, and SFRs used in this section are extracted from the catalogue of physical parameters presented in \citet{Q1-TP005}. These quantities were derived by the SED-fitting of \Euclid near-IR photometry and optical ground-based external data.

\subsection{\label{sc:results-nser-colours-bimod}Colour-morphology bimodality}

\begin{figure}[ht]
\centering
\includegraphics[width=\columnwidth]{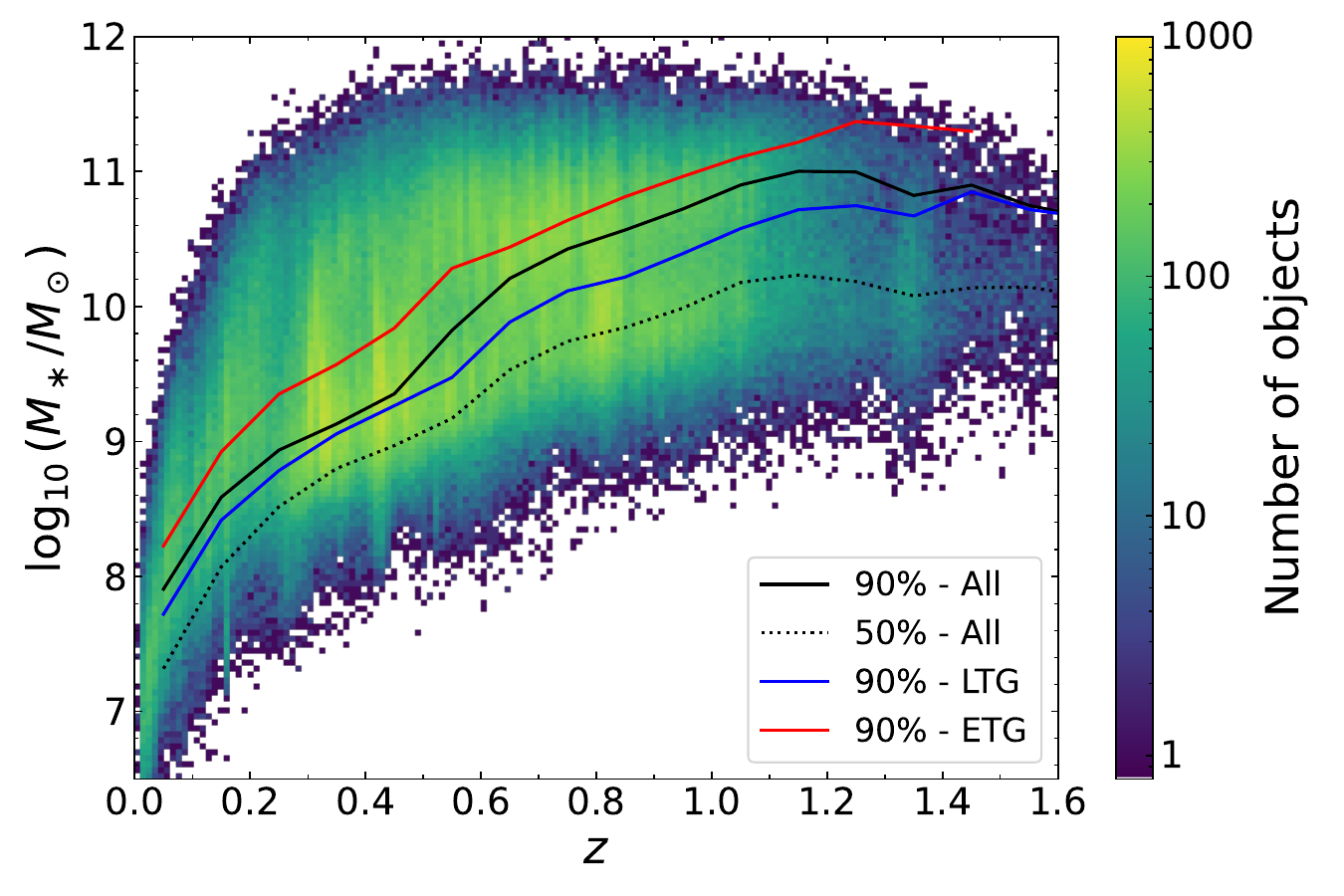}
\caption{Stellar mass versus photometric redshift of galaxies with \IE$\leq23$. Following the methodology of \cite{Pozzetti-2010-zCOSMOS-bimod-SFM}, we computed the mass completeness limits of the full sample at the 90\% and 50\% levels, shown as solid and dotted black lines, respectively, as well as for the sub-samples of late- and early-type galaxies, with their 90\% mass completeness limits shown in solid blue and red lines, respectively.}
\label{fig:mass-completeness-plot}
\end{figure}

The colour bimodality of galaxy populations \citep{Strateva-2001-color-bimodality, Baldry-2004-color-bimodality} shows a clear correlation between colour and morphology: early-type galaxies are predominantly red, whereas later types are mostly blue. However, further investigation by \cite{Quilley-2022-bimodality}, locating all the morphological types in the colour--mass diagram, showed that this connection goes beyond a simple bimodality, with the Hubble sequence monotonously spanning this plane.
Figure\,\ref{fig:colour-n-bimod} shows the $(u-r)$ distribution with $\logten(n)$. This colour is obtained from the absolute magnitudes derived in these bands from the SED-fitting of \Euclid and external photometry (see \citealp{Q1-TP005}). A colour--concentration bimodality emerges, with blue galaxies having mostly $n\sim$\,1, typical of the exponential profiles of discs, whereas red galaxies display $n\sim$\,4, more indicative of an early-type morphology (elliptical or lenticular). This is in agreement with the previous analysis of \citet{Allen-2006-MGC-BD-decomp}, which first highlighted this 2D colour--concentration bimodality. We recall that, due to the chromaticity of the VIS PSF not being taken into account (see \sct\ref{sc:Methodo}), redder (bluer) galaxies tend to have underestimated (overestimated) $n$, so the colour-S\'ersic index bimodality should appear even more separated (but likely by a non-significant amount).
We used \fg\ref{fig:colour-n-bimod} to define one sub-sample of late-type galaxies (LTGs) and another of early-type galaxies (ETGs), by following the criteria that ETGs and LTGs respectively have $(u-r)>(u-r)_{\rm lim}(n)$ and $(u-r)\leq(u-r)_{\rm lim}(n)$, where
\begin{equation}
    (u-r)_{\rm lim}(n)=2.32-1.32\logten(n)\,.
    \label{eq:colour-n-separation}
\end{equation}
We obtained this limit by locating the two density peaks, finding the minimum point density across this segment and similarly for parallel segments above and below it, and finally fitting a straight line through these density minima. For the rest of the analysis, we refer to ETG and LTG galaxies or sub-samples based on this definition.

\begin{figure*}[ht]
\centering
\includegraphics[width=\textwidth]{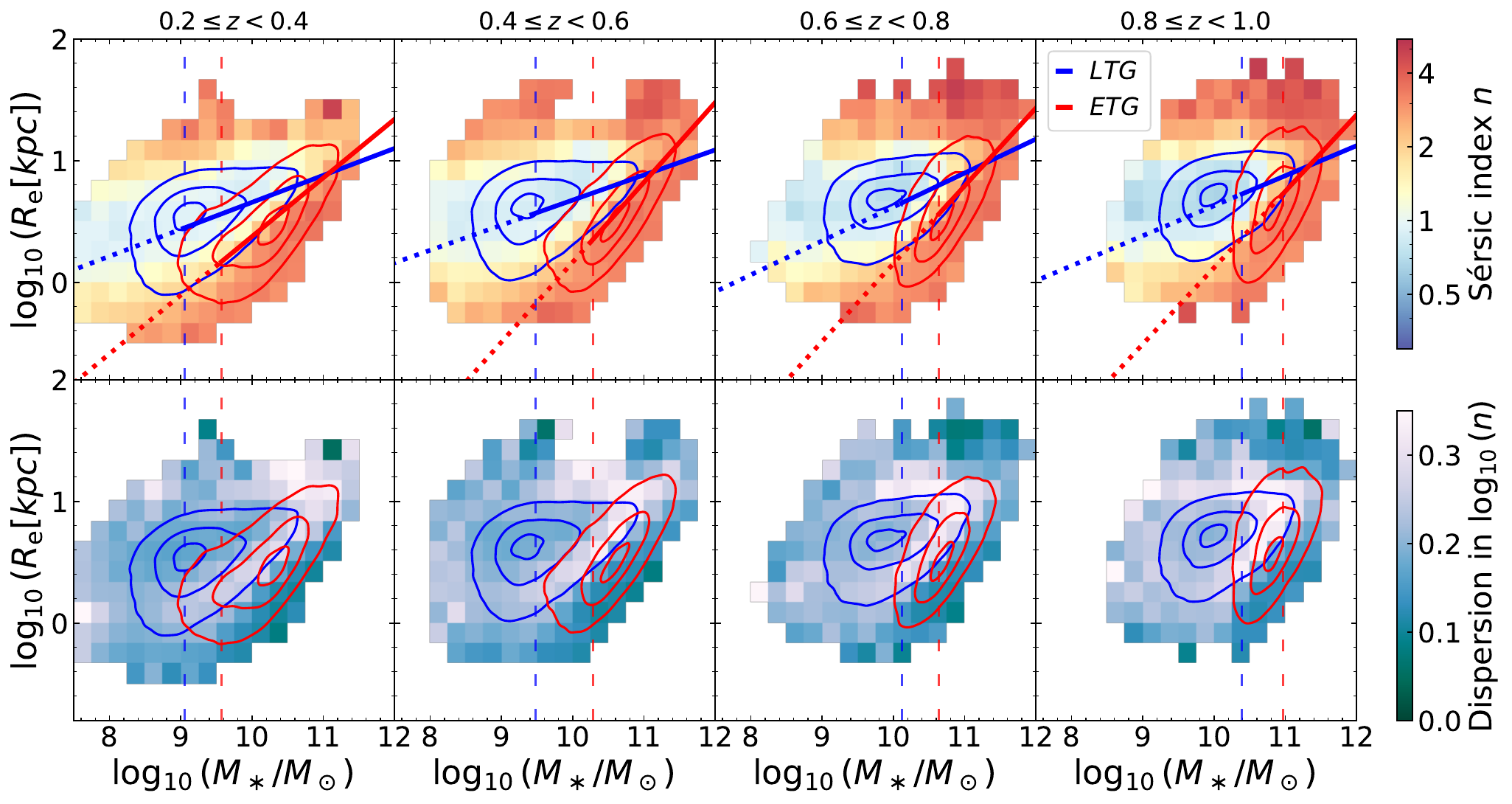}
\caption{Diagrams of $M_\ast$--$R_\mathrm{e}$ for the four consecutive intervals of $z$ colour-coded by the median $n$ (top row) and dispersion around it (bottom row) in each cell. We over-plot density contours at the 10\%, 50\%, and 90\% levels as well as vertical dashed lines indicating the 90\% mass completeness limit for both the ETG and LTG samples (red and blue, respectively). Power-law fits were performed on each sample above the 90\% mass completeness limit and are displayed as solid lines in their range of validity, while extended as dashed lines for lower masses. At all redshifts, the slope of the $M_\ast$--$R_\mathrm{e}$ relation is steeper for ETG than for LTG, indicating a transition between two regimes of stellar mass assembly. Low values of dispersion are found in areas of the diagram dominated by either ETG or LTG, whereas the cells with the highest dispersions correspond to the transition region between the two.}
\label{fig:size-mass}
\end{figure*}

\subsection{\label{sc:selection-effects}Completeness of the sample }

In order to evaluate the completeness of our sample, we plot in \fg\ref{fig:mass-completeness-plot} the $M_\ast$--$z$ plane for galaxies with \IE$\leq 23$ and reliable physical parameters (see \sct\ref{sc:Data}), as well as reliable $n$ fits.
We used the method developed in \cite{Pozzetti-2010-zCOSMOS-bimod-SFM} to compute the mass completeness limit at 90\%, and 50\% (solid and dotted black lines) in redshift bins with widths of 0.1. The \IE$\leq23$ limit leads to mass completeness values rapidly rising with redshift, with a 90\% completeness above $\logten(M_\ast/M_\odot)=9.35$ and $10.72$ at $z=$0.5 and 1.0, respectively.
We then repeated the same process for the sub-samples of ETGs (red line) and LTGs (blue line). Here, their 90\% completeness limits differ because galaxies of different colours (and morphologies) are characterised by different mass-to-light ratios. For instance, the 90\% completeness limits at $z=1.0$ are $\logten(M_\ast/M_\odot)=$10.59 and 10.96 for LTG and ETG galaxies, respectively.

\subsection{\label{sc:results-nser-size-mass}Morphology in the stellar mass--size relation}

Over the past two decades, a multitude of studies have consistently shown clear trends between $M_\ast$ and $R_\mathrm{e}$ for both star-forming and quiescent galaxies of $\logten(M_\ast/M_\odot)\gtrsim9$ and up to $z\sim3$ \citep[e.g.][]{shenSizeDistributionGalaxies2003, Trujillo06sizeevolutionfromz3, guoStructuralPropertiesCentral2009, williamsEvolvingRelationsSize2010,  moslehEvolutionMassSizeRelation2012,  vanderwel3DHSTCANDELSEvolution2014, langeGalaxyMassAssembly2015, faisstConstraintsQuenchingMassive2017, royEvolutionGalaxySizestellar2018, Tortora-2018-dark-matter-assembly-KiDS, Mowla-2019-size-mass-CANDELS, matharuHSTWFC3Grism2019, matharuHSTWFC3Grism2020, kawinwanichakijHyperSuprimeCamSubaru2021,  mercierScalingRelations2512022,George-2024-two-rest-frame-wavelengths-galaxy-sizes-emerging-bulges}. The parameters of the $M_\ast$--$R_\mathrm{e}$ relation
differ between the two galaxy populations, with star-forming galaxies exhibiting a single power-law relation between galaxy size and stellar mass: 
\begin{equation} 
    \logten\left(\frac{R_\mathrm{e}}{\rm kpc}\right)=\logten\left(\frac{R_\mathrm{0}}{\rm kpc}\right)+\alpha\logten\left(\frac{M_\ast}{5\times 10^{10}M_\odot}\right)\,,
    \label{eq:size-mass}
\end{equation}
with $\alpha\sim0.2$ and a characteristic size, $R_{\rm 0}$, that increases with decreasing redshift \citep[e.g.][]{shenSizeDistributionGalaxies2003,  williamsEvolvingRelationsSize2010, vanderwel3DHSTCANDELSEvolution2014, Mowla-2019-size-mass-CANDELS, kawinwanichakijHyperSuprimeCamSubaru2021, baroneLEGACSAMIGalaxy2022,George-2024-two-rest-frame-wavelengths-galaxy-sizes-emerging-bulges}.  

On the contrary, the quiescent population exhibits a more complex relation between $M_\ast$ and $R_\mathrm{e}$, parametrised by a broken power law. 
Below a pivot mass at $\sim3\times10^{10}M_{\odot}$, quiescent systems follow a trend similar to the equivalent relation for star-forming galaxies \citep[e.g.][]{morishitaGrismLensamplifiedSurvey2017, kawinwanichakijHyperSuprimeCamSubaru2021}. In  contrast, the exponent of the power-law $M_\ast$--$R_\mathrm{e}$ relation for quiescent galaxies above this pivot mass is significantly higher ($\alpha\sim0.7$) and their characteristic size decreases with redshift much faster \citep[e.g.][]{shenSizeDistributionGalaxies2003, williamsEvolvingRelationsSize2010, Bernardi-2011-size-mass-ETG-curvature-dry-mergers, newmanCanMinorMerging2012, Huertas-Company13sizevolutionearlytypez1,langeGalaxyMassAssembly2015, huangRelationsSizesGalaxies2017, Mowla-2019-size-mass-CANDELS, moslehGalaxySizesPerspective2020, kawinwanichakijHyperSuprimeCamSubaru2021, nedkovaExtendingEvolutionStellar2021, damjanovSizeSpectroscopicEvolution2023,George-2024-two-rest-frame-wavelengths-galaxy-sizes-emerging-bulges}.
This pivot stellar mass coincides with the equivalent parameter of the $M_\ast$--halo mass relation \citep{Stringer-2014-galaxy-sizes-consequence-cosmology, mowlaMassdependentSlopeGalaxy2019, kawinwanichakijHyperSuprimeCamSubaru2021}. 
The similarity between the pivot masses of both relations suggests that the change in the power-law exponent of the quiescent size--stellar mass relation indicates the stellar mass where ex situ driven growth via mergers and accretion starts to dominate over the in situ galaxy growth via star-formation in the progenitors of quiescent systems. 

\begin{figure*}[ht]
\centering
\includegraphics[width=\textwidth]{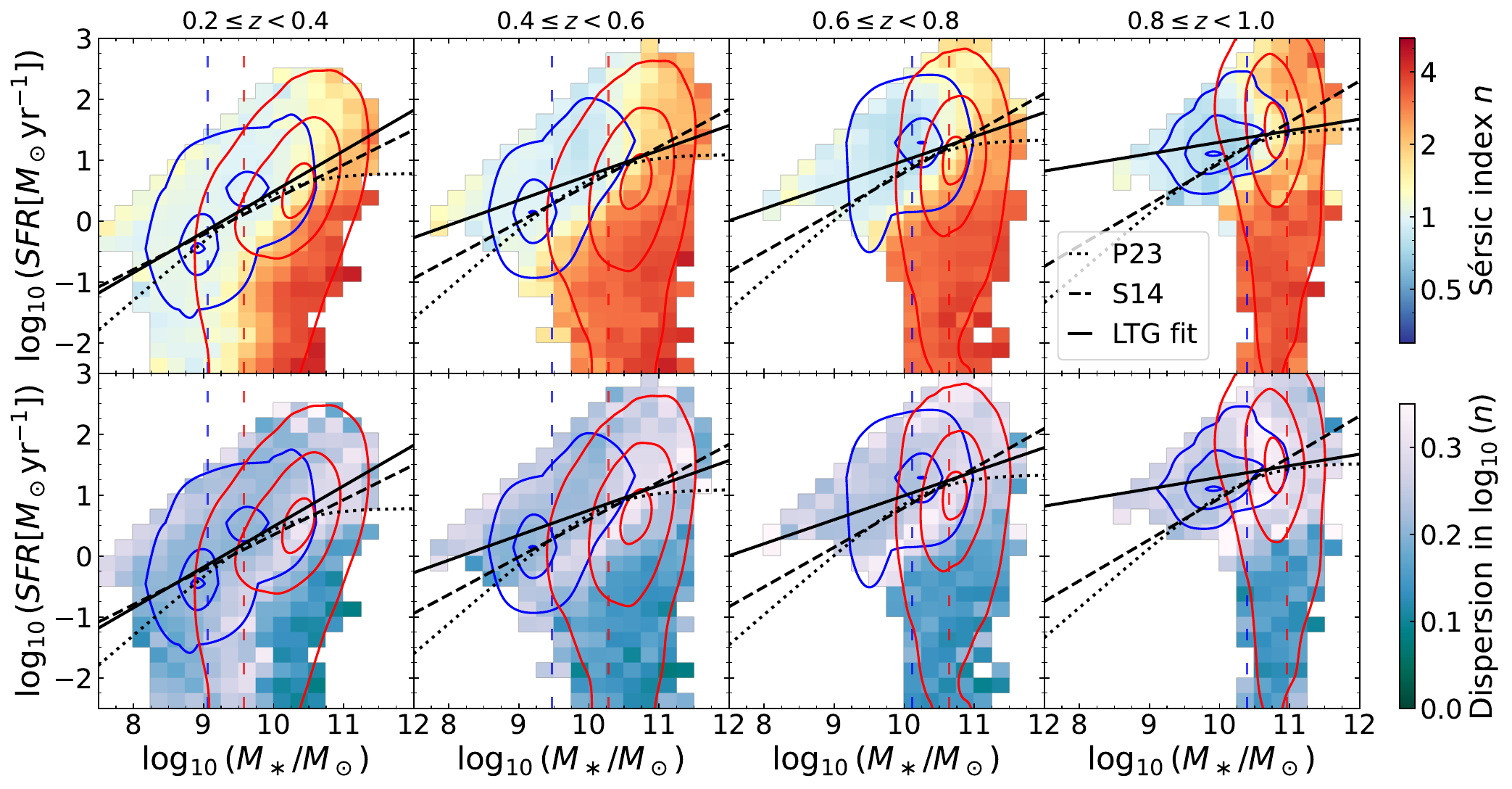}
\caption{Star-formation rate versus $M_\ast$ diagrams for the four consecutive intervals of redshift colour-coded by the median $n$ (top row) and associated dispersion around it (bottom row) in each cell of the plane. We over-plot density contours at the 10\%, 50\%, and 90\% levels as well as the 90\% mass completeness limit for the ETG and LTG samples (red and blue vertical dashed lines, respectively). The SFMS from \cite{Speagle-2014-Main-Sequence} and \cite{Popesso-2023-Main-Sequence} are displayed as black dashed and dotted lines, respectively, whereas a fit to the $M_\ast$--SFR relation of our sample of LTG galaxies is shown as a black solid line.
}
\label{fig:sfr-mass}
\end{figure*}

We explore how the change in the power-law exponent of the $M_\ast$--$R_\mathrm{e}$ relation varies with  $n$ for Q1 galaxies. Figure \ref{fig:size-mass} shows this relation for all galaxies up to $\IE<23$ in the previously studied redshift bins. The colour-maps indicate the median $n$ and dispersion around it, for the top and bottom plots respectively, in each cell of the $M_\ast$--$R_\mathrm{e}$ plane, whereas the contours indicate the distributions of the LTG and ETG samples defined in \fg\ref{fig:colour-n-bimod}, at the 10\%, 50\%, and 90\% levels. Cells outside the largest of the series of contours correspond to much lower statistics at the tails of distributions. We recall here that the PSF colour bias is such that redder (bluer) galaxies tend to have their $R_{\rm e}$ overestimated (underestimated). As these biases are only by a factor of $\sim 10^{-3}$, their impact on the $R_{\rm e}$ values displayed here by the red ETGs and blue LTGs can be neglected.
Power-law fits were performed for the $M_\ast$--$R_\mathrm{e}$ relation on both sub-samples, shown as solid-red and dotted-blue lines, respectively (for  galaxies with $M_\ast>90$\% completeness limit at a given $z$). All fits are performed on at least 10\,000 galaxies but due to the mass completeness limit, they were made on a restricted mass interval, which limits their quality. These fits allowed us to see that at all redshifts considered, the $M_\ast$--$R_\mathrm{e}$ relation displays a much steeper slope for the ETG sample than for the LTG one, with consecutive values in each $z$-bin of $\alpha$ of 0.48, 0.66, 0.64, and 0.63 for ETG against 0.22, 0.21, 0.28, and 0.25 for LTG. Furthermore, our sample traces the dramatic increase (by a factor of 1.5) in the characteristic size of ETGs from $z\sim0.9$ to $z\sim0.3$, with fitted values of $R_0$ of 3.6 and 5.2 kpc, respectively (see \eq\ref{eq:size-mass}). 
The fits are extended below the mass limits of each sub-sample (where we displayed them with dotted lines), showing agreement with the corresponding incomplete data for ETG and LTG in all redshifts intervals. Therefore the bias due to the mass limit completeness remains limited enough to reassert that, in agreement with many of the previous studies introduced above, these two galaxy populations are characterised by different $M_\ast$--$R_\mathrm{e}$ relations, hence suggesting different regimes of stellar mass assembly. 
However, these limitations prevent us from investigating variations in these relations across the full redshift range.

In the bottom panels of \fg\ref{fig:size-mass}, the low dispersions seen in the $M_\ast$--$R_\mathrm{e}$ cells dominated by either the LTG or ETG samples confirm that the dichotomy between these two populations can be observed in the $M_\ast$--$R_\mathrm{e}$ plane as well. The highest values of dispersions are seen on the cells where LTG and ETG distributions overlap (using their contours from \fg\ref{fig:size-mass}): these corresponds to the transition regions where galaxies transform from the former to the latter, hence a wide range of $n$ are observed there.

\subsection{\label{sc:results-nser-SFR-mass}Morphology in the stellar mass--SFR plane}

There is also a well-known relation between $M_\ast$ and SFR, usually referred to as the star-forming main sequence (SFMS; \citealt{Muzzin2013, Rodighiero2014, Saparre2015, CanoDíaz2016, Santini2017, Donnari2019}), where the SFR is approximately proportional to the $M_\ast$ of the galaxy. Additionally, the relation between galaxy morphology and SFR is also very strong: spiral galaxies show ongoing star-formation, while the majority of elliptical galaxies are quenched, falling below the SFMS in the $M_\ast$--SFR plane (e.g. \citealt{Pozzetti-2010-zCOSMOS-bimod-SFM, Lang-2014-bulge-growth-quenching-CANDELS}). More massive systems are usually quenched at earlier times, following the `downsizing' scenario (e.g. \citealt{Perez-Gonzalez-2008-mass-assembly-history-galaxies, Behroozi-2010-stellar-mass-halo-mass-relation, Thomas-2010-environment-self-regulation-galaxies}). 

We show in \fg\ref{fig:sfr-mass} the $M_\ast$--SFR plane colour-coded by the median $n$ and its scatter in the four redshift bins. Over-plotted are the distinct mass limits at each redshift (vertical dashed lines) for the ETG and LTG galaxies. At intermediate masses ($\sim10^{10}M_\odot$), there is a mix of the two populations, but some of the ETGs may be missing, due to their higher mass completeness limit.
We find a good agreement between our fit to LTG galaxies (solid black line) and the relations of either \citet[dashed line]{Speagle-2014-Main-Sequence} or \citet[dotted line]{Popesso-2023-Main-Sequence} for $z = 0.2 - 0.4$. But for higher redshifts, the mass completeness limits of our sample are such that we do not probe the low-mass part of the SFMS, leading to a very narrow mass interval, which in turns flattens the fit. However, the SFMS from previous studies still goes convincingly through the bulk of our LTG sample. The difference between a selection of LTGs versus ETGs, and star-forming versus quiescent galaxies, as well as different methods to compute stellar masses and SFRs can lead to the observed small differences.
A more thorough investigation of the SFMS using \Euclid Q1 data products, and additional IRAC bands, is performed in \cite{Q1-SP031}. Using a deeper sample ($\HE < 24$) than in the current study, and hence less impacted by selection effects, the authors confirm that their SFMS from \Euclid data agrees with the previously published results \citep[e.g.][]{Popesso-2023-Main-Sequence} at all redshifts from 0.2 to 3.0.

Figure \ref{fig:sfr-mass} shows that the median $n$ values of the galaxies located in the SFMS are around one (corresponding to disc-like LTGs), while larger $n$ values around four dominate the massive quenched distribution. This is in agreement with the trend observed for galaxies in the $0.5<z<2.5$ interval by \cite{Wuyts-2011-SFR-mass-plane-morpho-mode-SF} using HST imaging, and by \cite{Martorano-2025-Sersic-index-mass-SFR} using both HST and JWST imaging. It is worth noting that most of the $\logten(M_\ast/M_\odot) \lesssim 9.5$ galaxies below the SFMS also show low $n$ values ($n\sim1$). We note, however, that the fraction of such objects is small (only 3.8\% of galaxies in $0.2\leq z<0.4$ have $\logten(M_\ast/M_\odot)\lesssim 9.5$ and $\logten(${\rm SFR}$[M_\odot\,\rm{yr}^{-1}]) < 1.5$) and is affected by incompleteness. Due to the higher mass completeness limit for the ETG sample, these objects are more likely to be missed at $\logten(M_\ast/M_\odot)\lesssim 9.5$, which biases the median $n$ towards lower values. Moreover, resolution issues could affect the $n$ estimates at faint magnitudes (see \fg\ref{f7}). Modulo these selection effects, this trend may also suggest that the quenching of the low-mass galaxy population is not so strongly coupled with morphological transformations. Regarding the associated dispersion around the median shown in \fg\ref{fig:sfr-mass} bottom plots, the same observation as for \fg\ref{fig:size-mass} can be made with the highest dispersion values found for the transition region in between LTG and ETG, whereas cells dominated by only one of these populations show comparatively low dispersion.

\cite{Dimauro-2022-bulge-growth} found that the drop-out of the main sequence was occurring for galaxies with $B/T\sim0.2$, demonstrating that quenching was concomitant with morphological transformations, more specifically with the emergence of a prominent bulge (see also \citealt{Lang-2014-bulge-growth-quenching-CANDELS, Bluck-2014-bulge-mass, Bluck-2022-quenching-bulge-disk-ML, Quilley-2022-bimodality}). Even though $n$ is not as valuable a morphological indicator as the $B/T$ (Q25a), we do observe an increase of $n$ as galaxies drop below the SFMS.

However, we also noticed in \fg\ref{fig:sfr-mass} that this increase of $n$ partly occurs for massive galaxies along the SFMS, with an increase from a median $n$ of 1 below $\logten(M_ \ast/M_\odot) \sim 10.5$ to a median of $n=2$ above it. From the bottom panels of \fg\ref{fig:sfr-mass}, we note, however, that the increased dispersion at the massive end of the SFMS indicates a mix of morphologies. This could be interpreted as morphological transformations starting prior to the shutdown of star-formation. The intermediate median $n$ ($\sim 2$) are typical of early spirals or lenticular galaxies with both bulge and disc. This would be in agreement with scenarios in which the growth of the bulge triggers the quenching of galaxies, either by morphological quenching \citep{Martig-2009-morphological-quenching} or AGN feedback \citep[see e.g.][]{Bluck-2022-quenching-bulge-disk-ML, Brownson-2022-quenching-kinematics-bulge-AGN}. We note, however, that the full morphological transition from $n=1$ to $n=4$ is only completed by quenched systems below the SFMS, so we remain careful in our interpretation and do not conclude that morphological transformations occur prior to quenching, but seem to only start before it. Single-S\'ersic models are not tailored to fit galaxies with both a bulge and a disc, and are prone to entirely leaving out low flux bulges, as seen in \fg\ref{fig:sersic-vs-zoobot} bottom-left panel. As a result, they did not allow us to trace the full evolution of galaxy properties along the Hubble sequence, characterised by bulge growth \citep{Quilley-2022-bimodality}, so the transition from late and intermediate to early spirals (and/or lenticular) requires further investigation.

\section{\label{sc:Ccl} Summary and perspectives}

In this study, we have presented the results of S\'ersic fits available within the MER catalogue \citep{Q1-TP004} as part of the Q1 data release of \Euclid.
The fits were performed using two sets of structural parameters, one for the VIS image and another for the three NISP images together.
We explored the distribution of the parameters derived at different magnitude limits, mainly up to \IE$\leq 23$, which is the limit recommended by the EMC to ensure reliable single-S\'ersic fits \citep{Bretonniere-EP26}. 

In \sct\ref{sc:compare-other-methods}, we provided an extensive series of comparisons between the S\'ersic parameters and other morphological measurements. We obtained a convincing consistency between them and all the other morphological products of the MER pipeline. We showed the isophotal and S\'ersic parameters are well correlated, and sources showing concerning differences were identified as spurious, thus further cleaning the sample (\sct\ref{sc:sersic-vs-iso}). We found  $n$ is correlated to the non-parametric concentration measure, as expected (\sct\ref{sc:sersic-vs-CAS}), and the distribution of S\'ersic parameters varies with the morphological properties predicted by \texttt{Zoobot} in a physically meaningful manner (\sct\ref{sc:sersic-vs-Zoobot}). Furthermore, there is overall agreement between the \Euclid S\'ersic fits and those from the HST-based analysis of \cite{Griffith-2012-ACS-catalog-sersic-fits}, with small differences possibly being due to different image resolutions and depths.

A first investigation of galaxy colour gradients resulting from stellar population age and metallicity gradients as well as varying dust content was carried out in \sct\ref{sc:results-color-variation} by measuring the variations of ($R_{\rm e}$) and $n$ with an observing band. On the one hand, we found that $R_{\rm e, NISP}$ values are systemically lower than $R_{\rm e, VIS}$, becoming more prominent for higher $n$. However, these differences disappear by \IE$>21$, for which the bulge and disc components cannot be resolved. On the other hand, $n_{\rm NISP}$ values are systemically higher than $n_{\rm VIS}$, with a slight increase of this difference at fainter magnitudes and a clear decreasing trend with $n_{\rm VIS}$. The presence of a red bulge within a bluer disc can explain both the $R_{\rm e}$ decrease and $n$ increase with redder observing bands, but more specific trends require further investigation based on bulge and disc decomposition to properly decipher how they result from bulge and disc colours and colour gradients. In view of the steeply decreasing variations in the bulge-to-total ratios of nearby spiral galaxies from the UV to the NIR, the absence of variations with redshift of these NISP-to-VIS ratios across the $z=$0.2--1  range and the corresponding rest-frame ranges does not exclude redshift variations. Further study is required to obtain a more detailed understanding.

In order to assert the key role of morphology in galaxy evolution and to demonstrate that the \Euclid S\'ersic parameters enable a wide variety of science cases, we showed how the value of $n$ is connected to the physical parameters of galaxies. This was first done by showing that the colour bimodality of galaxy populations can be seen as a projection of the 2D colour--$n$ bimodality, which allowed us to define robust LTG and ETG sub-samples. These two galaxy populations display markedly different behaviours in their $M_ \ast$--$R_{\rm e}$ relations, ETGs having a steeper slope compared to LTGs, which is indicative of different regimes of $M_ \ast$ assembly. The populations also occupy different loci in the $M_ \ast$--SFR plane, with ETGs at higher masses and further below the main sequence of star-forming galaxies, indicating that morphological transformations occur in parallel to the quenching of galaxies.

The implementation of S\'ersic fits within the \Euclid processing function ensures that S\'ersic parameters will be available for all galaxies observed by \Euclid over its years of operation. Therefore, \Euclid is progressively building the largest morphological catalogue available, with an estimated total of 450\,000\,000 galaxies with \IE$<23$ reliably fitted by the end of the mission. With this first study of S\'ersic parameters on the Q1 area, we have demonstrated the quality, robustness, and usefulness of these data to study galaxy evolution, thus paving the way to many more analyses that will benefit from leveraging this information. The EWS and EDS will enable the study of morphological transformations across the large-scale structures and over cosmic times. However, due to the limitations of a single-component model that we highlighted here and which are more extensively discussed in Q25a, further work remains to be done to obtain in-depth characterisation of the morphology of galaxies. 

One pressing issue, and a direct follow-up of S\'ersic fits, is to perform reliable bulge and disc decomposition on galaxies from Q1. This would allow one to investigate the history of galaxy morphology beyond the ETG-LTG dichotomy and hopefully trace how galaxies evolved and transformed to build up the present-day Hubble sequence. Such an analysis would also raise new challenges in terms of methodology to account for varying selection effects for different morphological types or differential chromatic effects impacting bulges and discs \citep{Papaderos-2023-chromatic-SB-modulation}. Additionally, more complex models could also be built for accordingly more specific studies, such as the addition of bars to study the barred galaxies identified in \Euclid \citep{Q1-SP043} or of point sources to account for AGN contribution.

\begin{acknowledgements}  
LQ acknowledges funding from the CNES postdoctoral fellowship program.
\AckEC  
\AckQone\,
Based on data from UNIONS, a scientific collaboration using three Hawaii-based telescopes: CFHT, Pan-STARRS, and Subaru \url{https://www.skysurvey.cc/}.
Based on data from the Dark Energy Camera (DECam) on the Blanco 4-m Telescope at CTIO in Chile \url{https://www.darkenergysurvey.org/}.
\end{acknowledgements}

\bibliographystyle{aa}
\bibliography{biblio}

\begin{appendix}

\onecolumn

\section{\label{sc:appendix-photometry}Comparison of model photometry with isophotal and adaptive aperture magnitudes}

\begin{figure*}[h!]
\centering
\includegraphics[width=0.45\columnwidth]{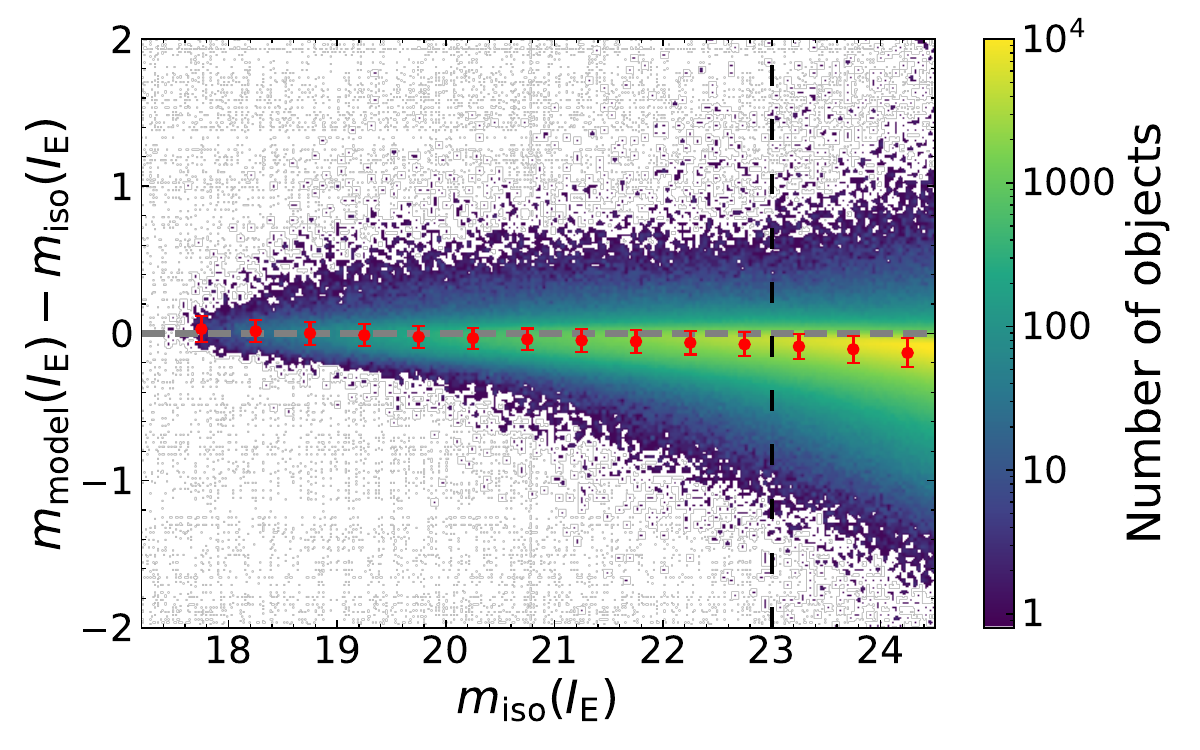}
\includegraphics[width=0.45\columnwidth]{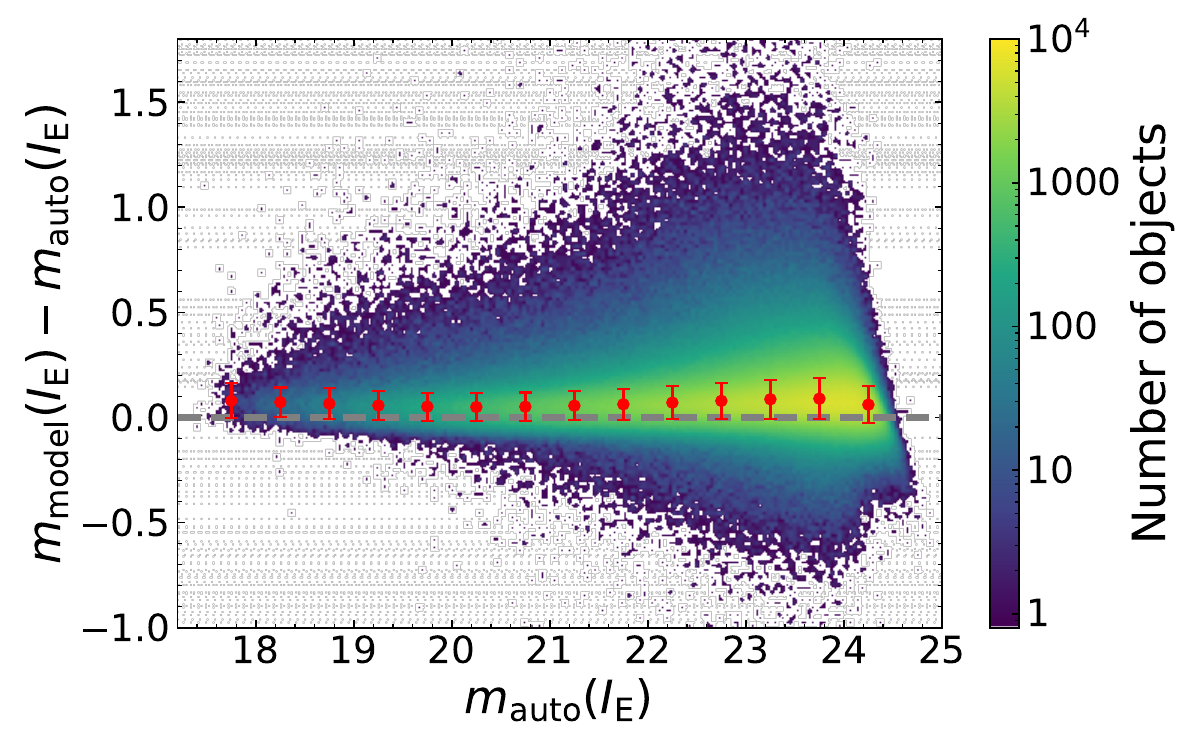}
\includegraphics[width=0.45\columnwidth]{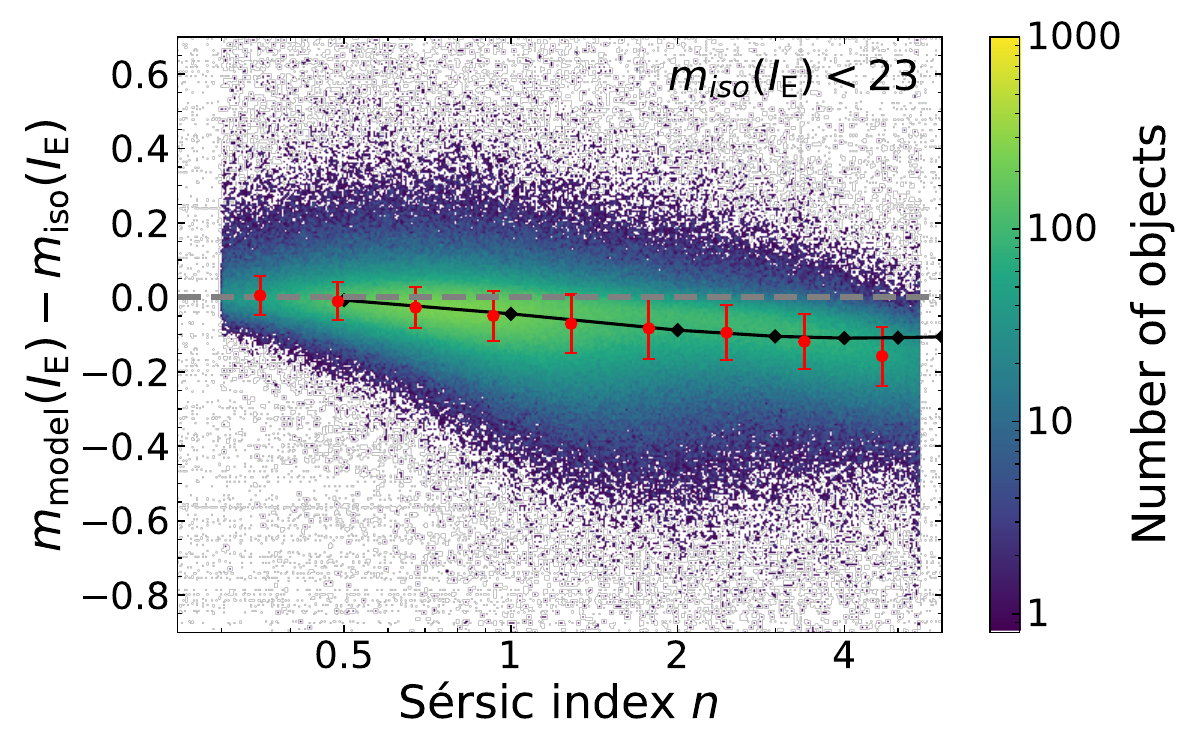}
\includegraphics[width=0.45\columnwidth]{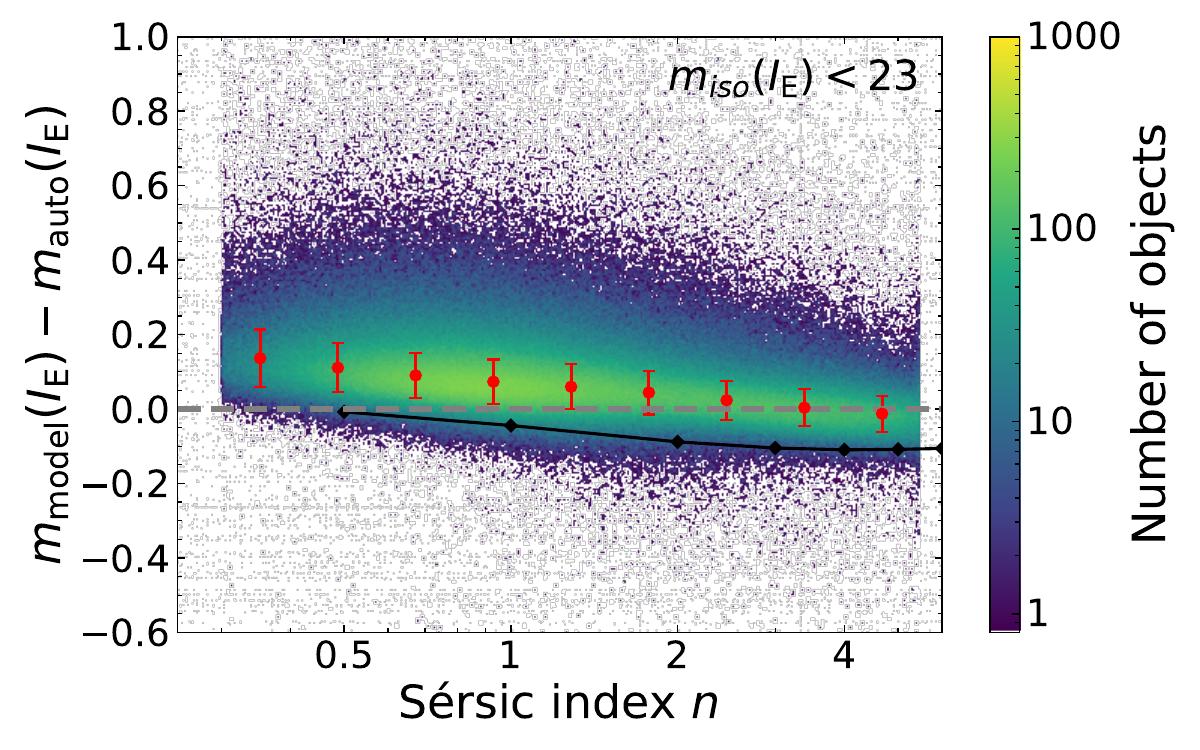}
\caption{Difference between the isophotal $m_{\rm iso}$ and model magnitude $m_{\rm model}$ (left) and between the adaptive aperture $m_{\rm auto}$ and model magnitude (right) as a function of either the magnitude compared to model photometry (top) or $n$ (bottom). Red points and error bars indicate median differences and associated dispersion around them. The black line indicates the mathematical prediction of $m_{\rm model} - m_{\rm auto}$ for a pure S\'ersic profile, and unexpectedly matches the values of $m_{\rm model} - m_{\rm iso}$ whereas it is shifted by 0.15 mag below the measured $m_{\rm model} - m_{\rm auto}$ that it aims at predicting.}
\label{fig:mag-comparison}
\end{figure*}

When fitting S\'ersic profiles, a model magnitude is computed for each fitted galaxy, which can be compared with other methods to determine photometry. Figure\,\ref{fig:mag-comparison} shows the difference between the model photometry $m_{\rm model}$ computed in the \IE band and both the isophotal magnitude $m_{\rm iso}$ (corresponding to {\tt FLUX$\_$SEGMENTATION} in the MER catalogue) and the adaptive aperture photometry within 2.5 times the Kron radius $m_{\rm auto}$ (corresponding to {\tt FLUX$\_$DETECTION$\_$TOTAL} in the MER catalogue), as a function of either magnitude or as a function of the $n$ measured in the \IE band, in the top and bottom panels respectively. Median differences in bins of either magnitude or $n$ are displayed as red points, and the error bars indicate $\pm 1 \sigma$ ($\sigma$ being the 0.68 quantile of the absolute differences to the median). The top-left panel shows the difference between the model and isophotal magnitudes, whose median value is 0.00 in the two brightest magnitude intervals, which then continuously decreases to $-0.13$ in the faintest magnitude interval. All differences for \IE$<23$ are below 1$\sigma$. The dispersion is very stable, with values between 0.07 and 0.10 across the seven magnitudes of $m_{\rm iso}$, and between 0.07 and 0.08 for \IE$<23$. The number of outliers with large differences increase with fainter magnitudes as a result of the increasing number of sources and decreasing signal-to-noise ratio. The top-right panel shows the difference between the model and adaptive aperture photometry, whose medians are all positive, with values ranging from 0.05 to 0.09, hence of the same order as the median differences between model and isophotal magnitudes. All median differences are below the 1$\sigma$ level with again a very stable dispersion between 0.07 and 0.10.

For both magnitude differences, the bottom panels of \fg\ref{fig:mag-comparison} indicate that while there is no magnitude effect, there is a morphological trend of differences decreasing with $n$. The median differences between model and isophotal magnitudes start at zero for the lowest values of $n$ and decreases continuously to reach $-0.15$ for $n > 4$. The median differences between model and adaptive aperture magnitudes follow a very similar decrease, but with an offset, since values decrease from $0.14$ to $-0.01$ for the lowest to highest bins of $n$. The black diamond symbols and solid line in both plots show the difference computed mathematically for a pure circular S\'ersic profile between its total magnitude and the adaptive aperture magnitude computed within 2.5 times the Kron radius, which only depends on $n$ (taken from \citealt{Graham-2005-sersic-considerations}). Because the fitted galaxies are not perfectly approximated by S\'ersic profiles, we would expect the difference between the model and adaptative aperture photometry to follow this theoretical behaviour, with some dispersion around it. However, we observe that this prediction matches instead the median difference between isophotal and model magnitudes. The behaviour of decreasing median differences for higher $n$ remains for the difference between the model and adaptative aperture photometry, but with a 0.15 magnitude offset.

\section{\label{sc:appendix-n-colour-mass}Colour--stellar mass--S\'ersic index space}

\begin{figure*}[h!]
\centering
\includegraphics[width=\textwidth]{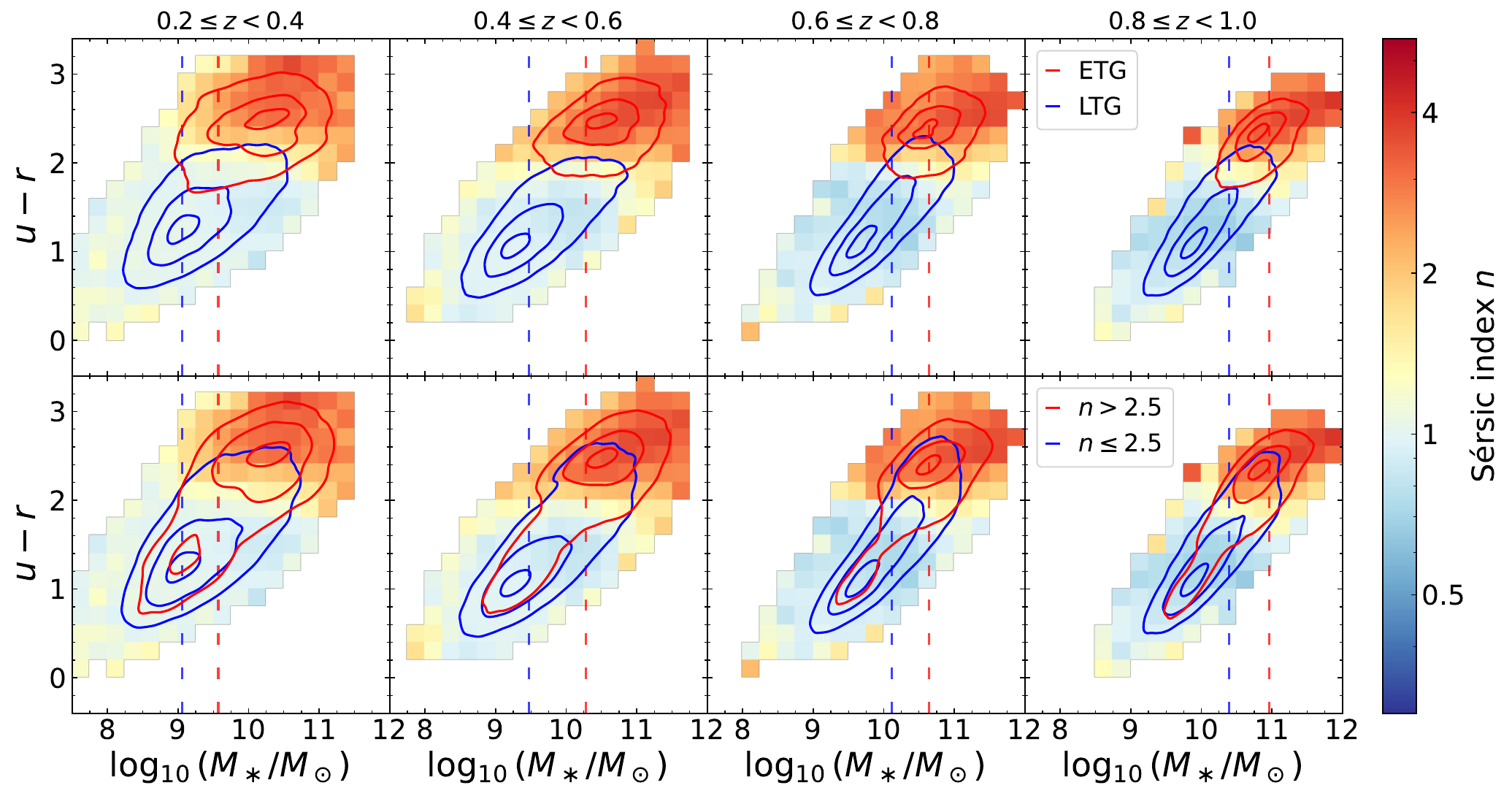}
\caption{Redshift evolution of $M_ \ast$ versus ($u-r$) colour, colour-coded by the median $n$ in each cell. In the top row, we over-plotted density contours at the 10\%, 50\%, and 90\% levels, for the ETG and LTG samples, as in previous figures, whereas the bottom row the blue and red contours correspond to galaxies with $n\leq2.5$ and $n >2.5$ respectively, displaying significant overlaps in both colour and mass. In all panels the vertical dashed lines indicate the 90\% mass completeness limit of the ETG and LTG samples. The colour--concentration bimodality evidenced in \fg\ref{fig:colour-n-bimod} of \sct\ref{fig:colour-n-bimod} is also present in the $M_ \ast$ --colour plane.}
\label{fig:colour-mass-diagram}
\end{figure*}

We show in \fg\ref{fig:colour-mass-diagram} the $(u-r)$ colour versus $M_ \ast$, colour-coded with the median $n$ as in \fgs\ref{fig:size-mass} and \ref{fig:sfr-mass}. This allowed us to see that the colour--$n$ bimodality can also be extended as a 3D colour--$M_ \ast$--$n$ bimodality, with red galaxies having steeper profiles since they are more massive than their bluer counterparts. The top row shows the density contours at the 10\%, 50\%, and 90\% levels for the ETG and LTG sub-samples defined in \sct\ref{sc:results-nser-colours-bimod}, whereas the bottom row uses a cut at $n_{\rm lim} = 2.5$ in order to divide the sample into two new classes. This leads to a more spread out distribution of the two new sub-samples, with significant overlaps at all redshifts. This illustrates that it is preferable to not use $n$ alone to separate galaxy populations but rather use joint conditions, such as colour or stellar mass. We checked that using a different criterion on $n$ only, such as using a different $n_{\rm lim}$ or two limits, i.e. having $n\leq n_1$ and $n\geq n_2$ with $n_1 < n_2$, leads to similar results as shown in \fg\ref{fig:colour-mass-diagram} with $n_{\rm lim} = 2.5$. The suggested selection of \eq(\ref{eq:colour-n-separation}), or a similar method adapting the presented process to the available data, therefore allows the different behaviours of early and late galaxies to be highlighted more significantly.

\end{appendix}

\end{document}